\documentclass[twocolumn, longauth]{aastex631}

\usepackage{graphicx}
\usepackage{txfonts}

\usepackage{subfigure}
\usepackage{natbib}
\usepackage{threeparttable}
\usepackage{gensymb}

\usepackage{longtable}
\DeclareUnicodeCharacter{2212}{-}
\usepackage{comment}

\begin{document}

\title{A Swarm of WASP Planets: Nine giant planets identified by the WASP survey}

\correspondingauthor{Nicole Schanche}
\email{nicole.e.schanche@nasa.gov}

\author{N.~Schanche}
\affiliation{Department of Astronomy, University of Maryland, College Park, MD 20742.}
\affiliation{NASA Goddard Space Flight Center, 8800 Greenbelt Road, Greenbelt, MD 20771, USA}

\author{G.~Hébrard} 
\affiliation{Institut d'astrophysique de Paris, UMR7095 CNRS, Universit\'e Pierre \& Marie Curie, 98bis boulevard Arago, 75014 Paris, France}
\affiliation{Observatoire de Haute-Provence, CNRS, Universit\'e d'Aix-Marseille, 04870 Saint-Michel-l'Observatoire, France}

\author{K.G.~Stassun} 
\affiliation{Department of Physics \& Astronomy, Vanderbilt University, 6301 Stevenson Center Ln., Nashville, TN 37235, USA}

\author{B.~Hord} 
\affiliation{NASA Goddard Space Flight Center, 8800 Greenbelt Road, Greenbelt, MD 20771, USA}
\affiliation{NASA Postdoctoral Program Fellow}

\author{K.~Barkaoui} 
\affiliation{Astrobiology Research Unit, Universit\'e de Li\`ege, All\'ee du 6 ao\^ut 19, Li\`ege, 4000, Belgium}
\affiliation{Department of Earth, Atmospheric and Planetary Science, Massachusetts Institute of Technology, 77 Massachusetts Avenue, Cambridge, MA 02139, USA}
\affiliation{Instituto de Astrof\'isica de Canarias (IAC), Calle V\'ia L\'actea s/n, 38200, La Laguna, Tenerife, Spain}

\author{A.~Bieryla} 
\affiliation{Center for Astrophysics | Harvard \& Smithsonian, 60 Garden Street, Cambridge, MA, 02138, USA}

\author{D.~Ciardi}
\affiliation{NASA Exoplanet Science Institute-Caltech/IPAC, Pasadena, CA 91125, USA}

\author{K.A.~Collins}
\affiliation{Center for Astrophysics | Harvard \& Smithsonian, 60 Garden Street, Cambridge, MA, 02138, USA}

\author{A.~Collier Cameron}
\affiliation{Centre for Exoplanet Science, SUPA, School of Physics and Astronomy, University of St Andrews, St Andrews KY16 9SS, UK}

\author{J.~Hartman} 
\affiliation{Department of Astrophysical Sciences, Princeton University, 4 Ivy Lane, Princeton, NJ 08544, USA}

\author{N.~Heidari}
\affiliation{Institut d'astrophysique de Paris, UMR7095 CNRS, Universit\'e Pierre \& Marie Curie, 98bis boulevard Arago, 75014 Paris, France}

\author{C.~Hellier} 
\affiliation{Astrophysics Group, Keele University, Staffordshire ST5 5BG, UK}

\author{S.B.~Howell} 
\affiliation{NASA Ames Research Center, Moffett Field, CA 94035, USA}

\author{M.~Lendl}
\affiliation{Observatoire de Gen\`eve, Département d’Astronomie, Universit\'e de Gen\`eve, Versoix, Switzerland}

\author{J.~McCormac} 
\affiliation{Department of Physics, University of Warwick, Gibbet Hill Road, Coventry CV4 7AL, UK}
\affiliation{Centre for Exoplanets and Habitability, University of Warwick, Gibbet Hill Road, Coventry CV4 7AL, UK}

\author{K.K.~McLeod} 
\affiliation{Department of Astronomy, Wellesley College, Wellesley, MA 02481, USA}

\author{H.~Parviainen}
\affiliation{Departamento de Astrofísica, Universidad de La Laguna, E-38206 La Laguna, Tenerife, Spain}
\affiliation{Instituto de Astrof\'isica de Canarias (IAC), Calle V\'ia L\'actea s/n, 38200, La Laguna, Tenerife, Spain}

\author{D.~Radford} 
\affiliation{Brierfield Observatory Bowral, NSW 2576, Australia}

\author{A.S.~Rajpurohit} 
\affiliation{Astronomy \& Astrophysics Division, Physical Research Laboratory, Ahmedabad 380009, India}

\author{H.~Relles}
\affiliation{Center for Astrophysics | Harvard \& Smithsonian, 60 Garden Street, Cambridge, MA, 02138, USA}

\author{R.~Sharma}
\affiliation{Astronomy \& Astrophysics Division, Physical Research Laboratory, Ahmedabad 380009, India}
\author{S.~Baliwal}
\affiliation{Astronomy \& Astrophysics Division, Physical Research Laboratory, Ahmedabad 380009, India}
\affiliation{Department of Physics, Indian Institute of Technology, 382335 Gandhinagar, India}


\author{G.~Bakos}
\affiliation{Department of Astrophysical Sciences, Princeton University, 4 Ivy Lane, Princeton, NJ 08544, USA}

\author{S.C.C.~Barros}
\affiliation{Instituto de Astrofísicae Ciências do Espaço, Universidade do Porto, CAUP, Rua das Estrelas, 4150-762 Porto, Portugal}

\author{F.~Bouchy} 
\affiliation{Observatoire de Gen\`eve, Département d’Astronomie, Universit\'e de Gen\`eve, Versoix, Switzerland}

\author{A.~Burdanov} 
\affiliation{Department of Earth, Atmospheric and Planetary Science, Massachusetts Institute of Technology, 77 Massachusetts Avenue, Cambridge, MA 02139, USA}

\author{P.A.~Budnikova} 
\affiliation{Lomonosov Moscow State University, Sternberg Astronomical Institute, Universitetsky pr., 13, 119992, Moscow, Russian Federation}

\author{A.~Chakaraborty} 
\affiliation{Astronomy \& Astrophysics Division, Physical Research Laboratory, Ahmedabad 380009, India}

\author{C.~Clark}
\affiliation{Jet Propulsion Laboratory, California Institute of Technology, Pasadena, CA 91109, USA}
\affiliation{NASA Exoplanet Science Institute-Caltech/IPAC, Pasadena, CA 91125, USA}

\author{Laetitia Delrez}
\affiliation{Astrobiology Research Unit, Universit\'e de Li\`ege, All\'ee du 6 ao\^ut 19, Li\`ege, 4000, Belgium}
\affiliation{Space Sciences, Technologies and Astrophysics Research (STAR) Institute, Universit\'e de Li\`ege, All\'ee du 6 Ao\^ut 19C, B-4000 Li\`ege, Belgium.}
\affiliation{Institute of Astronomy, KU Leuven, Celestijnenlaan 200D, 3001 Leuven, Belgium}

\author{O.D.S.~Demangeon}
\affiliation{Instituto de Astrofísicae Ciências do Espaço, Universidade do Porto, CAUP, Rua das Estrelas, 4150-762 Porto, Portugal}

\author{R.F.~Díaz}
\affiliation{International Center for Advanced Studies (ICAS) and ICIFI (CONICET), ECyT-UNSAM, Campus Miguelete, 25 de Mayo y Francia, (1650) Buenos Aires, Argentina}

\author{J.~Donnenfield}
\affiliation{Department of Astrophysical Sciences, Princeton University, 4 Ivy Lane, Princeton, NJ 08544, USA}
\affiliation{Division of Plastic Surgery, Department of Surgery, University of California, Los Angeles, CA, USA}

\author{M.~Everett} 
\affiliation{NSF’s National Optical-Infrared Astronomy Research Laboratory, 950 N. Cherry Ave., Tucson, AZ 85719, USA}

\author{M.~Gillon}
\affiliation{Astrobiology Research Unit, Universit\'e de Li\`ege, All\'ee du 6 ao\^ut 19, Li\`ege, 4000, Belgium}

\author{C.~Hedges}
\affiliation{Center for Space Sciences and Technology, University of Maryland, Baltimore County, Baltimore, MD 21250}
\affiliation{NASA Goddard Space Flight Center, 8800 Greenbelt Road, Greenbelt, MD 20771, USA}

\author{J.~Higuera} 
\affiliation{NSF’s National Optical-Infrared Astronomy Research Laboratory, 950 N. Cherry Ave., Tucson, AZ 85719, USA}

\author{E.~Jehin}
\affiliation{Space Sciences, Technologies and Astrophysics Research (STAR) Institute, Universit\'e de Li\`ege, All\'ee du 6 Ao\^ut 19C, B-4000 Li\`ege, Belgium.}

\author{J.M.~Jenkins}
\affiliation{NASA Ames Research Center, Moffett Field, CA 94035, USA}

\author{F.~Kiefer}
\affiliation{LESIA, Observatoire de Paris, Universit\'e PSL, CNRS, Sorbonne Universit\'e, Universit\'e de Paris, 5 place Jules Janssen, 92195 Meudon, France}

\author{D.~Laloum}
\affiliation{Société Astronomique de France, 3 Rue Beethoven, 75016 Paris, France}

\author{M.~Lund}
\affiliation{NASA Exoplanet Science Institute-Caltech/IPAC, Pasadena, CA 91125, USA}

\author{P.~Magain}
\affiliation{Institut d’Astrophysique et de Géophysique, Universit\'e de Li\'ege, 17 All\'ee du 6 Ao\^ut, Bat. B5C, 4000 Li\`ege, Belgium}

\author{P.~Maxted}
\affiliation{Astrophysics Group, Keele University, Staffordshire ST5 5BG, UK}

\author{I.~Mireles}
\affiliation{Department of Physics and Astronomy, University of New Mexico, 1919 Lomas Blvd NE, Albuquerque, NM 87131, USA}

\author{K.J.~Nikitha}
\affiliation{Astronomy \& Astrophysics Division, Physical Research Laboratory, Ahmedabad 380009, India}

\author{C.~Opitom}
\affiliation{Institute for Astronomy, University of Edinburgh, Edinburgh, UK}

\author{Y.G.~Patel} 
\affiliation{NSF’s National Optical-Infrared Astronomy Research Laboratory, 950 N. Cherry Ave., Tucson, AZ 85719, USA}

\author{M.~Rose}
\affiliation{NASA Ames Research Center, Moffett Field, CA 94035, USA}

\author{S.G.~Sousa}
\affiliation{Instituto de Astrofísicae Ciências do Espaço, Universidade do Porto, CAUP, Rua das Estrelas, 4150-762 Porto, Portugal}

\author{I.A.~Strakhov}
\affiliation{Lomonosov Moscow State University, Sternberg Astronomical Institute, Universitetsky pr., 13, 119992, Moscow, Russian Federation}

\author{P.~Strøm}
\affiliation{Department of Physics, University of Warwick, Gibbet Hill Road, Coventry CV4 7AL, UK}

\author{A.~Tuson}
\affiliation{Center for Space Sciences and Technology, University of Maryland, Baltimore County, Baltimore, MD 21250}
\affiliation{NASA Goddard Space Flight Center, 8800 Greenbelt Road, Greenbelt, MD 20771, USA}

\author{R.~West}
\affiliation{Astronomy \& Astrophysics Group, Department of Physics, University of Warwick, Gibbet Hill Road, Coventry CV4 7AL, UK}

\author{J.~Winn}
\affiliation{Department of Astrophysical Sciences, Princeton University, 4 Ivy Lane, Princeton, NJ 08544, USA}

    \date{Received ; accepted }

\begin{abstract}
{The Wide Angle Search for Planets (WASP) survey provided some of the first transiting hot Jupiter candidates. With the addition of the Transiting Exoplanet Survey Satellite (TESS), many WASP planet candidates have now been revisited and given updated transit parameters. Here we present 9 transiting planets orbiting FGK stars that were identified as candidates by the WASP survey and measured to have planetary masses by radial velocity measurements. Subsequent space-based photometry taken by TESS as well as ground-based photometric and spectroscopic measurements have been used to jointly analyze the planetary properties of WASP-102\,b, WASP-116\,b, WASP-149\,b WASP-154\,b, WASP-155\,b, WASP-188\,b, WASP-194\,b/HAT-P-71\,b, WASP-195\,b, and WASP-197\,b. These planets have radii between 0.9 R$_{\rm{Jup}}$ and 1.4 R$_{\rm{Jup}}$, masses between 0.1 M$_{\rm{Jup}}$ and 1.5 M$_{\rm{Jup}}$, and periods between 1.3 and 6.6 days.}
\end{abstract}

\keywords{Planets and satellites: detection}

\section{Introduction}\label{sec:intro}
Scorching-hot, massive planets in tight orbits around their stars were once the realm of science fiction. However, by the early 2000s, exoplanet surveys had begun to discover many of these `hot Jupiter' systems. While these planets are comparatively rare, their frequent, deep transits make them accessible to wide field ground-based surveys. In fact, there are now more than 600 known planets larger than half the radius of Jupiter with orbits under 10 days. Even with a growing sample of these strange planets, their formation history remains a mystery - did these planets form in their current location or further beyond the snow line and migrate inwards to their present location? While observing an individual planet's history is not possible, we can explore the influence of different physical drivers that shape planetary formation and evolution through the careful study of population-level demographics \citep[see][for a review]{Fortney2021}. 

One early contributor to the sample of hot Jupiter planets was the Wide Angle Search for Planets (WASP) survey \citep{Pollacco2006}. From locations in the northern and southern hemisphere, the full sky was monitored for stars showing transit dips. While this strategy provides a vast sample of stars to search, it also presents challenges in terms of data processing and follow-up efforts. Due to the combination of systematic noise, scatter, and pixel size leading to blending, many genuine transit signals appear inconclusive, while many false positives are flagged for follow-up. Nonetheless, the WASP survey has led to the discovery of nearly 200 planets including the actively in-spiraling WASP-12\,b \citep{Hebb2009, Yee2020}, the close-in planet orbiting a giant $\delta$-Scuti star WASP-33\,b \citep{CollierCameron2010}, and the planet with a tail WASP-69\,b \citep{Anderson2014, Tyler2024}. In the discovery process, WASP has also identified 1041 false positives in the northern hemisphere alone \citep{Schanche2019FalsePositives}. 

The launch of the Transiting Exoplanet Survey Satellite \citep[TESS;][]{TESS} in April 2018 provided an opportunity to better characterize existing planetary systems as well as rule out false alarms due to systematics for the remaining WASP planet candidates without the need to schedule and coordinate extensive follow-up from the ground. TESS' $\sim$27-day observing windows, referred to as Sectors, provide longer continuous observing intervals than is possible to achieve from the ground, thereby alleviating the challenge of observing the transits with drifting periods and ephemerides. 

In this paper, we present 9 planets that 1) were initially identified as candidates in the WASP survey, 2) have sufficient radial velocity followup to establish a planetary mass, 3) have not been published in an accepted refereed paper, and 4) have been identified as TESS Objects of Interest (TOIs). These planets fall in the hot Jupiter regime, with periods ranging from 1.3-6.6 days and radii roughly that of Jupiter. In Section \ref{sec:obs} we highlight the wide array of observations used to characterize the objects, while in Section \ref{sec:stellar} we describe the methods used to refine the properties of the host stars. Section \ref{sec:fit} presents the final models that jointly fit the transit and radial velocity data in order to characterize the planets. Section \ref{sec:discussion}
provides discussion of the new planets in the context of the exoplanet population at large. Finally, we provide a summary of the work in Section \ref{sec:conclusion}.


\section{Observations} \label{sec:obs}
All planets reported here were discovered via their transit signals. However, further observations were obtained to establish the planetary interpretation of the data and rule out potential false positives. In this section we describe the facilities used to obtain photometric (Sec. \ref{sec:photometry}), spectroscopic (Sec. \ref{sec: spectroscopy}), and high resolution imaging (Sec. \ref{sec:highres}) data. Many of these contributions were made by collaborators in the Tess Follow-up Observing Program (TFOP).

\subsection{Photometry} \label{sec:photometry}
All planets presented here were originally identified as transit candidates in WASP data. In addition, TESS observed each star for one or more sectors. A variety of additional ground-based data was taken to further constrain the transit timing and depth, and when possible to confirm that a consistent planet-star radius ratio is measured at different wavelengths in order to exclude eclipsing binary systems. A full description of photometric data available for each star can be found in Tables \ref{tab:follow_up_obs_tess} (TESS) and \ref{tab:follow_up_obs_phot} (ground facilities). Additional information on the observation of each target can be found in Section \ref{sec:fit}.

\subsubsection{WASP}
With regular operations starting in 2006, WASP was among the first ground-based surveys dedicated to searching for exoplanets via the transit method. To achieve a large sky coverage, the WASP consortium consists of instruments at two observatory sites. The northern skies are surveyed by SuperWASP, located at the Observatorio del Roque de los Muchachos on La Palma, while the southern skies are probed by WASP-South at the Sutherland station of the South African Astronomical Observatory. The telescopes at each site are composed of eight commercial camera lenses (Canon 200mm f/1.8) with 2k x 2k E2V CCD cameras. 

Once WASP data are collected, images are processed to provide lightcurves for all stars in the field. In this study, we use the ORion transit search Combining All data on a given target with TAMuz and TFA decorrelation (ORCA\_TAMTFA) product, which as the lengthy name suggests, removes common patterns of systematic error using a combination of the Trend Filtering Algorithm \citep[TFA;][]{Kovacs2005} and the SysRem algorithm \citep{Tamuz2005}. A Box-Least-Squares \citep[BLS;][]{Kovacs2002} method is then applied to search for transit signals in the detrended lightcurves. Originally, all candidates were searched for by eye, but later a machine learning model was applied to the lightcurves \citep{Schanche2019_machine_learning}, which highlighted the candidates WASP-194\,b, WASP-195\,b, and WASP-197\,b as strong candidates for further charactarization. 

\subsubsection{TESS} 
TESS, launched in April of 2018, is a space-based all-sky survey with a primary goal to search for exoplanet transits around nearby, bright stars. TESS' four broad band, red-sensitive cameras stare at a 24\degree\, $\times$ 90\degree\, strip of the sky in ~27-day blocks, called Sectors. The spacecraft then reorients itself to point at another patch of sky. In the 6 years since launch, TESS has surveyed more than 95\% of the sky, and will continue to fill in the remaining observational gaps in future Sectors. 

This observing strategy makes TESS a great compliment to WASP's legacy. For WASP targets already characterized by existing follow-up observations, the new TESS observations help to refine the orbital parameters of the system. For the candidates that still require additional observations, TESS' near-all sky coverage provided the ability to quickly search for corresponding transit signals and identify astrophysical or systematic false alarms

The 9 planetary systems presented in this paper were flagged as planet candidates in the WASP data archive and were also independently identified as Targets of Interest (TOIs) by the TESS Science Office \citep{Guerrero2021}. Four of the stars have `postage stamp' data from TESS, meaning that 120-second cadence, systematic error-corrected light curves produced by the Science Processing Operations Center \citep[SPOC;][]{Jenkins2016} are available. The remaining 5 stars were observed in the TESS Full Frame Images (FFIs) with a cadence of 1800-s (primary mission), 600-s (first mission extension), or 200-s (second mission extension). While the SPOC did not automatically extract lightcurves for these 5 stars, there are a number of community-created High Level Science Products (HLSPs) available that produce detrended lightcurves from FFI data. In this work, we use lightcurves produced by either the Quick-Look Pipeline 
\citep[QLP;][]{Huang2020a, Huang2020b, Kunimoto2022QLP} or the TESS-SPOC \citep{Caldwell2020} HLSPs, which are available at MAST (See Table \ref{tab:follow_up_obs_tess}). When FFI lightcurves were available with multiple cadences, we chose to use only Sectors with the shortest cadence available.

\begin{table*}[]
    \centering
    \caption{Summary of TESS observations used for analysis. When possible, 120-second PDCSAP lightcurves produced by the TESS pipeline were used \citep{Stumpe2012, Stumpe2014, Smith2012}. When the star was observed only in FFIs, lightcurves produced by the High Level Science Products TESS-SPOC or QLP were used, as indicated in the table. }
    \begin{tabular}{l c c c}
    \hline
    \hline
         Target & Sector & Cadence (s) & Source \\
         \hline
         WASP-102/TOI-6170          & 56                             & 200 & TESS-SPOC \\
         WASP-116/TOI-4672          & 31                             & 600 & TESS-SPOC \\ 
         WASP-149/TOI-6101          & 61                             & 120 & SPOC \\
         WASP-154/TOI-5288          & 42                             & 600 & TESS-SPOC \\
         WASP-155/TOI-6135          & 56                             & 200 & QLP   \\
         WASP-188/TOI-5190          & 40, 53, 54                     & 200 & TESS-SPOC  \\
         WASP-194/HAT-P-71/TOI-3791 & 40, 41, 50, 54, 55, 56, 57, 60 & 120 & SPOC  \\
         WASP-195/TOI-4056          & 50, 52                         & 120 & SPOC \\
         WASP-197/TOI-5385          & 48                             & 120 & SPOC \\

         \hline
    \end{tabular}

    \label{tab:follow_up_obs_tess}
\end{table*}

\begin{table*}[]
    \centering
    \caption{Summary of ground-based photometric time-series observations for the 9 planets. See text for further details on these observations.  }
    \begin{tabular}{l c c c c }
    \hline
    \hline
         Target & Filter & Facility & Date & Transit Coverage  \\
         \hline
        WASP-102/TOI-6170           & broadband      & WASP        & 2009-07-14 - 2011-11-10 & --    \\
         --                         & blue blocking  & TRAPPIST    & 2013-08-13 & Full    \\
         --                         & \textit{r-Gunn}         & EulerCam    & 2013-08-13 & Full    \\
         --                         & blue blocking  & TRAPPIST    & 2013-09-20 & Full    \\
         --                         & \textit{I$_C$}          & EulerCam    & 2013-09-20 & Full    \\
         --                         & blue blocking  & TRAPPIST    & 2013-10-09 & Full    \\
         WASP-116/TOI-4672          & broadband      & WASP        & 2008-07-30 - 2010-12-27 & -- \\
         --                         & blue-blocking  & TRAPPIST    & 2013-11-05 & Ingress \\
         --                         & NGTS           & EulerCam    & 2013-11-05 & Ingress \\
         --                         & blue-blocking  & TRAPPIST    & 2013-11-25 & Ingress   \\
         --                         & NGTS           & EulerCam    & 2013-11-25 & Ingress \\
         --                         & SDSS \textit{i'} & LCO/SAAO    & 2023-10-15 & Ingress \\
         --                         & SDSS \textit{i'} & LCO/HAL     & 2023-10-22 & Ingress \\
         WASP-149/TOI-6101          & broadband      & WASP        & 2009-11-20 - 2012-03-31 & -- \\
         --                         & Sloan \textit{z'}       & TRAPPIST    & 2015-02-06 & Full    \\
         --                         & \textit{z-Gunn} & EulerCam    & 2015-04-07 & Full    \\
         --                         & Sloan \textit{z'}       & TRAPPIST    & 2015-05-05 & Full    \\
         --                         & Sloan \textit{z'}       & TRAPPIST    & 2015-11-26 & Full    \\
         --                         & \textit{z-Gunn} & EulerCam    & 2015-12-20 & Full    \\
         WASP-154/TOI-5288          & broadband      & WASP        & 2008-06-06 - 2010-10-28 & -- \\
         --                         & NGTS           & EulerCam    & 2016-08-04 & Full    \\
         --                         & \textit{I}              & NITES       & 2016-08-08 & Full    \\
         --                         & blue-blocking  & TRAPPIST    & 2015-10-08 & Full    \\
         --                         & Sloan \textit{z'}       & TRAPPIST    & 2017-07-13 & Full    \\
         --                         & SDSS \textit{i'} & LCO/CTIO    & 2022-06-21 & Full    \\
         --                         & \textit{R}              & Brierfield  & 2022-06-13 & Full    \\
         WASP-155/TOI-6135          & broadband      & WASP        & 2004-05-25 - 2007-11-22 & -- \\
         --                         & none           & NITES       & 2016-08-11 & Full    \\
         --                         & \textit{g', r', zs}     & MuSCAT2     & 2019-08-18 & Full    \\
         WASP-188/TOI-5190          & broadband      & WASP        & 2004-05-14 - 2010-08-24 & -- \\
         --                         & \textit{g', r', i', zs} & MuSCAT2     & 2018-06-10 & Ingress \\
         --                         & Sloan \textit{i'} & KeplerCam   & 2022-04-25 & Egress  \\
         WASP-194/HAT-P-71/TOI-3791 & broadband      & WASP        & 2007-05-10 - 2010-09-22 & -- \\
         --                         & \textit{Sloan r', Cousins R$_C$ }& HATNet      & 2008-08-06 - 2012-12-20 & -- \\
         --                         & CBB            & OPM         & 2021-07-04 & Full    \\
         --                         & \textit{g', r', i', zs} & MuSCAT2     & 2021-07-04 & Full    \\
         --                         & \textit{Sloan i'} & KeplerCam   & 2017-05-24 & Full    \\
         WASP-195/TOI-4056          & broadband      & WASP        & 2007-03-30 - 2011-08-04 & --  \\
         --                         & Sloan \textit{r'} & Whitin      & 2022-06-06 & Full    \\
         WASP-197/TOI-5385          & broadband      & WASP        & 2006-12-22 - 2007-05-05  & -- \\

         \hline
    \end{tabular}

    \label{tab:follow_up_obs_phot}
\end{table*}

\subsubsection{HATNet}
WASP-194/HAT-P-71\,b was independently identified as a candidate transiting planet system by the Hungarian-made Automated Telescope Network (HATNet) project \citep{HATNet} based on time-series observations gathered by all six of the instruments in the network. Four of these instruments are located at Fred Lawrence Whipple Observatory in Arizona, while the other two are located at Mauna Kea Observatory in Hawaii.  

The star was observed in two separate HATNet fields: G081, and G115. A total of 11,124 observations of WASP-194/HAT-P-71 in field G081 were gathered through a Sloan $r^{\prime}$ filter between 2012 July 20 and 2012 December 20 using an exposure time of 3\,min. Field G115 was observed using both a Cousins $R_{C}$ filter and a Sloan $r^{\prime}$ filter. A total of 2298 $R_{C}$ observations were obtained between 2008 August 6 and 2008 September 14 using an exposure time of 5\,min, while 7141 $r^{\prime}$ observations were obtained between 2008 September 15 and 2010 July 25 using an exposure time of 3\,min.  The G081 $r^{\prime}$, G115 $R_{C}$ and G115 $r^{\prime}$ observations were independently reduced to aperture photometry light curves following the procedure described by \citet{bakos:2010:hat11}. Instrumental systematic variations were removed from the light curves using TFA, and the light curves were searched for transiting planet signals using the BLS algorithm. Following this process, WASP-194/HAT-P-71/TOI-3791\,b was identified as a transiting planet candidate system on 2013 October 7, and follow-up observations were carried out to establish the planetary nature of the system.

\subsubsection{NITES}

We observed a transit of WASP-154\,b on 2016 August 8 and a transit of WASP-155\,b on 2016 August 11 using the Near Infra-red Transiting ExoplanetS telescope \citep[NITES]{McCormac14}, La Palma. A total of 525 30 second cadence images were obtained for WASP-154\,b using an \textit{I}-band filter and 604 30 second images of WASP-155\,b were obtained using an \textit{R}-band filter. The data were reduced in {\scshape Python} using {\scshape ccdproc} \citep{Craig15}. A master bias, dark and flat was created using the standard process on each night. A minimum of 21 of each frame was used in each master calibration frame. Non-variable nearby comparison stars were selected by hand and aperture photometry extracted using {\scshape sep} \citep{Barbary16, Bertin96}. The shift between each image was measured using the {\scshape donuts} algorithm \citep{McCormac13} and the photometry apertures were recentered.

\subsubsection{MuSCAT2}
We observed WASP-155\,b, WASP-188\,b, and WASP-194\,b using the MuSCAT2 multicolor imager \citep{Narita2018} installed on the 1.52-m Telescopio Carlos S\'anchez (TCS) at the Teide Observatory in Tenerife, Spain. MuSCAT2 is a four-color simultaneous imager with four e2v $1024 \times 1024$ pixel CCD detectors, providing a $7.4 \times 7.4$ arcmin field-of-view and a pixel scale of 0.43$\arcsec~{\rm px}^{-1}$.

A partial transit of WASP-188\,b was observed on 2018 June 10, a full transit of WASP-155\,b was observed on 2019 August 18, and a full transit of WASP-194\,b was observed on 2021 July 4. The observing conditions were generally good for all the observations, however some of the WASP-194 data had to be discarded due to saturation. The exposure times were optimized separately for each night and camera.

A dedicated MuSCAT2 photometry pipeline \citep{Parviainen2019} was used to perform standard reduction steps and aperture photometry. The pipeline calculates aperture photometry for various aperture sizes and comparison stars, generating the final relative light curves via global optimization of the photometry and a transit model calculated using PyTransit \citep{Parviainen2015_pytransit}.

\subsubsection{KeplerCam}
WASP-194/HAT-P-71 was observed through a Sloan $i^{\prime}$ filter with the KeplerCam imager on the Fred Lawrence Whipple Observatory (FLWO) 1.2-m telescope on the following eight nights: 2014 October 5,  2015 April 17, 2016 April 20, 2016 April 28, 2016 October 6, 2016 October 22, 2017 April 18, and 2017 May 23. The observations were gathered with an exposure time of 13 seconds, and the telescope was kept in focus. The data were reduced to ensemble-corrected aperture photometry light curves following the procedures described in \citet{bakos:2010:hat11}. The observations on the first two nights were fully out of transit, and were used to refine the transit ephemeris. The observations on 2016 October 6 were cut short due to weather, so only out-of-transit observations were gathered. An ingress was observed on 2016 April 20, and egresses were observed on 2016 April 28, 2016 October 22, and 2017 April 18. Finally, a full transit was observed on 2017 May 23, and we include only this transit in our final analysis of the system. 

An egress of WASP-188/TOI-5190\,b was observed with the KeplerCam instrument on the night of 2022 April 25 using a Sloan $i^{\prime}$ filter. A total of 442 observations were collected at an average cadence of 34 seconds. The images were calibrated, and aperture photometry was extracted for the target, and for a number of comparison stars, using the AstroImageJ package \citep{Collins2017}. We used a 5 pixel radius aperture, corresponding to 3.36$\arcsec$, for the photometry.

\subsubsection{TRAPPIST-South}
The TRAnsiting Planets and PlanetesImals Small Telescope \citep[TRAPPIST;][]{gillon2011,jehin2011}, located at ESA's La Silla Observatory in Chile observed four of the transiting planets here. The telescope has an FLI camera with an image scale of 0.64$\arcsec~{\rm px}^{-1}$, and is fitted with several optical filters, including the blue-blocking filter that has a transmittance $>90\%$ from 500 nm to above 1000 nm. The image data were extracted via the dedicated pipeline. 

The transits of WASP-102\,b, WASP-116\,b, and WASP-154 \,b were observed using this blue-blocking filter with exposure times of 22, 11, and 9 seconds, respectively. The transits of WASP-149 b were observed using the Sloan \textit{z'} filter with exposures of either 10 or 13 seconds.

\subsubsection{EulerCam}
We used EulerCam at the 1.2-m Euler telescope located at La Silla observatory to observe transits of WASP-102\,b, WASP-116\,b, WASP-149\,b and WASP-154\,b. The instrument is a back-illuminated 4k$\times$4k CCD imager, a pixel resolution of 0.215$\arcsec~{\rm px}^{-1}$ and a field-of-view of $14.76 \times 14.76$ arcmin. For all observations, photometry was extracted via differential aperture photometry, optimizing aperture size and reference stars iteratively to achieve minimal residual RMS on the final transit light curve. The instrument and associated data reduction procedures are described in detail in \citep{Lendl2012}. 

WASP-102 was observed throughout two full transits of planet b on 2013 August 13 and 2013 September 20 using an \textit{r}-Gunn and an \textit{I}-Cousins filter, respectively. Two partial transits of WASP-116 were observed on 2013 November 05 and 2013 November 25 through a broad NGTS filter (500 - 900 nm), while applying a defocus of 0.15\,mm to optimize the duty cycle. Two full transits of WASP-149\,b were observed through a \textit{z}-Gunn filter on 2015 April 07 and 2015 December 20, and a full transit of WASP-154\,b was observed through a NGTS filter on 2016 August 04, again applying a 0.1\,mm defocus. 

\subsubsection{LCOGT}
The Las Cumbres Observatory Global Telescope network \citep[LCOGT;][]{Brown:2013} is a globally distributed network of 0.4, 1, and 2-m telescopes. We make use of observations from three of the network telescopes as described below. 

WASP-116\,b was observed on 2023 October 15 by the 0.35-m telescope at the South African Astronomical Observatory (SAAO) and again on 2023 October 22 by the 0.35-m telescope at Halaeakala (HAL), both using the \textit{i'} filter. Both observations showed an on-time transit consistent with the expected depth. 

We observed a full transit of WASP-154\,b on 2022 June 21  from the LCOGT 0.4-m network node at Cerro Tololo Inter-American Observatory (CTIO). The telescope is equipped with 2048$\times$3072 SBIG STX6303 cameras having a pixel scale of 0.57$\arcsec~{\rm px}^{-1}$, resulting in a 19' x 29' field of view. The observation was carried out in the \textit{i'} filter with an exposure time of 180 seconds. The science images calibration was performed using the standard LCOGT BANZAI pipeline \citep{McCully:2018}, and photometric extraction was performed using AstroImageJ \citep{Collins2017}. Some data were affected because of the poor sky transparency. 

\subsubsection{Brierfield Private Observatory}

The Brierfield Observatory is located near Bowral, N.S.W. Australia. The 0.36-m Planewave CDK14 telescope is equipped with a 4096 $\times$ 4096 Moravian 16 803 camera. The image scale after binning 2 $\times$ 2 is 1.47$\arcsec~{\rm px}^{-1}$, resulting in a 50 arcmin $\times$ 50$\arcmin$ field of view. The photometric data for WASP-154 includes a single full transit with the \textit{R$_C$} filter, consisting of 120 200-second exposures extracted on 2022 June 13 using the \texttt{AstroImageJ} software package \citep{Collins2017}, utilizing a circular photometric aperture with an 11.8 arcsec radius and no detrending parameters.

\subsubsection{Whitin}

We observed a transit of WASP-195\,b as part of the TFOP program on 2022 June 06 using the Whitin Observatory 0.7-m PlaneWave telescope in Eastern Massachusetts, USA. Its FLI ProLine PL23042 CCD camera has a pixel scale of 0.68$\arcsec~{\rm px}^{-1}$ resulting in a 23 $\times$ 23 arcmin field of view. We collected 350 exposures of length 30 seconds with a Sloan-\textit{r'} filter, with a gap due to clouds. We used \texttt{AstroImageJ} \citep{Collins2017} to perform standard calibration and photometric extraction in a 5.4 arcsec radius circular aperture. 

\subsubsection{Observatoire Priv\'e du Mont}
The transit of WASP-194\,b was observed by the Observatoire Priv\'e du Mont (OPM), located in Saint-Pierre du Mont. The facility hosts a Ritchey-Chrétien GSO 0.2-m telescope with an Atik 383L+ camera. A full transit was observed on 2021 July 04 using a CBB filter. The lightcurve was extracted using a 2.2 arcsec aperture.

\subsection{Spectroscopy} \label{sec: spectroscopy}
All nine planets presented here have masses characterized through radial velocity (RV) measurements. Below we briefly describe these facilities and observations. A summary of the observations can be found in Table \ref{tab:follow_up_obs_rv}.

\begin{table*}[]
    \centering
    \caption{Summary of radial velocity observations used to characterize the planet masses. }
    \begin{tabular}{l c c c }
    \hline
    \hline
         Target & Facility & $N_{obs}$ & Observation span \\
         \hline
         WASP-102/TOI-6170          & SOPHIE     & 14                      & Sept 2012-Jan 2013  \\
         --                         & CORALIE    & 11                      & Sept 2013-Aug 2014  \\
         WASP-116/TOI-4672          & SOPHIE     & 15                      & Jan 2013-Nov 2013  \\
         --                         & CORALIE    & 24                      & Aug 2013-Nov 2014  \\
         WASP-149/TOI-6101          & SOPHIE     & 15                      & Nov 2014- Apr 2015 \\
         --                         & CORALIE    & 11                      & Dec 2013- Apr 2015 \\
         WASP-154/TOI-5288          & SOPHIE     & 14                      &  Nov 2014-Oct 2015 \\
         WASP-155/TOI-6135          & SOPHIE     & 22                       & Jul 2015-Dec 2015   \\
         WASP-188/TOI-5190          & SOPHIE     & 15                      & Jul 2015-Dec 2015 \\
         WASP-194/HAT-P-71/TOI-3791 & FLWO/TRES  & 29                      & Oct 2013-Sept 2016 \\
         WASP-195/TOI-4056          & SOPHIE     & 88                      & May 2014-Jul 2021  \\
         WASP-197/TOI-5385          & FLWO/TRES  & 16                      &  Apr 2022 - Apr 2024 \\
         --                         & SOPHIE     & 5                       &  Feb 2023 - Jan 2024\\
         --                         & PARAS-2    & 8                       &  Jan 2024 - March 2024\\

         \hline
    \end{tabular}

    \label{tab:follow_up_obs_rv}
\end{table*}

\subsubsection{SOPHIE}
Between 2014 and 2024, we observed eight of the nine objects presented here with the SOPHIE spectrograph at the 193-cm telescope at Observatoire de Haute-Provence, France. This is a stabilized \'echelle spectrograph dedicated to high-precision radial-velocity measurements in optical wavelengths \citep{Perruchot2008, Bouchy2009}. We used its high-efficiency
mode (resolving power $R=40,000$) and the slow-reading mode of its CCD for seven of them (WASP-102, 116, 149, 154, 155, 188, and 195). WASP-197 was observed with SOPHIE's high-resolution mode ($R=75,000$) and the fast-reading mode. Depending on the stars and the weather conditions, exposure times typically range between 15 and 45~minutes, for typical signal-to-noise ratios per pixel at 550~nm between 20 and 50. A few exposures with particularly low signal-to-noise ratios were excluded. 

The radial velocities were extracted with the SOPHIE pipeline, as presented by \cite{Bouchy2009} and refined by \cite{Heidari2024}, which derives cross correlation functions (CCF) from standard numerical masks corresponding to different spectral types. In particular, that version of the pipeline includes corrections for CCD charge transfer inefficiency, instrumental drifts, and pollution by moonlight. Moonlight is estimated and corrected using the second SOPHIE fiber aperture that is targeted on the sky, 2 arcmin away from the first one pointing toward the star.

The derived radial velocities and their uncertainties are available through the Digital Repository for the University of Maryland (DRUM)\footnote{https://doi.org/10.13016/iuro-f0od}. They show variations in phase with the periods and transit times derived from photometry. The amplitudes of those radial-velocity variations agree with planetary masses, and do not depend on the stellar type of the numerical mask used for the CCF, as it might be expected in cases where photometric transits are actually caused by blended binary stars of different spectral types. In addition, the bisector spans measured on the CCF show no significant variations nor correlations with the observed radial-velocity variations, as it might also be expected in cases of blended eclipsing binaries perturbing the profiles of the spectral lines \citep{Queloz2001}.

Thus, those measurements establish the planetary nature of the transiting events, and constrain the mass of the transiting planets as well as the eccentricity of their orbits.


\subsubsection{FLWO/TRES}
The Tillinghast Reflector Echelle Spectrograph \citep[TRES;][]{Furesz2008} was used to obtain reconnaissance spectra of WASP-197 and WASP-194. TRES is is a fiber-fed echelle spectrograph with a wavelength range of 390-910~nm and a resolving power of R $\approx$ 44,000 mounted on the 1.5-m Tillinghast Reflector telescope at the Fred Lawrence Whipple Observatory (FLWO) atop Mount Hopkins, Arizona. Thirty spectra of WASP-197 were obtained during 2013 October and 2016 September, and sixteen spectra of WASP-194 during 2022 April and 2024 April. The spectra were extracted as described in \cite{Buchhave2010} and a multi-order analysis was used to derive RVs. The multi-order analysis uses the strongest observed spectrum as a template and then cross-correlates each spectrum, order-by-order, against the template spectrum.

For the TRES observation of WASP-194, there was a single outlying data point of several hundred meters per second. This data point corresponded to an observation during a full moon, which could likely have contaminated the observation. We therefore remove this data point from analysis. 

\subsubsection{CORALIE}

The high resolution CORALIE spectrograph \citep{Queloz2000} is installed at the 1.2-m Leonhard Euler Telescope at ESO's La Silla Observatory in Chile. 11 observations of WASP-102 were taken by CORALIE between 2012 and 2014, all obtained during grey/dark time to reduce stray light from the moon. 
WASP-116 was observed by CORALIE on 24 nights. However, one of these observations was taken after the optical fibre of CORALIE was replaced in November of 2014. We therefore exclude the last data point from our analysis. Similarly, 9 of the 11 CORALIE data points for WASP-149 were taken after the optical fiber change. We use only these 9 data points in our final analysis. 

\subsubsection{PARAS-2}
RV follow-up of WASP-197 was done with the PARAS-2 spectrograph. The spectrograph is attached to the PRL 2.5-m telescope at Mount Abu Observatory and works at high resolution (R $\approx$ 107,000) in 380--690~nm. It uses the simultaneous referencing method using the Uranium Argon (UAr) hollow cathode lamp for wavelength calibration and precise RV calculations. More details of the spectrograph can be found in \cite{Chakraborty2018} and \cite{Chakraborty2024}. 

A total of 8 spectra were acquired between 2024 Jan 04 to 2024  Apr 08. The exposure time of each exposure was kept at 60 minutes, which resulted in a signal-to-noise ratio of 18-30 per pixel at a blaze peak wavelength of 550~nm. A custom-made PARAS2 pipeline is used for data extraction and RV calculations \citep{Baliwal2024}. 
In brief, before doing the optimal extraction of the spectra, the pipeline does various corrections, including bias and dark subtractions, cosmic ray rejection, scattered light corrections etc. The pipeline is written in IDL and based upon the PARAS-1 pipeline \citep{Chakraborty2014} and the algorithms of \cite{Piskunov2002}. The RVs are calculated by cross-correlating the extracted and wavelength-calibrated spectra with a template spectrum of G-type stars. The RV photon noise is found to be between 8.3--18.4 m s$^{-1}$, calculated using the techniques mentioned in \cite{Chaturvedi2016}.

\subsection{High Resolution Imaging} \label{sec:highres}

Close stellar companions (bound or line of sight) can confound exoplanet discoveries in a number of ways.  The detected transit signal might be a false positive due to a background eclipsing binary. Even real planet discoveries will yield incorrect stellar and exoplanet parameters if a close companion exists and is unaccounted for \citep{Ciardi2015, Furlan2017a}. Additionally, the presence of a close companion star leads to the non-detection of small planets residing within the same exoplanetary system \citep{Lester2021}. Given that nearly one half of solar-like stars are in binary or multiple star systems \citep{Matson2018}, high-resolution imaging provides crucial information toward our understanding of exoplanetary formation, dynamics and evolution.

The high resolution imaging used in this work are presented below. Plots showing a selection of the contrast curves can be found in Appendix \ref{app:highres}. We summarize the nearby ($<$10$\arcsec$) companions identified in Table \ref{tab:companions}.

\begin{table*}[]
    \centering
    \begin{threeparttable}
    
    \caption{Summary of the identified nearby ($<$ 10$\arcsec$) stellar companions to the planet host stars. }
    \begin{tabular}{l c c c}
    \hline
    \hline
         Primary Target & Nearby companion Gaia ID & $\Delta$mag (filter) & distance (arcsec)  \\
         \hline
        WASP-149 & 3062565055054980352 & 9.0 (G) & 7.8 \\
        WASP-154 & 2618260274649763840 & 8.4 (G) & 5.3 \\
        WASP-155 & 1910906408272899200 & 2.6 (G), 5.3$^\dag$ (832nm) & 2.9 \\
        --       & 1910906403977533184 & 6.4 (G) & 8.3 \\
        WASP-188 & 2095929373136748928 & 5.7 (G), 5.5 (I)& 1.8 \\
        --       & 2095930850605499136 & 5.2 (G) & 4.1 \\
        WASP-194 & 2139082485811741440 & 0.5 (G) & 9.7 \\
        WASP-197 & 734156214654777984 & 6.5 (G) & 6.7 \\
        
         \hline

    \end{tabular}
    \begin{tablenotes}
        \item $\dag$: See Section \ref{sec:fit_wasp155}. 
    \end{tablenotes}
    \label{tab:companions}
    \end{threeparttable}
\end{table*}

\subsubsection{WIYN/NESSI}
The stars WASP-116, WASP-155, WASP-194, WASP-195, and WASP-197 were observed using the NN-EXPLORE Exoplanet Stellar Speckle Imager \citep[NESSI;][]{Scott2018}, a speckle imager employed at the WIYN 3.5-m telescope on Kitt Peak, Arizona.  
NESSI is a dual-channel speckle imager that yields simultaneous speckle images in two filters. WASP-116, WASP-155, and WASP-194 were observed on 2023 January 28, and WASP-155 and WASP-194 were both observed on 2024 September 12 through two filters centered at $\lambda_c=562$~nm and 832~nm.  WASP-195 and WASP-197 were observed on 2022 April 21 and April 18 respectively, using only the 832~nm filter due to an alignment problem with the blue channel during this time.  Each observation consisted of a set of 9 1000-frame 40-ms speckle exposures.  The field-of-view was set by the sub-array readout region of 256 $\times$ 256 pixels to be 4.6 $\times$ 4.6 arcsec, although speckle measurements are limited to a smaller radial extent from the target star.  Alongside the observation of the science target, a single 1000-frame image set was taken of a nearby single star (point source) for calibration of the underlying PSF.

The speckle data were reduced using the pipeline described in \cite{Howell2011}. The pipeline produces, among other things, a reconstructed image of the field around each target and a contrast curve representing the relative magnitude limits for detecting nearby point sources between the diffraction-limited inner working angle ($0.04-0.06$~arcsec for these filters) and an outer angle of $1.2$~arcsec. No companion sources were detected within 1.2 arcseconds of any of the targets in the NESSI data, the angular extent over which the speckle data are most accurate.  In the case of WASP-155, we tentatively report a star 2.9 arcseconds from the target. More information about this companion can be found in Section \ref{sec:fit_wasp155}. 

\subsubsection{SAI/Speckle Polarimeter}
WASP-188 was observed on 2023 December 02, and WASP-197 was observed two nights later with the speckle polarimeter on the 2.5-m telescope at the Caucasian Observatory of Sternberg Astronomical Institute (SAI) of Lomonosov Moscow State University. A low--noise CMOS detector Hamamatsu ORCA--quest \citep{Strakhov2023} was used as a detector. WASP-195 was observed on 2021 July 20 with a previous, EMCCD--based version of the instrument. The atmospheric dispersion compensator was active, which allowed using the $I_\mathrm{c}$ band. The corresponding angular resolution is 0.083$\arcsec$. For WASP-188 and WASP-197 we have accumulated 5200 frames with 23-ms exposure. For WASP-195 4000 30-ms frames were accumulated.

For WASP-195 we did not detect any stellar companions, with detection limits of $\Delta I_\mathrm{c}=4.7$~mag and $6.0$~mag at distances $0.25$ and $1.0^{\prime\prime}$ from the star, respectively. The SAI observations of WASP-197 provided the detection limits of $\Delta I_\mathrm{c}=4.0$~mag and $\Delta I_\mathrm{c}=5.9$~mag at distances $0.25$ and $1.0^{\prime\prime}$ from the star, respectively.

For WASP-188 a companion was detected with a position and magnitude difference consistent with Gaia DR3 2095929373136748928 --- a star which is $\Delta G=5.6$ fainter than WASP-188 (=Gaia DR3 2095929368839731072). No other, closer companions were detected with the limits $\Delta I_\mathrm{c}=3.2$~mag and $5.6$~mag at distances $0.25$ and $1.0^{\prime\prime}$ from the star, respectively. 

\subsubsection{SOAR/HRCam}
We searched for stellar companions to WASP-154, WASP-149, and WASP-102 with speckle imaging on the 4.1-m Southern Astrophysical Research (SOAR) telescope \citep{Tokovinin2018}, observing in Cousins I-band, a similar visible bandpass as TESS. Observations were performed on 2022 February 22 (WASP-154), 2024 January 08 (WASP-149), and 2023 August 31 (WASP-102). More details of the speckle observations from the SOAR TESS survey are available in \cite{Ziegler2020}. No nearby stars were detected within 3$\arcsec$ of WASP-102, WASP-149, or WASP-154.

\subsubsection{Gemini/Alopeke and Zorro}

The Alopeke and Zorro instruments are identical speckle imagers located on the Gemini-North and Gemini-South telescopes respectively \citep{Scott2021}\footnote {https://www.gemini.edu/sciops/instruments/alopeke-zorro/}. The instruments provide simultaneous speckle imaging in two bands (562~nm and 832~nm) with output data products including a reconstructed image and robust contrast limits on companion detections \citep{Howell2011}.

WASP-116 was observed on 2023 January 09 using the Zorro speckle instrument. The two 5$-\sigma$ contrast curves result and our reconstructed 862~nm speckle image show that WASP-116 is a single star with no close companion brighter than 5 to 7 magnitudes from the diffraction limit (20 mas) out to 1.2$\arcsec$. At the distance of WASP-116 (d=560 pc) these angular limits correspond to spatial limits of 11 to 672 AU.
Alopeke was used to observe WASP-155 on 2024 August 13. The resulting contrast curve showed no companions with a contrast within 5 mag (562nm) or 6 mag (832nm) from the diffraction limit out to 1.2$\arcsec$, corresponding to 8 to 480 AU.

\subsection{Palomar/PHARO}
Observations of WASP-197 were made on 2024 February 17 with the PHARO instrument \citep{Hayward2001} on the Palomar Hale 5-m telescope in the narrow-band \textit{K}-cont filter $(\lambda_o = 2.29; \Delta\lambda = 0.035~\mu$m) and the \textit{H}-cont filter $(\lambda_o = 1.668; \Delta\lambda = 0.018~\mu$m) using the P3K natural guide star AO system \citep{Dekany2013}. The PHARO pixel scale is 0.025$\arcsec~{\rm px}^{-1}$. A standard 5-point quincunx dither pattern with steps of 5$\arcsec$ was performed three times with each repeat separated by 0.5$\arcsec$. The reduced science frames were combined into a single mosaic image with final resolutions of 0.099 and 0.90$\arcsec$, respectively.  The sensitivity of the final combined AO image were determined by injecting simulated sources azimuthally around the primary target every $20^\circ $ at separations of integer multiples of the central source's FWHM \citep{Furlan2017a}. The brightness of each injected source was scaled until standard aperture photometry detected it with 5$-\sigma $ significance.  The final 5$-\sigma $ limit at each separation was determined from the average of all of the determined limits at that separation and the uncertainty on the limit was set by the root mean square dispersion of the azimuthal slices at a given radial distance. The infrared imaging did detect a star within 6$\arcsec$ of the primary target but no other close-in stars were found, in agreement with the speckle observations.


\section{Stellar properties} \label{sec:stellar}
    
    \thispagestyle{empty}
    \floattable
    \rotate
    \begin{deluxetable}{l c c c c c }
    \label{tab:stellar_properties}
    \tablecaption{Stellar properties for the planets' host stars. Details on spectral line fitting and SED modeling can be found in Section \ref{sec:stellar}.}
    \tablehead{
        & WASP-102 & WASP-116 & WASP-149 & WASP-154 & WASP-155  \\
        & TOI-6170 & TOI-4672 & TOI-6101 & TOI-5288 & TOI-6135  \\
    }
    \startdata
        \emph{Identifiers} & & & & &  \\
        TIC ID   & 51637609 & 332911893 & 19342878 & 857186 & 100909102 \\
        2MASS    & J22255144+1551242 & J02205177-0149337 & J08161768-0841121 & J21505262-0838084 & J23115512+3302519 \\
        Gaia DR3 & 2736318583335936128 & 2493785078665571200 & 3062565055055954304 & 2618260274650146048 & 1910906408272899328 \\
        & & & & &  \\
        \emph{Magnitudes} & & & & &  \\
       TESS               & 12.1 & 11.9 & 10.8 & 12.2 & 12.0 \\
        B                  & 13.4 & 13.0 & 12.3 & 14.1 & 12.9 \\
        V                  & 12.7 & 12.5 & 11.7 & 13.2 & 12.4 \\
        Gaia               & 12.6 & 12.3 & 11.3 & 12.8 & 12.5 \\
        J                  & 11.5 & 11.3 & 10.2 & 11.3 & 11.3 \\
        H                  & 11.2 & 11.1 & 9.9 & 10.9 & 10.9 \\
        K                  & 11.1 & 11.0 & 9.8 & 10.8 & 10.9 \\
        WISE 3.4 $\mu m$   & 11.1 & 11.0 & 9.8 & 10.7 & 10.6 \\
        WISE 4.6 $\mu m$   & 11.1 & 11.0 & 9.8 & 10.7 & 10.7 \\
        WISE 12 $\mu m$    & 11.0 & 10.9 & 9.8 & 10.7 & 10.8 \\
        WISE 22 $\mu m$    & 9.0 & 8.6 & 8.4 & 9.1  &9.0  \\
        & & & & &  \\
        \emph{Properties} & & & & &  \\
        RA (J2000)              & 22:25:51.43 & 02:20:51.78 & 08:16:17.68 & 21:50:52.6 & 23:11:55.13 \\
        Dec (J2000)             & +15:51:23.9 & -01:49:33.75 & -08:41:11.69 & -08:38:09.4 & +33:02:51.4 \\
        pm (RA) mas yr$^{-1}$   & -13.6549 $\pm$ 0.0156 & 8.6868 $\pm$ 0.0189 & -1.5381 $\pm$ 0.0269 & -24.4276 $\pm$ 0.0188 &18.6723 $\pm$ 0.0118 \\
        pm (Dec) mas yr$^{-1}$  & -24.0709 $\pm$ 0.0145 & 1.7498 $\pm$ 0.0156 & 23.0363 $\pm$ 0.0216 & -57.1313 $\pm$ 0.0156 & -28.3813 $\pm$ 0.0109 \\
        Parallax mas            & 1.9564 $\pm$ 0.0150 & 1.9009 $\pm$ 0.0166 & 4.6465 $\pm$ 0.0231 & 4.4212 $\pm$ 0.0178 & 2.5810 $\pm$ 0.0115 \\
        Distance pc             & 535.689 $\pm$ 15.6205 & 559.243 $\pm$ 15.677 & 211.73 $\pm$ 1.688 & 215.501 $\pm$  1.962& 399.63 $\pm$ 4.884 \\
        T$_{\text{eff}}$ (K)             & 5990 $\pm$ 100  & 6250 $\pm$ 125 & 5750 $\pm$ 125 & 4774 $\pm$ 133 & 5660 $\pm$ 100 \\
        $[\text{Fe/H}]$                  & -0.2 $\pm$ 0.3  & 0.0 $\pm$ 0.3 & 0.0 $\pm$ 0.3 & 0.2 $\pm$ 0.1 & 0.0 $\pm$ 0.3 \\
        M$_*$ (M$_\odot$)         & 1.08 $\pm$ 0.07  & 1.25 $\pm$ 0.08 & 1.05 $\pm$ 0.06&0.80 $\pm$ 0.05 & 1.09 $\pm$ 0.07 \\
        R$_*$ (R$_\odot$)         & 1.375 $\pm$ 0.048  & 1.426 $\pm$ 0.064 & 1.080 $\pm$ 0.049&0.823 $\pm$ 0.047 & 1.240 $\pm$ 0.060 \\
        log \emph(g)       & 4.20 $\pm$ 0.15  & 4.25 $\pm$ 0.15 & 4.40 $\pm$ 0.15&4.47 $\pm$ 0.08 & 4.3 $\pm$ 0.15 \\
        F$_{\text{bol}}$ (erg\,s$^{-1}$cm$^{-2}$)   & 2.686 $\pm$ 0.035  & 3.23 $\pm$ 0.11 & 7.94 $\pm$ 0.18&1.985 $\pm$ 0.046 & 3.03 $\pm$ 0.20 \\
        Age (Gyr)               & 0.67 $\pm$ 0.07  & 1.7 $\pm$ 0.8 & 0.6 $\pm$ 0.3 & & 1.7 $\pm$ 0.8 \\
        RUWE$^\dag$ & 1.1096977 & 1.19287 & 1.2159479 & 1.0076779 & 0.9177554 
    \enddata
    
    \end{deluxetable}

    \thispagestyle{empty}
    \floattable
    \rotate
    \begin{deluxetable}{l c c c c }
    \label{tab:stellar_properties2}
    \tablehead{
         & WASP-188 & WASP-194          & WASP-195 & WASP-197  \\
         & TOI-5190 & TOI-3791 HAT-P-71 & TOI-4056 & TOI-5385  \\}
    \startdata
        \emph{Identifiers} & & & &  \\
        TIC ID   & 289574465 & 400432230 & 232567319 & 85266608 \\
        2MASS    & J18350767+3636562 & J19413306+5613043 & J16301192+4953446 & J10423138+2811550 \\
        Gaia DR3 & 2095929368839731072   & 2139082524468869248   & 1411707818360668416   & 734156218947945216 \\
        & & & &   \\
        \emph{Magnitudes} & & & & \\
       TESS               & 11.7 & 11.6 & 11.4 & 10.9 \\
        B                  & 12.5 & 12.7 & 12.3 & 11.9 \\
        V                  & 12.2 & 12.0 & 11.9 & 11.6  \\
        Gaia               & 12.0 & 11.9 & 11.8 & 11.2 \\
        J                  & 11.3 & 11.1 & 10.9 & 10.3 \\
        H                  & 11.1 & 10.9 & 10.7 & 10.1 \\
        K                  & 11.0 & 10.9 & 10.7 & 10.0 \\
        WISE 3.4 $\mu m$   & 11.0 & 10.8 & 10.6 & 10.0 \\
        WISE 4.6 $\mu m$   & 11.0 & 10.8 & 10.7 & 10.0 \\
        WISE 12 $\mu m$    & 10.9 & 10.8 & 10.6 & 10.0 \\
        WISE 22 $\mu m$    & 9.3 & 9.5 & 9.6 & 8.8  \\
        & & & &  \\
        \emph{Properties} & & & &   \\
        RA (J2000)              & 18:35:07.67       & 19:41:33.05       & 16:30:11.91       & 10:42:31.37       \\
        Dec (J2000)             & +36:36:56.28      & +56:13:04.1      & +49:53:44.85       & +28:11:55.05      \\
        pm (RA) mas yr$^{-1}$   & -0.2471 $\pm$ 0.0110     & -4.5138 $\pm$ 0.0113     & -3.8919 $\pm$ 0.0120     & -9.2100 $\pm$ 0.0190     \\
        pm (Dec) mas yr$^{-1}$  & -1.3574 $\pm$ 0.0117    & -14.8880 $\pm$ 0.0125    & 14.5456 $\pm$ 0.0138    & 0.8583 $\pm$ 0.0193    \\
        Parallax mas            & 1.4310 $\pm$ 0.0098 & 2.0635 $\pm$ 0.0089 & 2.0266 $\pm$ 0.0102 & 2.0927 $\pm$ 0.0209 \\
        Distance pc             & 675.281 $\pm$ 9.674     & 476.831 $\pm$ 3.860     & 484.852 $\pm$ 5.091     & 483.675 $\pm$ 12.2885     \\
        T$_{\text{eff}}$ (K)               & 6850 $\pm$ 125     & 6405 $\pm$ 200     & 6300 $\pm$ 125     &6050 $\pm$ 100      \\
        $[\text{Fe/H}]$                  & 0.0 $\pm$ 0.3      & 0.00 $\pm$ 0.25      & 0.0 $\pm$ 0.3      &0.0 $\pm$ 0.3       \\
        M$_*$ (M$_\odot$)           & 1.50 $\pm$ 0.09    & 1.29 $\pm$ 0.08    & 1.30 $\pm$ 0.08    &1.36 $\pm$ 0.08     \\
        R$_*$ (R$_\odot$)          & 1.830 $\pm$ 0.069    & 1.409 $\pm$ 0.090    & 1.578 $\pm$ 0.066    &2.112 $\pm$ 0.077  \\
        log \emph(g)       & 4.10 $\pm$ 0.2     & 4.27 $\pm$ 0.5     & 4.14 $\pm$ 0.25     &3.9 $\pm$ 0.2       \\
        F$_{\text{bol}}$ (erg\,s$^{-1}$cm$^{-2}$)   & 4.355 $\pm$ 0.050  & 4.101 $\pm$ 0.096 & 4.65 $\pm$ 0.11&7.55 $\pm$ 0.18      \\
        Age (Gyr)               &       & 0.75 $\pm$ 0.55      & 0.6 $\pm$ 0.2      &2.6 $\pm$ 1.5$^*$        \\
        RUWE$^\dag$ & 0.8776204 & 0.95639163 & 0.8457882 & 0.99257046 \\
    \enddata
    
    \begin{tablenotes}
        \item \dag: Gaia renormalized unit weight error (RUWE). Values $\approx$1 are expected for single sources, values $\gtrsim$1.4 suggest extended or binary sources. 
     \end{tablenotes}
     
    \end{deluxetable}

\begin{figure*}
\centering
\includegraphics[width=0.3\linewidth,trim=80 70 50 50,clip]{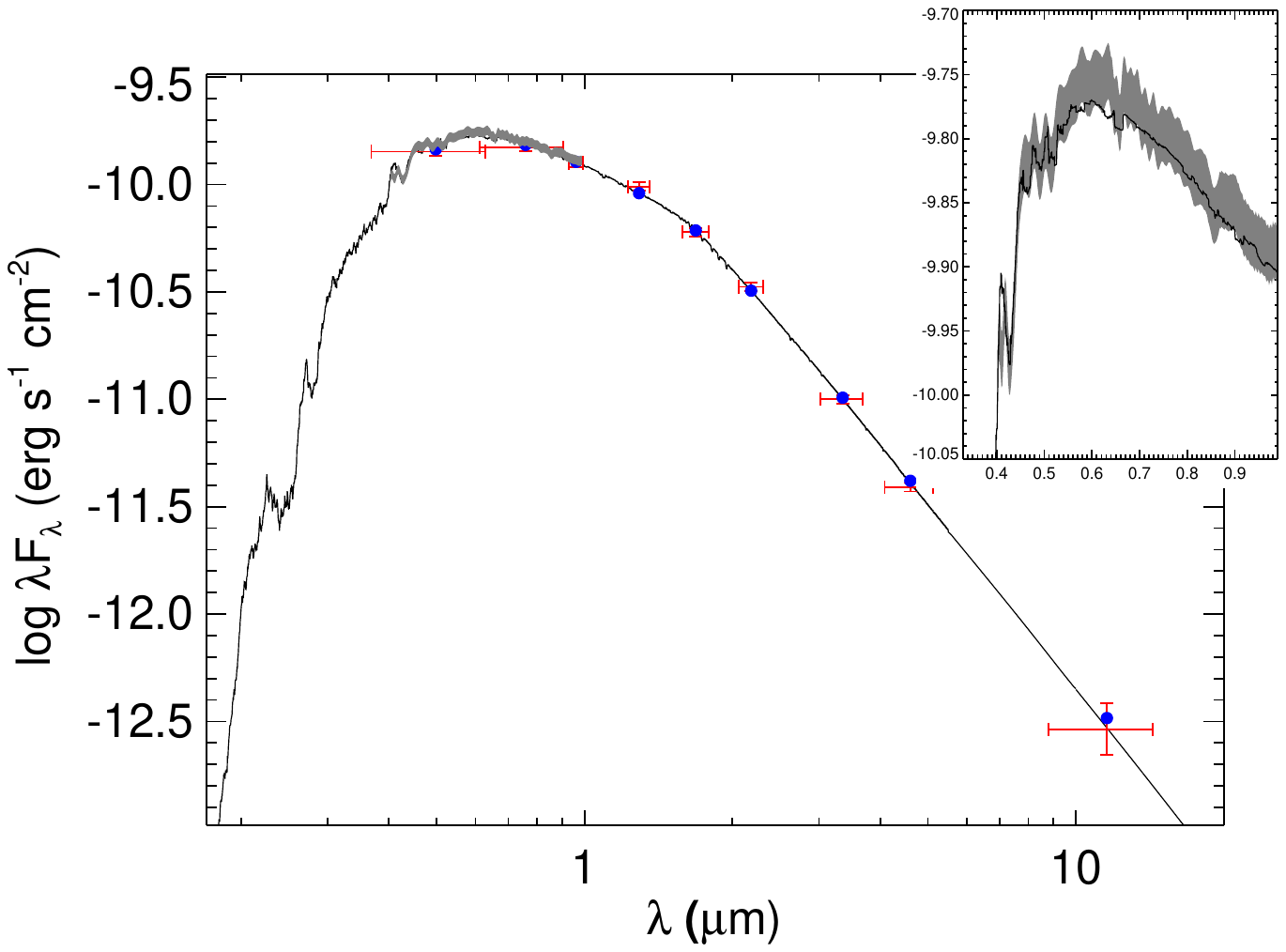} 
\includegraphics[width=0.3\linewidth,trim=80 70 50 50,clip]{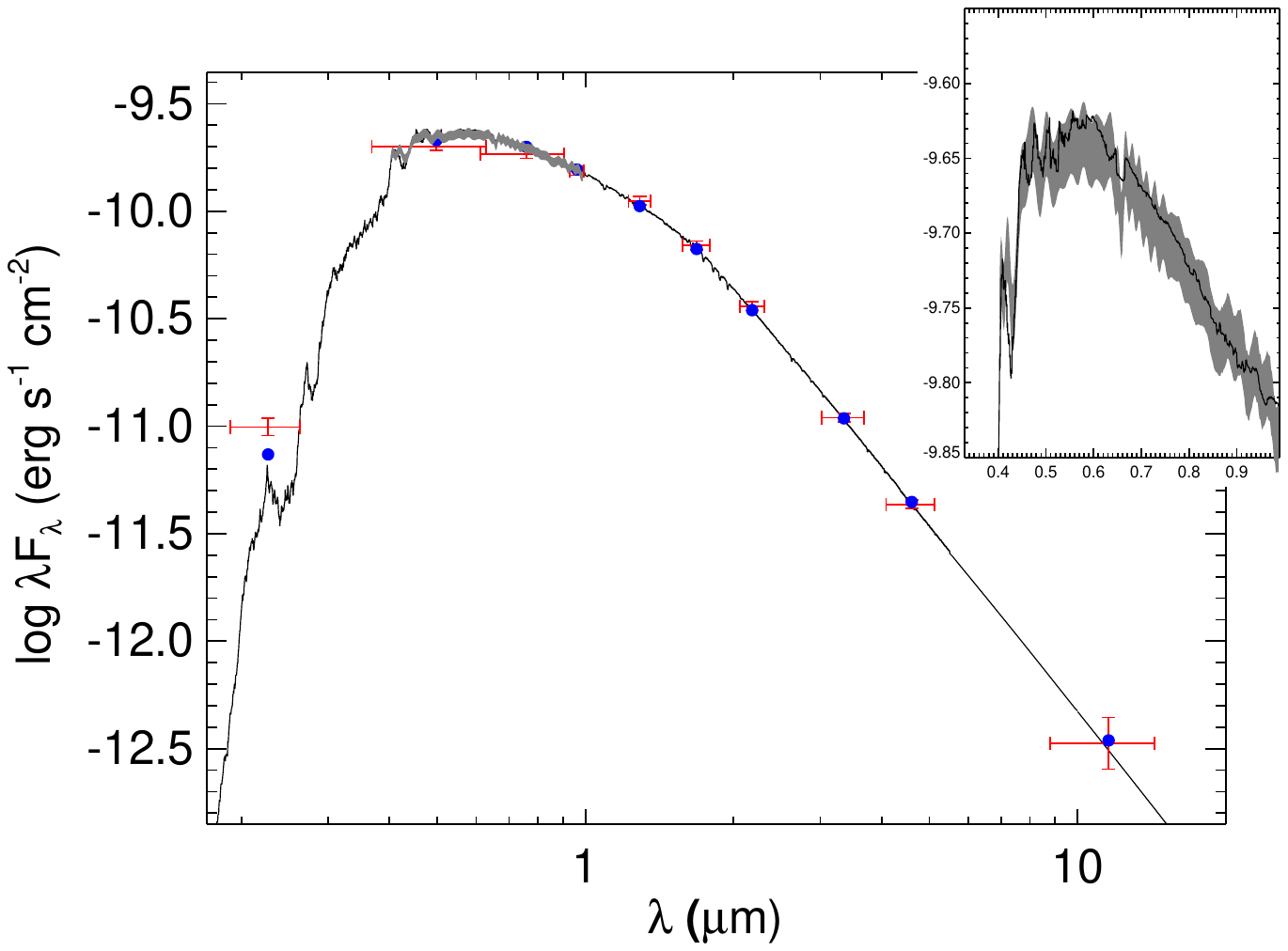} 
\includegraphics[width=0.3\linewidth,trim=80 70 50 50,clip]{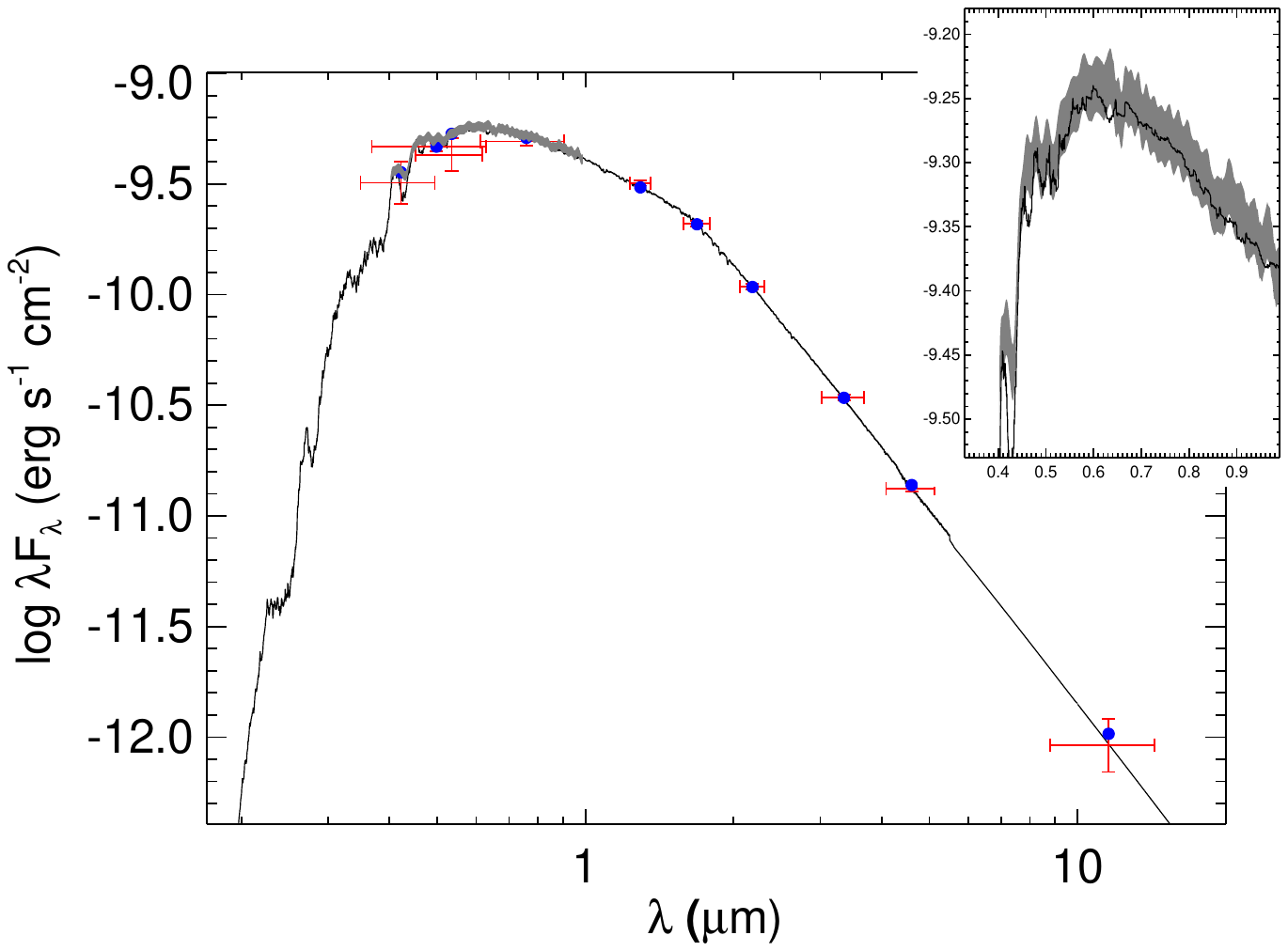} 
\includegraphics[width=0.3\linewidth,trim=80 70 50 50,clip]{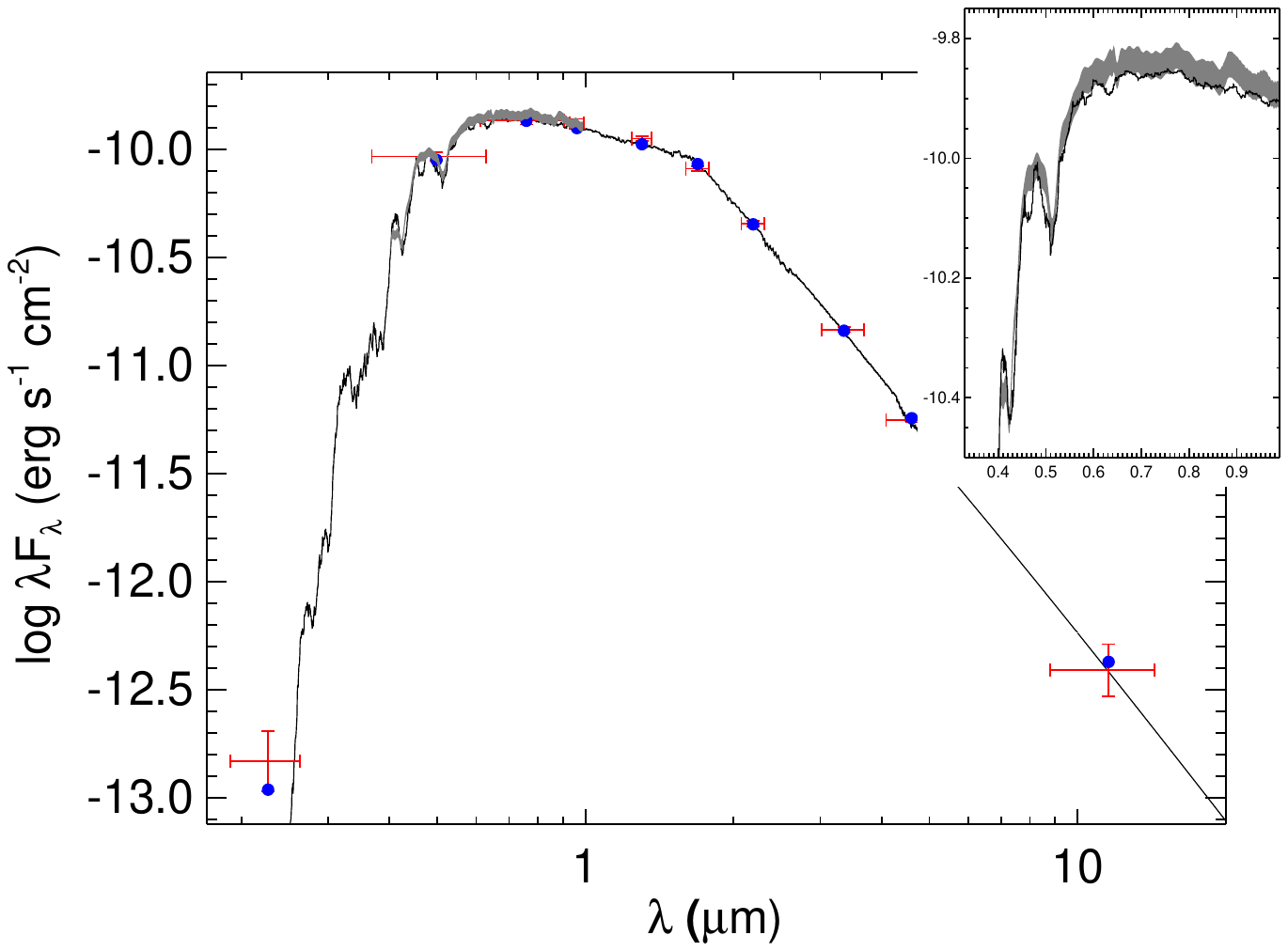} 
\includegraphics[width=0.3\linewidth,trim=80 70 50 50,clip]{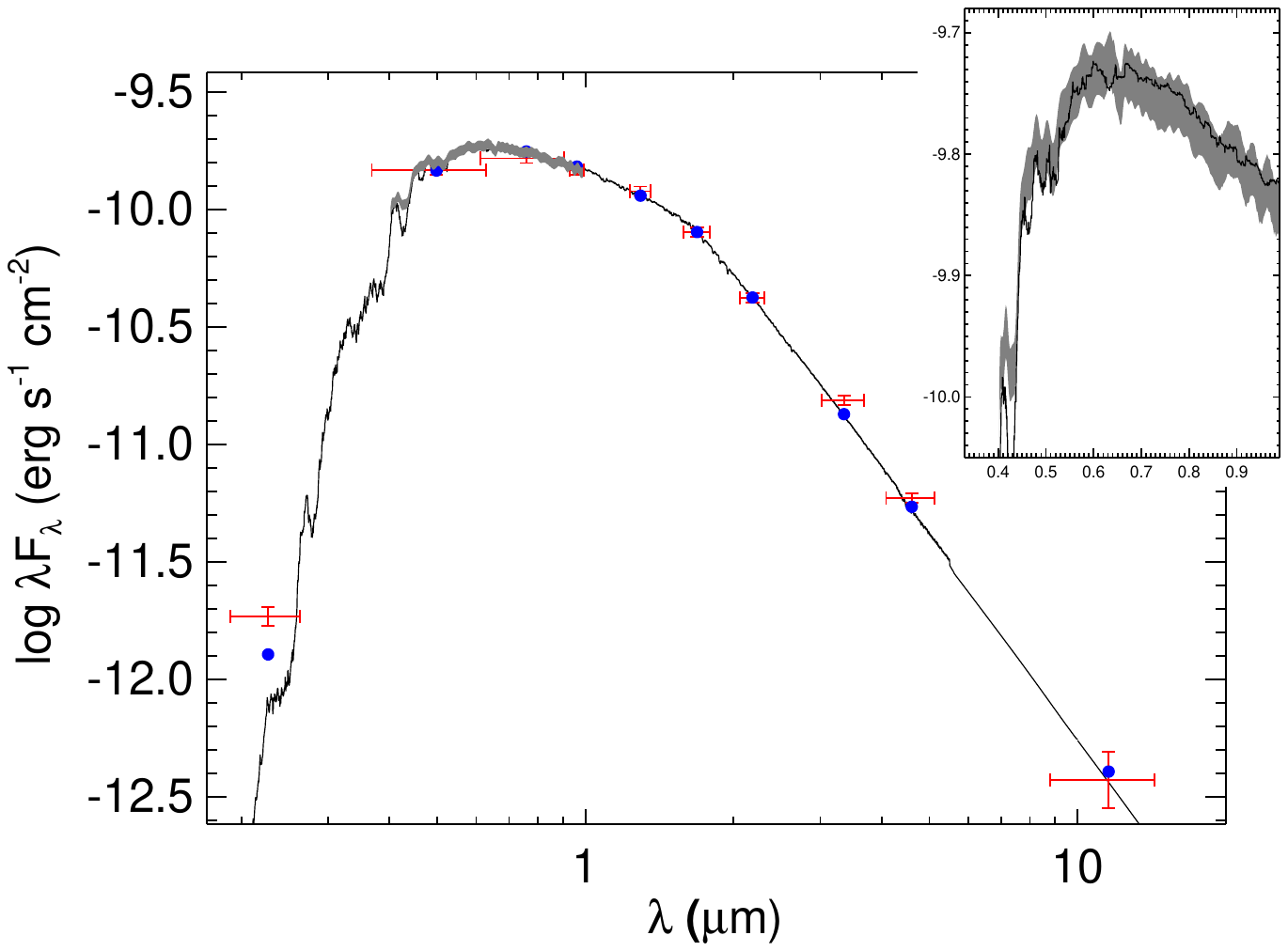} 
\includegraphics[width=0.3\linewidth,trim=80 70 50 50,clip]{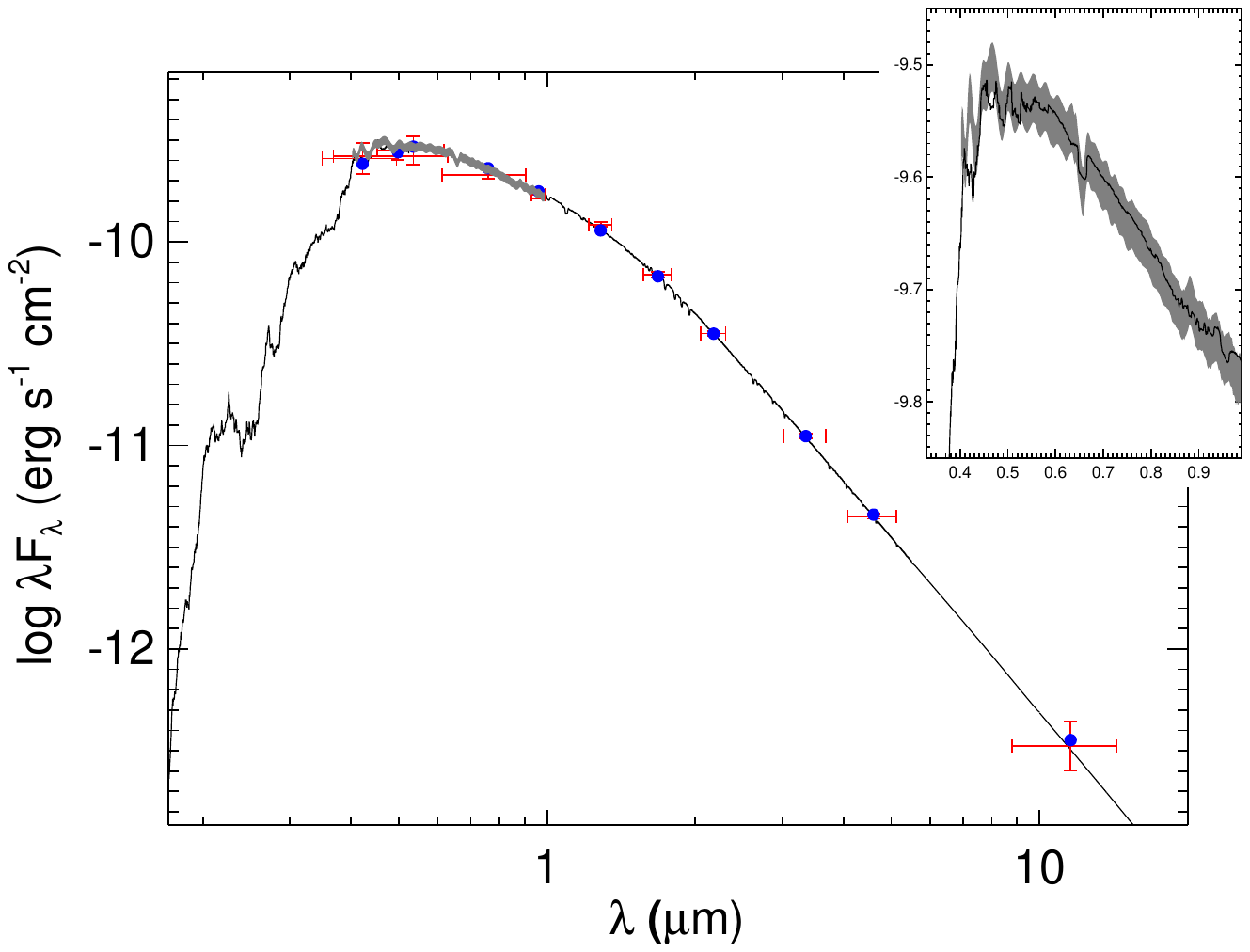} 
\includegraphics[width=0.3\linewidth,trim=0 0 0 0,clip]{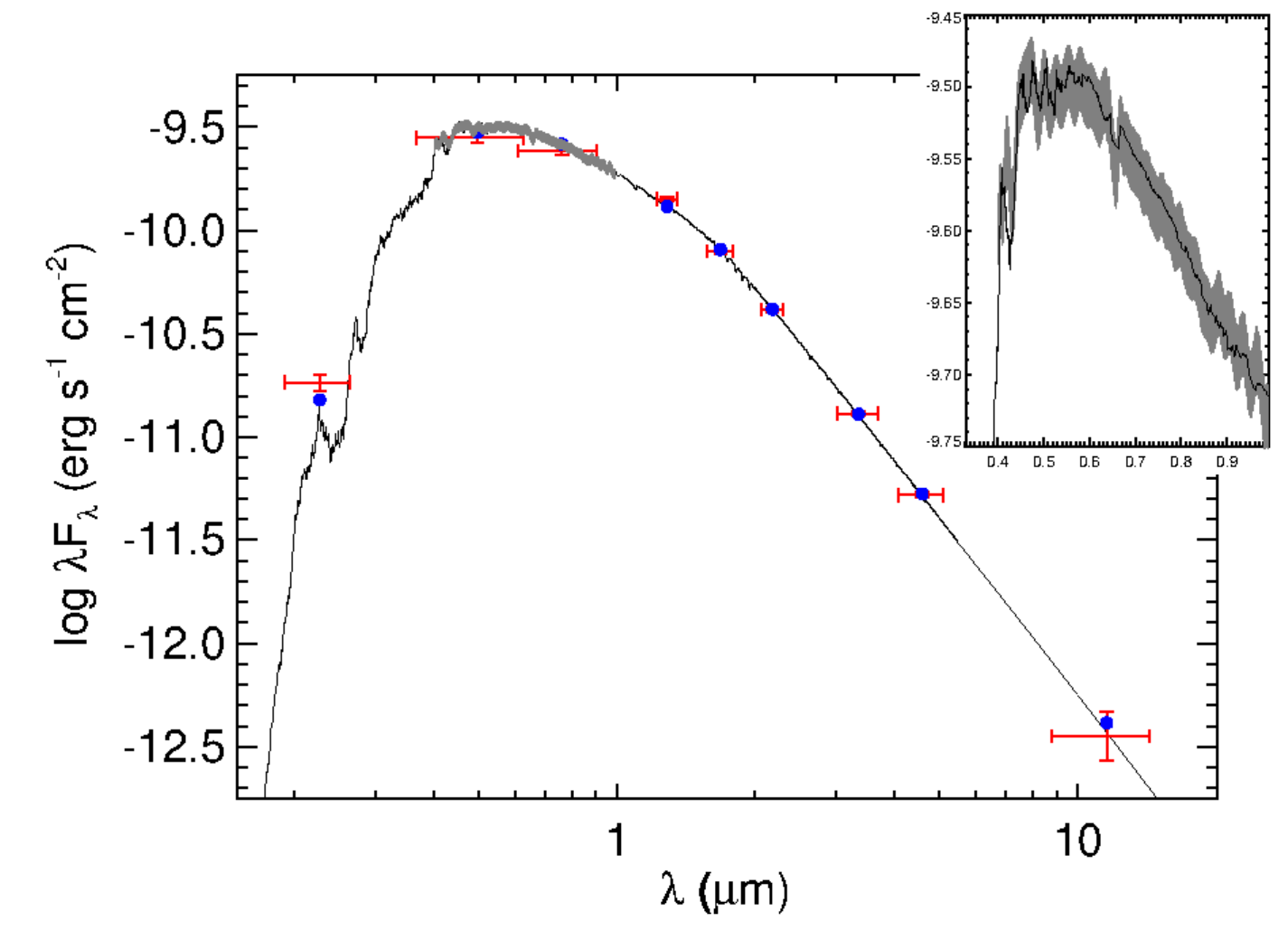} 
\includegraphics[width=0.3\linewidth,trim=80 70 50 50,clip]{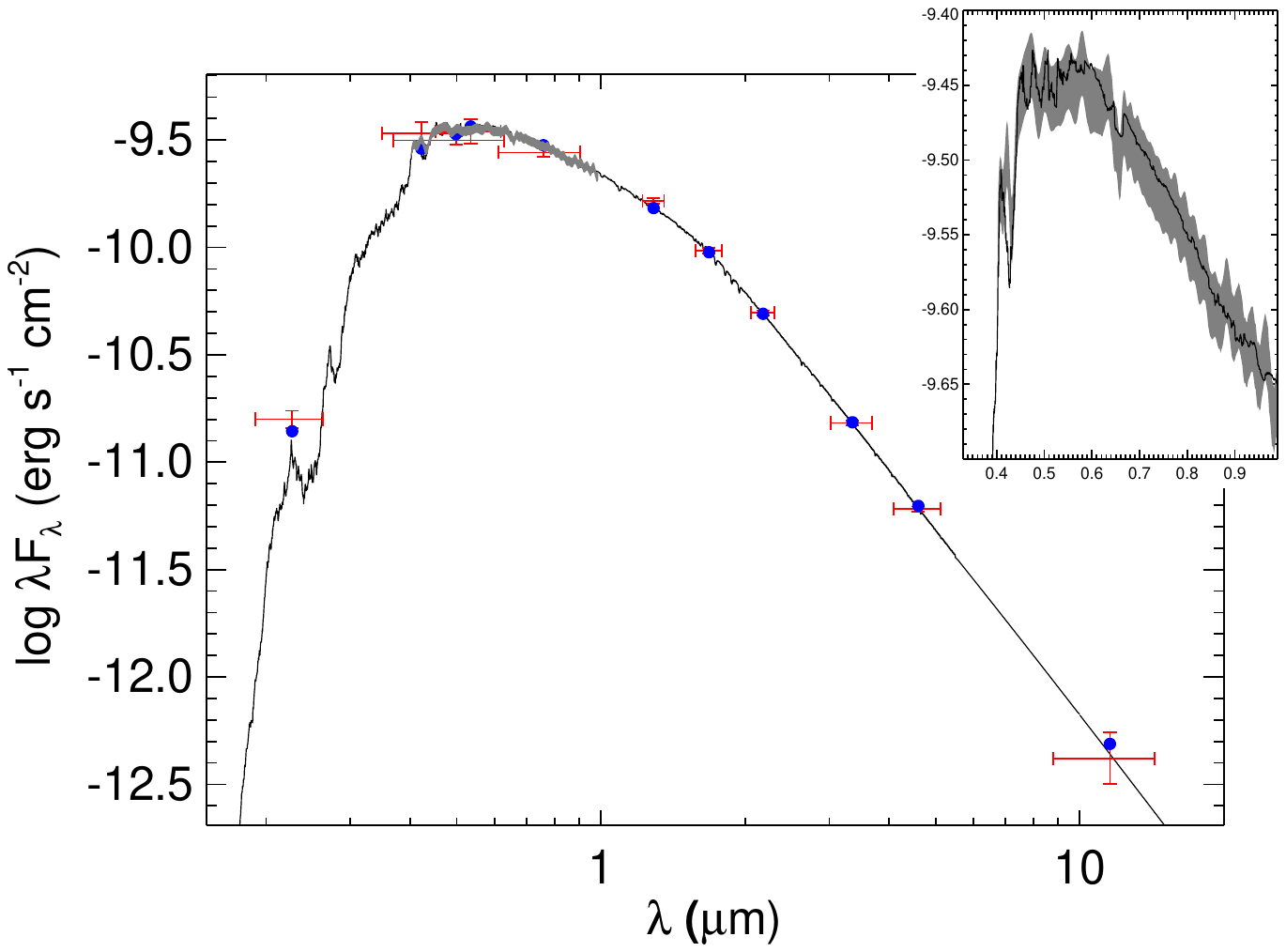} 
\includegraphics[width=0.3\linewidth,trim=80 70 50 50,clip]{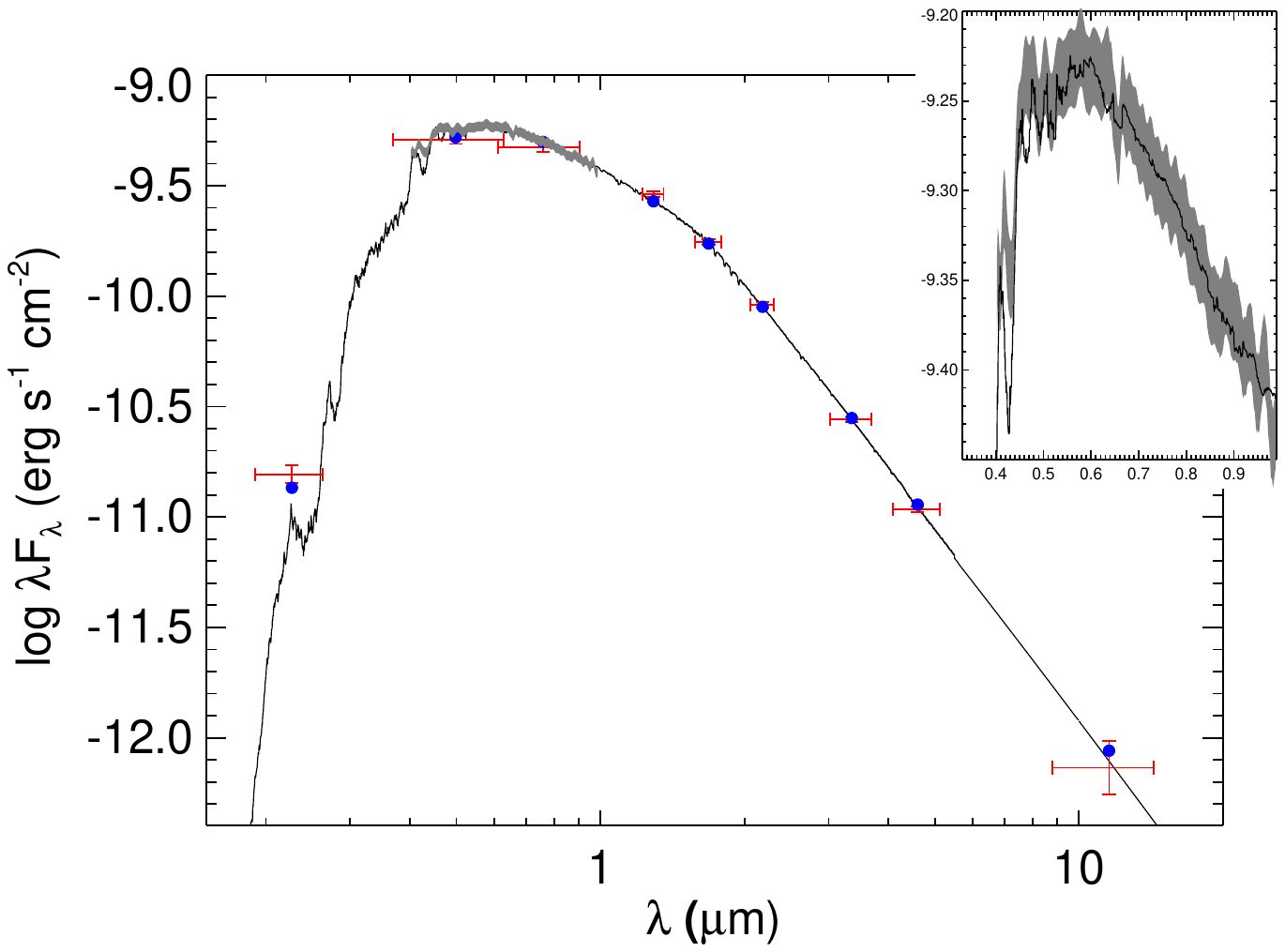} 
\caption{Spectral energy distributions of WASP-102 (top row left), WASP-116 (top row middle), WASP-149 (top row right), WASP-154 (second row left), WASP-155 (second row middle), WASP-188 (second row right), WASP-194 (bottom row left), WASP-195 (bottom row middle), and WASP-197 (bottom row right). Red symbols represent the observed photometric measurements, where the horizontal bars represent the effective width of the pass-band. Blue symbols are the model fluxes from the best-fit PHOENIX atmosphere model (black). The absolute flux-calibrated {\it Gaia\/} spectrum is shown as a grey swathe in the inset figure. \label{fig:sed}}
\end{figure*}

TRES spectra were used to derive stellar parameters using the Stellar Parameter Classification tool \citep[SPC;][]{Buchhave2012} for all stars with the exception of WASP-154. SPC cross correlates each observed spectrum against a grid of synthetic spectra based on Kurucz atmospheric models \citep{Kurucz1992} and derives effective temperature, surface gravity, metallicity, and the rotational velocity of the star. 

Stellar parameters of WASP-154 were derived from the combined SOPHIE spectra unpolluted by moonlight. We used the ARES+MOOG methodology described e.g. in \citep{Sousa2021}, together with ARES\footnote{https://github.com/sousasag/ARES} \citep{Sousa2015} to measure the equivalent widths of iron lines selected following \cite{Sousa2008}. A minimization process is used to find the ionization and excitation equilibrium and converge to the best set of spectroscopic parameters, using a grid of Kurucz model atmospheres \citep{Kurucz1993} and the radiative transfer code MOOG \citep{Sneden1973}.

An analysis of the broadband spectral energy distribution (SED) of each star was performed together with the {\it Gaia\/} DR3 parallax \citep{GaiaEDR3}, in order to determine an empirical measurement of the stellar radii \citep{Stassun:2016,Stassun:2017,Stassun:2018}. Where available, the $JHK_S$ magnitudes were sourced from {\it 2MASS}, the W1--W4 magnitudes from {\it WISE}, the $G_{\rm BP} G_{\rm RP}$ magnitudes from {\it Gaia}, and the NUV magnitude from {\it GALEX}. The absolute flux-calibrated {\it Gaia\/} spectrum was also utilized, when available. Together, the available photometry spans the full stellar SED over the wavelength range of at least 0.4--10~$\mu$m and as much as 0.2--20~$\mu$m (see Figure~\ref{fig:sed}). 

A fit using PHOENIX stellar atmosphere models \citep{Husser:2013} was performed, adopting from the spectroscopic analysis the effective temperature ($T_{\rm eff}$), metallicity ([Fe/H]), and surface gravity ($\log g$). The extinction $A_V$ was fitted for, limited to the maximum line-of-sight value from the Galactic dust maps of \citet{Schlegel:1998}. 
Integrating the (unreddened) model SED gives the bolometric flux at Earth, $F_{\rm bol}$. 
Taking the $F_{\rm bol}$ together with the {\it Gaia\/} parallax directly gives the bolometric luminosity, $L_{\rm bol}$. 
The Stefan-Boltzmann relation then gives the stellar radius, $R_\star$. 
In addition, the stellar mass was estimated using the empirical relations of \citet{Torres:2010}. 
Finally, the system age may be estimated from the observed rotation period of the star (or, if the rotation period is not known, from the projected rotation period $P_{\rm rot}/\sin i$ determined from the spectroscopically determined $v\sin i$ and the radius determined as above) and the empirical gyrochronology relations of \citet{Mamajek:2008}.
The resulting fits are shown in Figure~\ref{fig:sed}. The derived parameters are listed in Table~\ref{tab:stellar_properties}.


\section{System Modelling} \label{sec:fit}

    \thispagestyle{empty}
    \begin{table*}
    \centering
    \caption{Summary of fit and derived planetary system parameters for the nine giant planets. A full list of fitted parameters and prior values can be found in Appendix \ref{app:prior_posterior}, and corner plots for key system parameters are presented in Appendix \ref{app:corner_plots}. }
    \begin{tabular}{l c c c c c c}
    \hline
    \hline
         & WASP-102 & WASP-116 & WASP-149 & WASP-154 & WASP-155  \\
         & TOI-6170 & TOI-4672 & TOI-6101 & TOI-5288 & TOI-6135  \\
     \hline
\emph{Fit Parameters } \\
P (days)                           & 2.709813 $\pm$ 3e-7          & 6.61320 $\pm$ 2e-06          & 1.332813 $^{6e-7}_{5e-7}$          & 3.811678 $\pm$ 1e-06          & 3.110413 $\pm$ 1e-06  \\
t0 (BJD-2450000)                   & 7109.45577 $^{+1.6e-4}_{-1.7e-4}$         & 7092.22528 $^{+5.3e-4}_{-4.9e-4}$         & 7757.62450 $\pm$ 8e-5         & 9465.89195 $^{+2.8e-4}_{-2.7e-4}$         & 9852.08494 $\pm$ 4.1e-4 \\
b                                  & 0.11 $^{+0.06}_{-0.04}$          & 0.07 $^{+0.06}_{-0.05}$          & 0.58 $\pm$ 0.005          & 0.31 $\pm$ 0.04          & 0.43 $\pm$ 0.02 \\
R$_{\text{pl}}$/R$_*$                   & 0.0997 $\pm$ 0.0004 & 0.0881 $\pm$ 0.0004 & 0.1297 $^{+0.0008}_{-0.0009}$ & 0.12 $\pm$ 0.001 & 0.0997 $^{+0.0015}_{-0.0014}$ \\
$\rho_*$ (g cm$^{-3}$)             & 0.684 $^{+0.011}_{-0.016}$        & 0.404 $^{+0.007}_{-0.010}$        & 1.180 $\pm$ 0.009        & 2.012 $^{+0.067}_{-0.066}$        & 0.803 $^{+0.014}_{-0.011}$ \\
K (m s$^{-1}$)                     & 86.16 $^{+4.37}_{-4.28}$          & 59.67 $^{+3.09}_{-3.01}$          & 175.25 $^{+5.19}_{-5.31}$          & 94.55 $^{+2.46}_{-2.44}$          & 114.29 $^{+2.50}_{-2.53}$ \\
eccentricity                       & 0 (fixed) & 0 (fixed) & 0 (fixed) &  0 (fixed) & 0 (fixed) \\
        \hline
\emph{Derived Parameters} \\
Rp (R$_{\text{Jup}}$)                        & 1.33 $\pm$ 0.05        & 1.22 $\pm$ 0.06        & 1.36 $\pm$ 0.06        & 0.96 $\pm$ 0.06        & 1.20 $\pm$ 0.06 \\
Mp (M$_{\text{Jup}}$)                        & 0.622 $\pm$ 0.133        & 0.64 $\pm$ 0.14        & 0.991 $\pm 0.196$        & 0.626 $\pm$ 0.129        & 0.871 $\pm$ 0.18 \\
pl density (g cm$^{-3}$)              & 0.32 $\pm$ 0.08   & 0.43 $\pm$ 0.12   & 0.49 $\pm$ 0.12   & 0.88 $\pm$ 0.25   & 0.62 $\pm$ 0.17 \\
depth (ppm)                         & 9949 $\pm$ 986     & 7763 $\pm$ 989     & 16834 $\pm$ 2170     & 14400 $\pm$ 2338     & 9940 $^{1393}_{-1388}$ \\
a/R$_s$                             & 6.43 $^{+0.04}_{-0.05}$ & 9.77 $^{+0.05}_{-0.08}$ & 4.8 $\pm$ 0.01 & 11.56 $\pm$ 0.13 & 7.43 $\pm$ 0.04 \\
a (AU)                              & 0.041 $\pm$ 0.001         & 0.065 $\pm$ 0.003         & 0.024 $\pm$ 0.001         & 0.044 $\pm$ 0.003         & 0.043 $\pm$ 0.002 \\
i (deg)                             & 89.06 $^{+0.57}_{-0.39}$         & 89.57 $^{+0.05}_{-0.28}$         & 83.02 $^{+0.38}_{-0.37}$         & 88.46 $^{+0.3}_{-0.27}$         & 86.7 $\pm$ 0.29 \\
T$_{\text{eq}}$ (K)                      & 1671 $\pm$ 65      & 1414 $\pm$ 70       & 1855 $\pm$ 93       & 994 $\pm$ 63       & 1468 $\pm$ 76 \\
S$_{\text{pl}}$ (S$_{\oplus}$) & 1293 $\pm$ 155 & 663 $\pm$ 100 & 1966 $\pm$ 305 & 162 $\pm$ 31 & 771 $\pm$ 119 \\
TSM$^\dag$                   & 86 $\pm$ 20       & 54 $\pm$ 13       & 189 $\pm$ 43       & 57 $\pm$ 14       & 54 $\pm$ 13 \\
        \hline

\\
\\
    \hline
    \hline
         & WASP-188 & WASP-194 & WASP-195  & WASP-197  \\
         & TOI-5190 & TOI-3791 & TOI-4056  & TOI-5385  \\
         &          & HAT-P-71 &           &           \\
     \hline
\emph{Fit Parameters} \\
P (days)                           & 5.748916 $\pm$ 3e-06          & 3.183387 $^{4e-7}_{5e-7}$          & 5.051928 $\pm$ 4e-6          & 5.167228 $\pm$ 3e-06  \\
t0 (BJD-2450000)                   & 7033.12141 $\pm$ 0.001         & 7449.0511 $\pm$ 0.0003         & 7357.23855 $^{+0.0022}_{-0.00185}$         & 6885.10428 $^{+0.00166}_{-0.00173}$ \\
b                                  & 0.61 $\pm$ 0.01          & 0.84 $\pm$ 0.040          & 0.55 $\pm$ 0.01          & 0.43 $\pm$ 0.02 \\
R$_{\text{pl}}$/R$_*$                   & 0.0742 $\pm$ 0.0005 & 0.1007 $\pm$ 0.0004 & 0.0600 $^{+0.0018}_{-0.0016}$ & 0.0627 $\pm$ 0.0007 \\
$\rho_*$ (g cm$^{-3}$)               & 0.345 $\pm$ 0.002        & 0.683 $^{+0.027}_{-0.016}$        & 0.466 $^{+0.004}_{-0.003}$        & 0.204 $\pm$ 0.002 \\
K (m s$^{-1}$)                     & 124.63 $^{+5.82}_{-6.49}$          & 135.90 $^{+13.42}_{-13.26}$          & 10.32 $^{+2.16}_{-2.19}$          & 121.32 $^{+3.7}_{-3.59}$ \\
eccentricity                       & 0 (fixed) & 0 (fixed) & 0 (fixed) & 0 (fixed) \\
        \hline
\emph{Derived Parameters} \\
Rp (R$_{\text{Jup}}$)                        & 1.322 $\pm$ 0.051        & 1.381 $\pm$ 0.088        & 0.92 $\pm$ 0.05        & 1.289 $\pm$ 0.049 \\
Mp (M$_{\text{Jup}}$)                        & 1.443 $^{+0.296}_{-0.298}$        & 1.174 $\pm 0.265 $        & 0.104 $^{0.03}_{0.031}$         & 1.269 $\pm$ 0.254 \\
pl density (g cm$^{-3}$)              & 0.78 $\pm$ 0.20   & 0.55 $\pm$ 0.18   & 0.16 $\pm$ 0.06   & 0.74 $\pm$ 0.18 \\
depth (ppm)                         & 5509 $\pm$ 593     & 10143 $\pm$ 1835     & 3602 $^{+477}_{-466}$     & 3931 $^{+414}_{-415}$ \\
a/R$_s$                             & 8.46 $\pm$ 0.01 & 7.15 $\pm$ 0.09 & 8.57 $\pm$ 0.02 & 6.6 $\pm$ 0.03 \\
a (AU)                              & 0.072 $\pm$ 0.003         & 0.047 $\pm$ 0.003         & 0.063 $\pm$ 0.003         & 0.065 $\pm$ 0.002 \\
i (deg)                             & 85.88 $\pm$ 0.20         & 83.23 $^{+0.47}_{-0.46}$         & 86.30 $\pm$0.25         & 86.29 $\pm$ 0.32 \\
T$_{\text{eq}}$ (K)                      & 1666 $\pm$ 70       & 1693 $\pm$ 122       & 1522 $\pm$ 71       & 1665 $\pm$ 67 \\
S$_{\text{pl}}$ (S$_{\oplus}$) & 1278 $\pm$ 165 & 1365 $\pm$ 303 & 890 $\pm$ 127 & 1276 $\pm$ 157 \\
TSM$^\dag$                        & 23 $\pm$ 5       & 59 $\pm$ 16       & 153 $\pm$ 48       & 28 $\pm$ 6 \\
        \hline

    \end{tabular}

    \begin{tablenotes}
        \item \dag: Transmission Spectroscopy Metric \citep{Kempton2018}. 
     \end{tablenotes}
    \label{tab:planet_properties}
    \end{table*}

We fit each planetary system using the \texttt{juliet} code package \citep{Espinoza2019} using the \texttt{dynesty} sampler \citep{Speagle2020}. \texttt{juliet} allows for joint fitting between photometric and radial velocity datasets. For each system, we first fit two models to the RV data, one with eccentricity fixed to 0 and another which allowed the orbit to be eccentric using the $\sqrt{e}\sin{\omega}$ and $\sqrt{e}\cos{\omega}$ parameterization. We then compared the log-evidence to determine which model is supported given the data. We used the Jeffrey's scale \citep{Jeffreys1939} to interpret the resulting odds ratio for each planet. For 7 systems, there was moderate evidence to support the circular orbit, with odds ratios ranging between 7.3 and 19.4. The remaining two planets, WASP-149 and WASP-195, showed weak evidence supporting the circular orbit, with odds ratios of 3.7 and 2.9 respectively. For these two planets, we jointly fit the photometry and radial velocity data with and without eccentricity. In both cases, the circular model was preferred, with an odds ratio of 252 tor WASP-149 and an odds ratio of 6239 for WASP-195. Therefore, in all cases we held the eccentricity fixed for the final global (transit + RV) modeling. 

For all models, we use the approximate Mat\'ern kernel Gaussian Process (GP) to account for the presence of stellar or systematic variability in TESS data. We used the pipeline-produced detrended lightcurves from WASP and HATNet, binned to a 5-minute cadence. For other ground-based transit observations, we fit models with and without linear detrending to airmass. When the log evidence favors the detrending, we incorporate it into the final model. Appendix \ref{app:prior_posterior} provides a full list of the parameters used for each system, including any detrending parameters used. 

The final fit parameters were period (P), epoch (t0), impact parameter (b), planet-to-star radius ratio ($R_p$/$R_*$), stellar density ($\rho_*$), and RV semi-amplitude (K). All datasets are fit with an additional baseline offset parameter, as well as a jitter term. We use the $q_1, q_2$ parameterization for quadratic limb darkening \citep{exoplanet:kipping13}, with priors set from values determined using \texttt{ldtk} \citep{Parviainen2015_ldtk, Husser:2013}. Limb darkening parameters are shared for observations taken using the same photometric filter. For WASP and HAT, the data does not highly constrain limb darkening, so we share the parameters with that of TESS.

The TESS and TESS-SPOC pipelines account for the contribution of nearby sources in the \texttt{PDCSAP} lightcurve product, and the QLP pipeline applies deblending for additional sources within the target aperture. These contamination correction methods rely on brightness estimates of nearby Gaia sources in the TESS input catalog, so we include the additional dilution term to account for any resulting over or under corrections. This term is defined in Juliet as $D = \frac{1}{1+\sum_{n} F_n/F_T}$ where the dilution (D) is a number between 0 and 1, 1 being no additional dilution. 
More details on the model fits can be found in the sections below, and a summary of the results of the final fits can be found in Table \ref {tab:planet_properties}. All datasets used in the final models are available through DRUM\footnote{http://hdl.handle.net/1903/33819}.

\subsubsection{WASP-102}
\begin{figure}[!htbp]
    \centering
    \includegraphics[width=.9\linewidth]{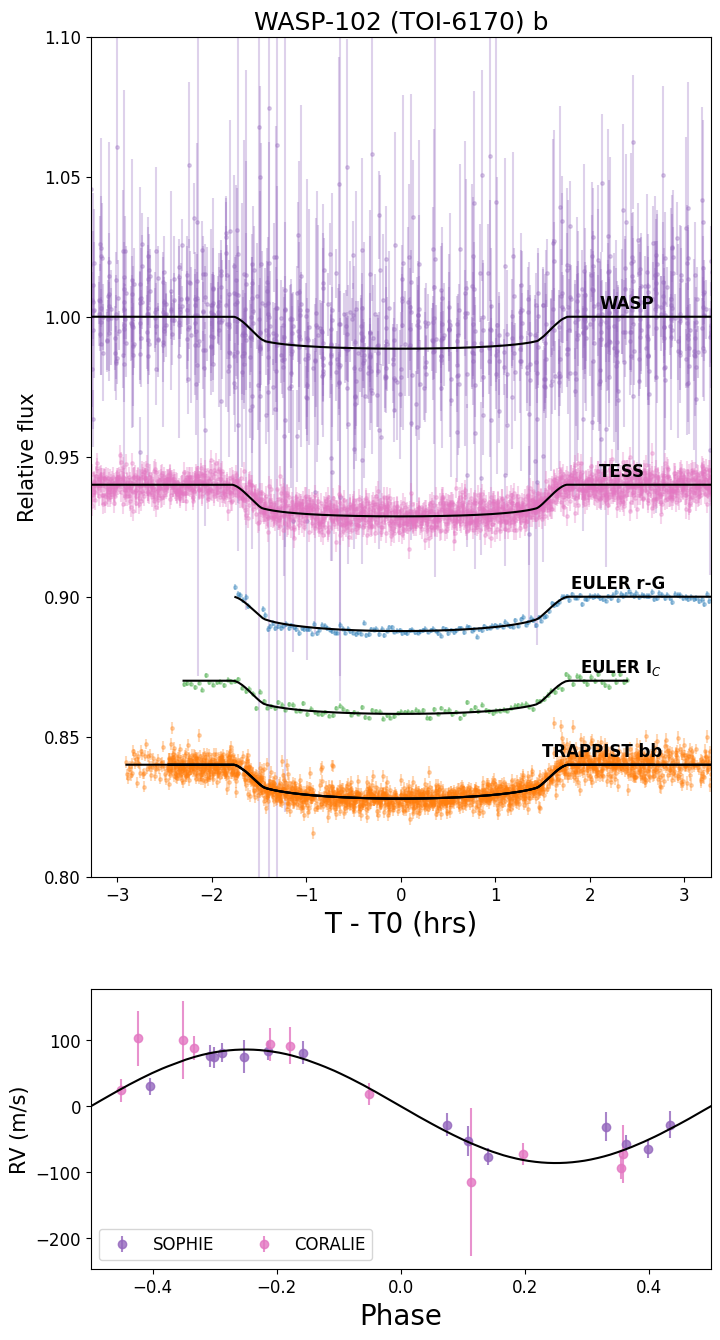}
    \caption{Transit and RV fit for WASP-102\,b. Transit observations are offset for clarity. The three TRAPPIST observations were taken with the same filter, and are plotted together. }
    \label{fig:WASP102global}
\end{figure}

WASP-102\,b was first publicly alerted in \cite{Faedi2016}. We present here a reanalysis of the data used in this work and updated with new TESS observations (See Fig. \ref{fig:WASP102global}). WASP-102 is a well-isolated star with low contamination in TESS and WASP data. High contrast imaging of the star by SOAR rules out a close contaminant within 4.9 mag at a separation of 1". 

After WASP identified this planet candidate, ground based photometric observations were made by EulerCam (Gunn-\textit{r} filter) as well as TRAPPIST (blue blocking filter) in 2013, obtaining a total of 5 full transits. Radial velocity measurements were also collected by SOPHIE and CORALIE. TESS observed the star in Sector 56 in the full-frame images at a 200-s cadence. The QLP faint star search identified this as a planet candidate in March of 2023. The transit was also detected in the TESS-SPOC FFI light curve search of this sector, and the location of WASP-102 was located within 0.967 ± 2.5$\arcsec$ of the transit source location by the difference image centroiding analysis in the TESS-SPOC Data Validation(DV) report \citep{Twicken2018}.

We jointly fit the WASP, TESS, TRAPPIST, EulerCam, SOPHIE, and CORALIE data. We include GP detrending for TESS and linear detrending model with airmass for the ground based follow up observations. Consistent with \cite{Faedi2016} which reported M$_{\text{pl}}$=0.62$\pm$ 0.05 M$_{\text{Jup}}$ and R$_{\text{pl}}$=1.26 $\pm$ 0.02 R$_{\text{Jup}}$, we find the planet to have a mass of 0.62$\pm$ 0.13 M$_{\text{Jup}}$ and a radius of 1.33$\pm$ 0.05 R$_{\text{Jup}}$. 

\subsubsection{WASP-116}

WASP-116 was observed by both SuperWASP and WASP-South from August 2008 through December 2010. A planet transit was identified in this combined dataset, leading to spectroscopic follow-up by both the SOPHIE and CORALIE instruments. Additional photometric follow-up was taken by TRAPPIST and EulerCam, with the two instruments simultaneously observing two different transit ingress events. WASP-116\,b was subsequently publicly announced by \cite{Brown2014}. 

\begin{figure}
    \includegraphics[width=.9\linewidth]{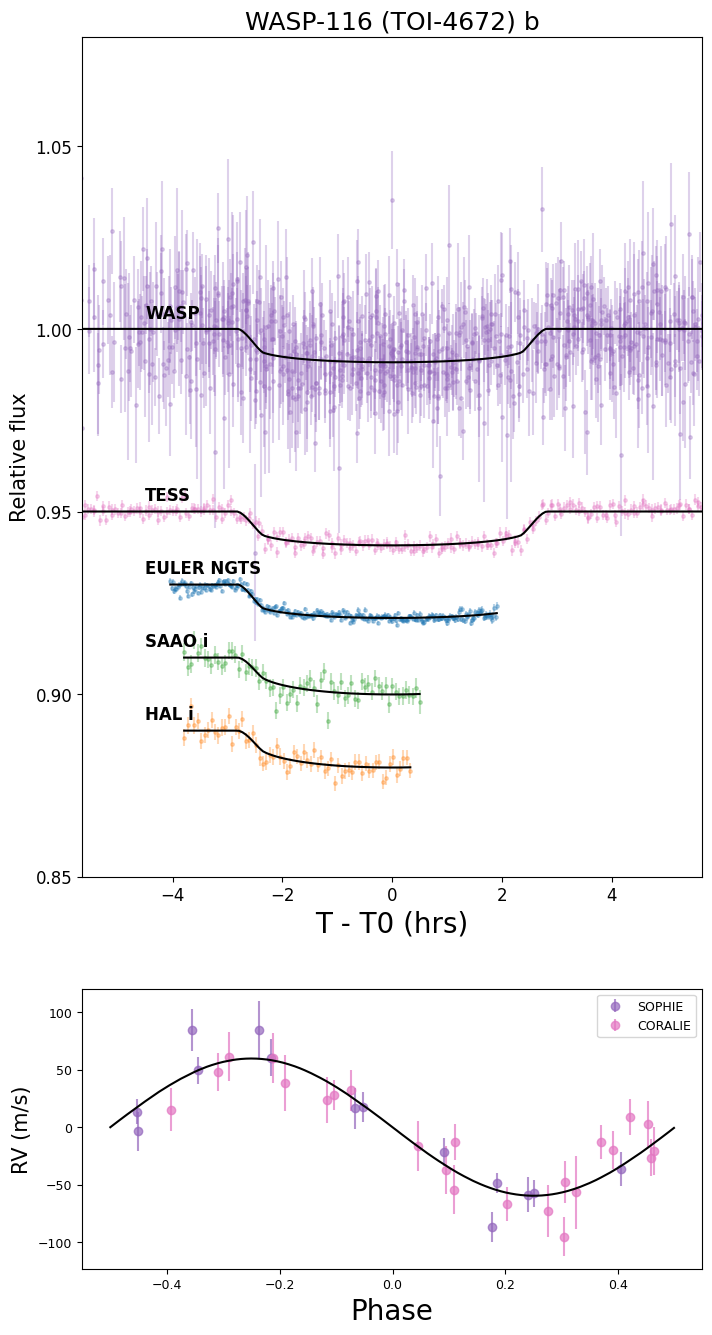}

    \caption{Global fit for WASP 116\,b. Photometric lightcurves have been offset for visual clarity. }
    \label{fig:WASP116global}
\end{figure}

TESS observed this target in the 30-minute cadence Full Frame Images (FFIs) in the primary mission. The target was revisited by TESS in its first mission extension, in which the FFI cadence was reduced to 10 minutes. The Quick-Look Pipeline Faint Star Search \citep{Kunimoto2022QLP} identified this target as a TOI in December 2021 following these observations. After being given a TOI designation, several observations were submitted through TFOP including LCOGT observations from SAAO and Haleakala taken with the \textit{i'} filter. We note that an additional observation was submitted from Hazelwood observatory using a Sloan \textit{g'} filter. However, an ill-timed gap due to clouds immediately following the meridian flip coinciding with the transit egress make the transit depth difficult to constrain. We therefore remove this dataset from our analysis. The final best-fit transit and RV models can be seen in Figure \ref{fig:WASP116global}. We find the planet is in a 6.61 day orbit and has a mass of 0.64 $\pm$ 0.14 M$_{\text{Jup}}$ and a radius of 1.22 $\pm$ 0.06 R$_{\text{Jup}}$, which is consistent to the values of 0.59 $\pm$ 0.05 M$_{\text{Jup}}$ and 1.43 $\pm$ 0.07 R$_{\text{Jup}}$ found in \cite{Brown2014} within 2$\sigma$.

\subsubsection{WASP-149}

Like WASP-116\,b, WASP-149\,b was introduced in \cite{Brown2014}. The star was observed by both SuperWASP and WASP South from November 2009 through March 2012. A joint analysis of the WASP datasets identified a planetary candidate, leading to SOPHIE and CORALIE radial velocity measurements. Photometric followup from NITES, TRAPPIST, and EulerCam were also collected; however, NITES data from the initial analysis was not recovered and could not be included in this analysis. 

WASP-149 was later observed by TESS FFIs with 30-minute cadence in sector 7 and 10 minute cadence in Sector 34. When it was again re-observed in Sector 61, the star was selected for 2-minute cutout data, and was processed by the SPOC pipeline. The transit was identified by SPOC, with centroiding analysis locating the transit source to 0.47 $\pm$ 2.4". 

For TESS, we include only the 2-minute data in our final analysis. As part of the TFOP program, MuSCAT2 observations containing an egress event were uploaded to the ExoFOP webpage. The multi-band observations are consistent in depth supporting the planet interpretation, however the lack of a pre-transit baseline makes a definitive depth analysis impossible, and we do not include the data in the global analysis. However, deep eclipses from nearby targets can be ruled out from this dataset. 

\begin{figure}
    \includegraphics[width=.9\linewidth]{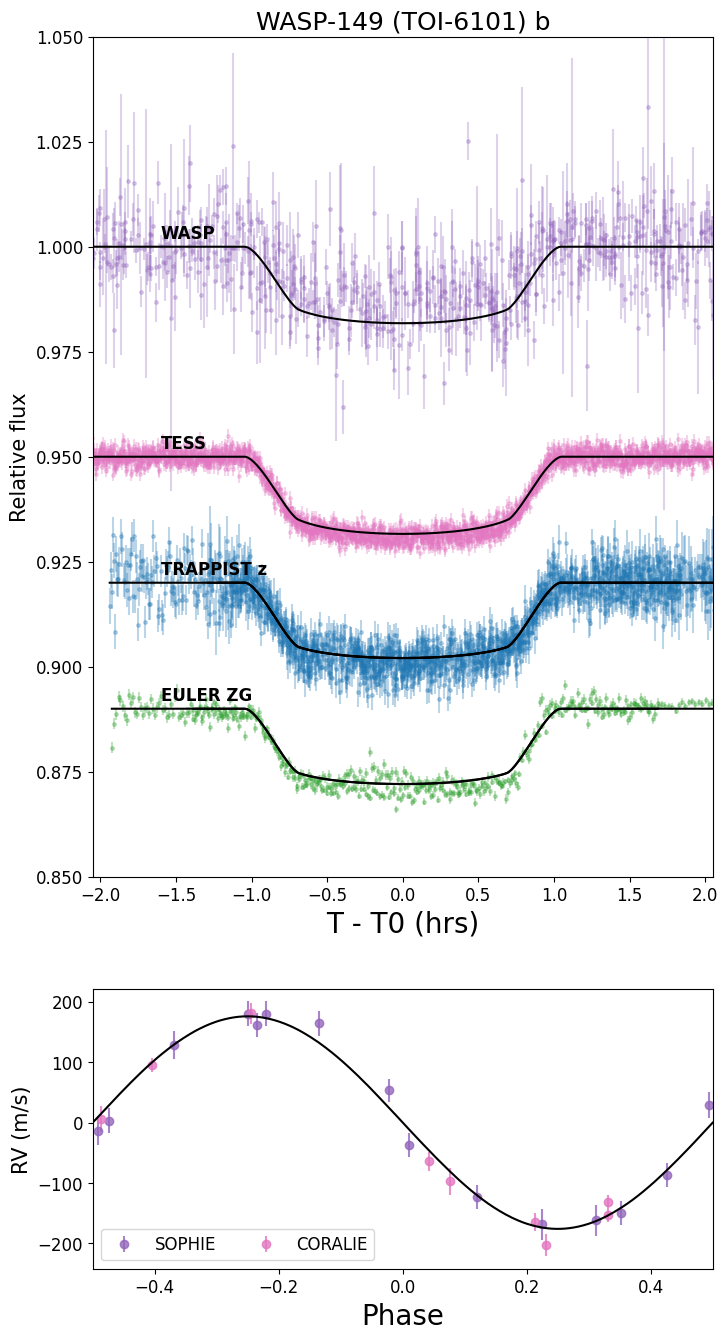}

    \caption{Final joint transit and RV model fit for WASP-149\,b. The top panels show the available photometric data, offset for clarity. The bottom panel shows the final RV fit for SOPHIE and CORALIE data. }
    \label{fig:WAsP149global}
\end{figure}

The final planet model along with the data used can be seen in Figure \ref{fig:WAsP149global}. Orbiting with a period of 1.33 days, WASP-149\,b has the shortest period of the planets presented in this work. The mass of 0.99 $\pm$ 0.20 M$_{\text{Jup}}$ and a radius of 1.36 $\pm$ 0.06 R$_{\text{Jup}}$ is consistent within 1-$\sigma$ to the values of 1.02 $\pm$ 0.04 M$_{\text{Jup}}$ and 1.32 $\pm$ 0.04 R$_{\text{Jup}}$ found in \cite{Brown2014}.

\subsubsection{WASP-154}
WASP-154 was observed by both SuperWASP and WASP South facilities from June 2008 - October 2010. After being identified as a planet candidate in 2013, radial velocity follow-up from SOPHIE began. In addition, observations of the full transit were captured by TRAPPIST, EulerCam, and NITES. 

\begin{figure} 
    \includegraphics[width=0.9\linewidth]{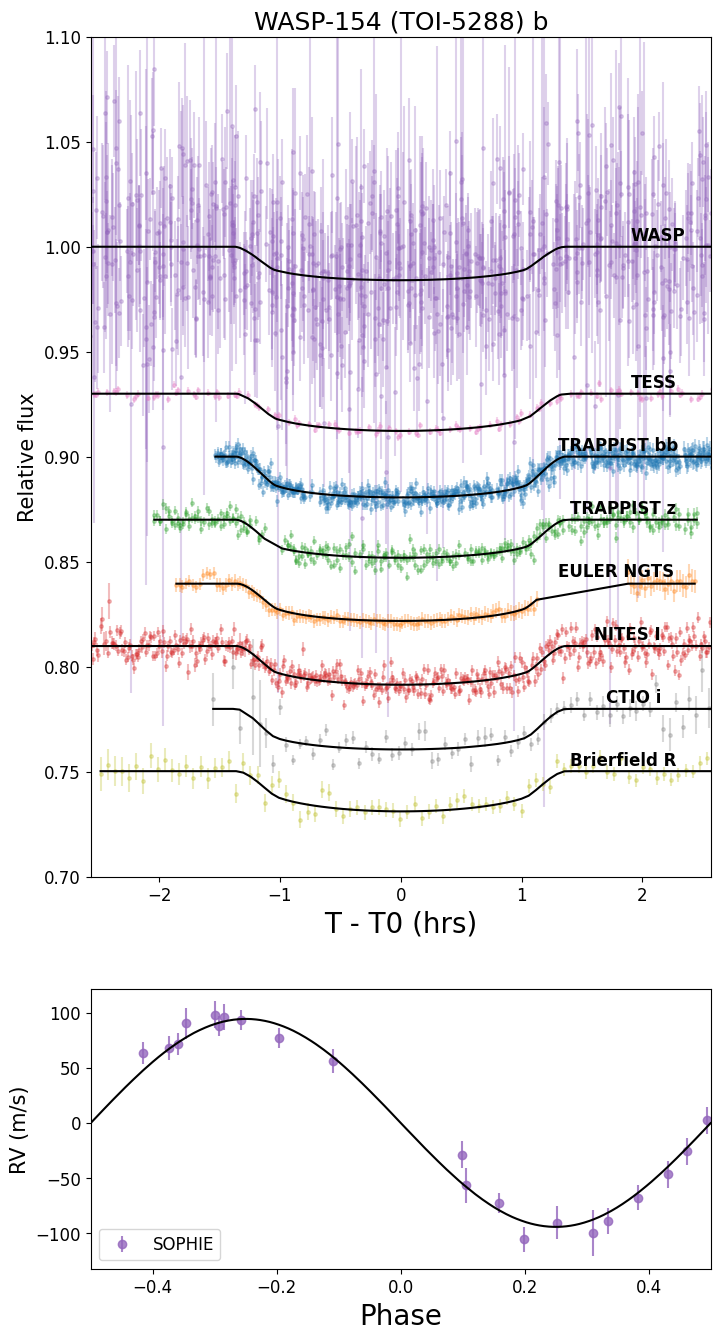}
    \caption{Final joint transit and RV model fit for WASP-154\,b. The top panel shows the photometric datasets, offset for clarity, while the bottom panel shows the fit to the radial velocity data. }
    \label{fig:WASP154global}
\end{figure}

This target was subsequently observed in TESS Sector 42 FFIs with a 10 minute cadence. The TESS-SPOC identified the transit, with centroiding analysis showing the transit location within 0.82 $\pm$ 2.50" from the star. The QLP faint target search \citep{Kunimoto2022QLP} identified this as a TOI in 2022. This triggered additional photometric follow-up observations at the LCO/CTIO and Brierfield facilities. The final transit and RV models along with the data used in the model are found in Figure~\ref{fig:WASP154global}. The results show this is a nearly Jupiter sized planet (R$_{\text{pl}}$=0.96 $\pm$ 0.06 R$_{\text{Jup}}$) but with a mass of 0.63 $\pm$ 0.13 M$_{\text{Jup}}$ in a 3.81 day orbit.

\subsubsection{WASP-155}\label{sec:fit_wasp155}
WASP-155 was observed by SuperWASP from May 2004 through November 2007. In 2014, the WASP team identified the lightcurve as a potential planet transit and initiated SOPHIE radial velocity measurements. Initial observations showed the RVs were in phase and suggested a planetary mass, and a photometric follow-up observation by NITES further confirmed the transit. 

\begin{figure}
    \includegraphics[width=.9\linewidth]{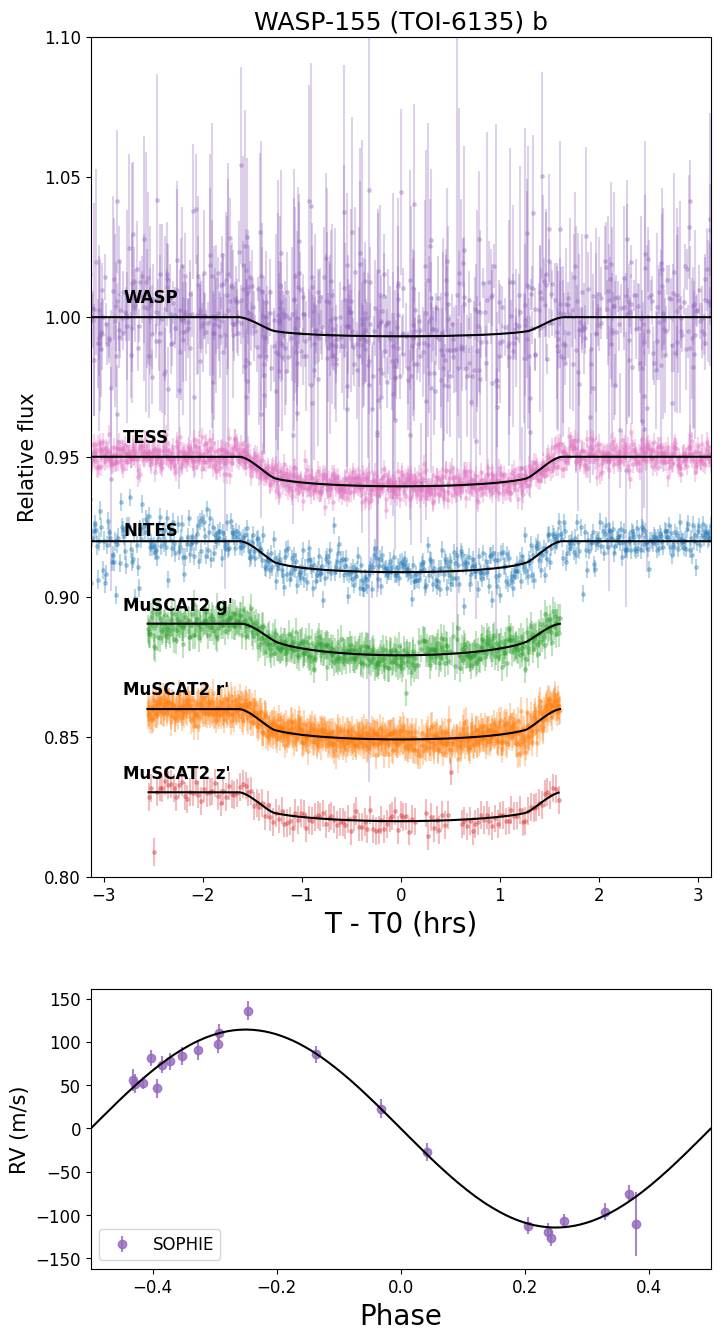}
    \caption{Final joint transit and RV model fit for WASP-155\,b. Photometric observations are offset for clarity.  }
    \label{fig:WASP155global}
\end{figure}

The star was included in the FFI images from TESS Sector 56 at a 200-s cadence and identified as a TOI in the faint star search using the QLP pipeline \citep{Kunimoto2022QLP}. We use the detrended and deblended data (\texttt{det\_flux}) from the QLP pipeline in our fit. The MuSCAT2 instrument made four multi-band observations of WASP-155 and found achromatic transits across the $g$, $r$, $i$, and $z_\mathrm{s}$ bands. In our final fit we include only one of these full transit multi-wavelength observations (See Figure \ref{fig:WASP155global}). 

Speckle images taken as part of the TFOP tentatively suggests a faint companion $\sim$3" from the target with a $\Delta$mag of 5.26 using the 832 nm filter. Gaia also reports a companion at the same separation, with a $\Delta$mag of 2.55 in the Gaia passband. Due to the wide separation, the $\Delta$mag value measured in the speckle data is likely to be overestimated. There are two effects to consider, both of which would lead to an anomalously large $\Delta$mag. One is that the companion star fell close to the edge of the CCD readout region, causing some of the companion star's speckle patterns to lie outside of this region and go undetected. Additionally, the correlation of adjacent speckle patterns decreases with increasing angular separation and this has not been accounted for in the NESSI photometry. The reported Gaia parallax and proper motion for both stars are similar, suggesting that the stars may be bound. More observations could determine whether this is truly a binary pair. The proximity to the nearby star leads to blending within the target aperture in the photometric observations. We therefore center the prior for the dilution factor to the expected dilution resulting from a source with $\Delta$mag 2.55 as seen in Gaia. 

We find WASP-155\,b to be a 1.91 R$_{\text{Jup}}$, 0.86 M$_{\text{Jup}}$ planet orbiting with a period of 3.11 days. The estimated temperature of the planet is just shy of 1500 K. 

\subsubsection{WASP-188}

WASP-188 was observed by SuperWASP from May 2004 through August 2010 and flagged as a planet candidate in 2014. RV observations with SOPHIE showed variations in phase with the transit period. Subsequent multi-band photometric observations by MuSCAT2 and KeplerCam were made to refine the ephemeris as well as check for achromatic depths. The observations for KeplerCam and MuSCAT2 were binned to a 2-minute cadence to reduce scatter during the model fit.  

\begin{figure}
    \includegraphics[width=.9\linewidth]{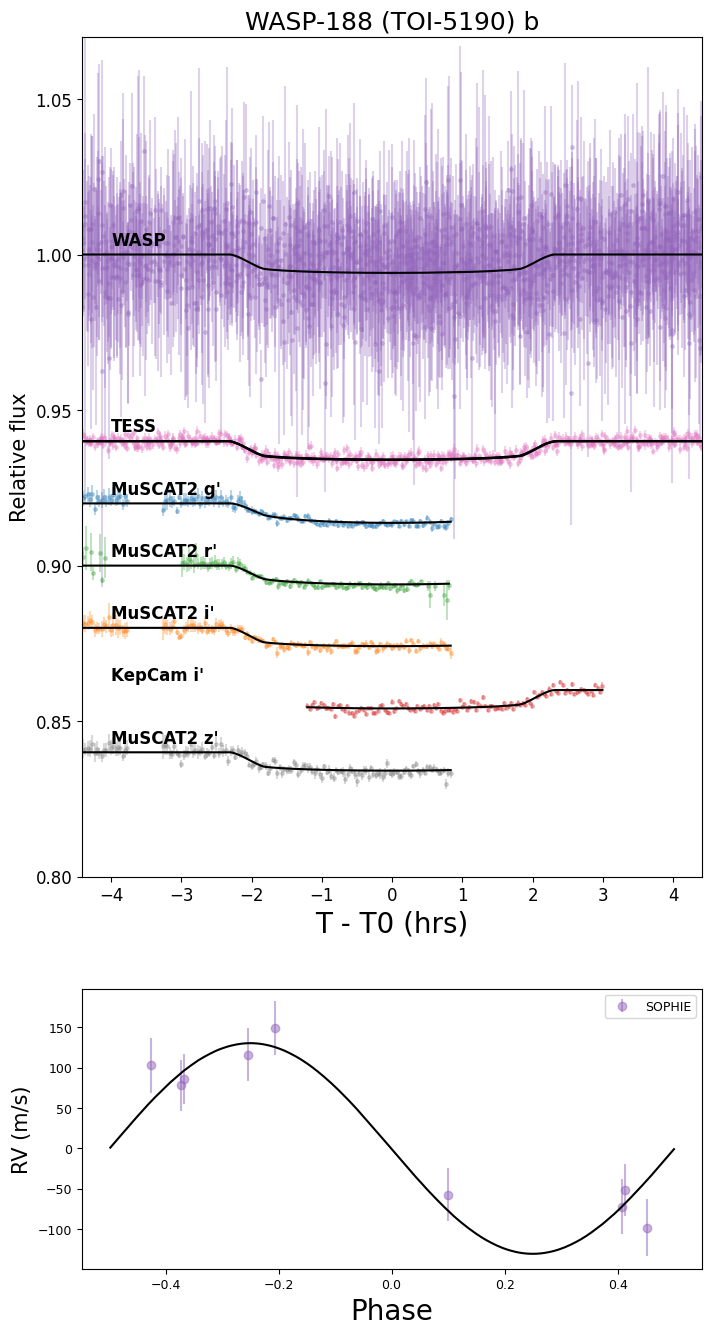}

    \caption{Global fit for WASP 188\,b. Photometric lightcurves have been offset and the TESS-SPOC lightcurves for sectors 40, 53, and 54 are combined in the final plot for visual clarity. The KeplerCam and MuSCAT2 data shown here are binned to a 2-minute cadence, which was used when fitting the model. }
    \label{fig:WASP188global}
\end{figure}

Gaia observations revealed that WASP-188 has a close companion ($\Delta$mag $\approx$5.5) at 1.81$\arcsec$, which was confirmed by speckle images taken at SAI, contributing flux to the photometric apertures. The TESS-SPOC PDCSAP lightcurve does take into account the contamination of this and other nearby stars in the aperture using the CROWDSAP (crowding) and FLFRCSAP (completeness) measures. However, there is still blending in the MuSCAT2 and KeplerCam lightcurves even with the smaller pixel scales. We set the dilution prior for these datasets based on the expected dilution for a target with the reported magnitude difference. The resulting fit can be found in Figure \ref{fig:WASP188global}. We find the planet to have a radius of 1.33 $\pm$ 0.05 R$_{\text{Jup}}$ and a mass of 1.52$^{+0.32}_{-0.31}$ M$_{\text{Jup}}$. 

\subsubsection{WASP-194/HAT-P-71}
WASP-194\,b was independently identified as a planet candidate with a 3.32 day period by both the WASP and HATNet \citep{HATNet} surveys. TRES radial velocity observations of this target began in 2013, showing that the object was consistent with a planetary mass. KeplerCam confirmed and refined the transit with 8 observations spanning from 2014-2017. TESS observed this target in FFIs in Sectors 14, 15, 16, and 20. The star was elevated to a TOI through the QLP faint star search. Subsequent TESS Sectors observed this star with 2-minute cutouts. Additional ground-based follow-up observations were taken by MuSCAT2 and RC8GSO. 

\begin{figure}
    \includegraphics[width=.9\linewidth]{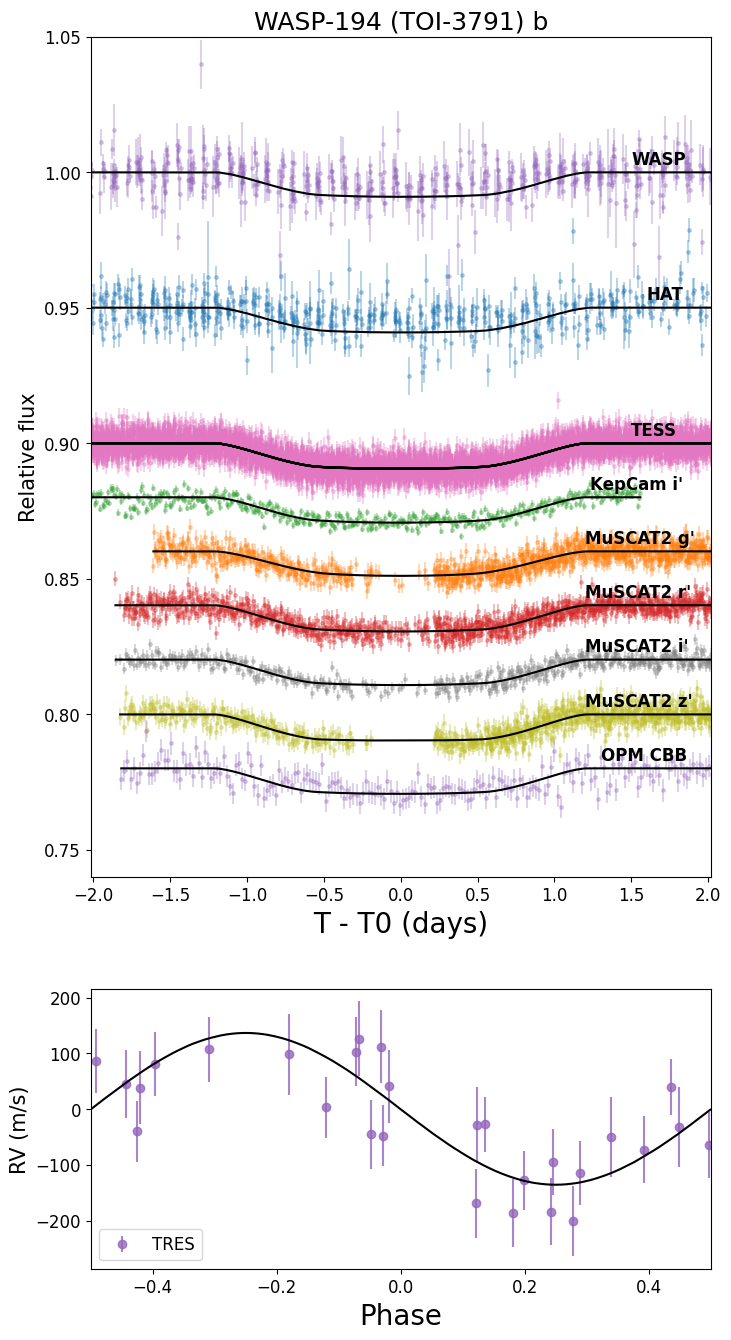}

    \caption{Final joint transit model fit for WASP-194\,b. Photometric observations have been offset and the 8 TESS sectors of observations are combined for clarity.}
    \label{fig:WASP194global}
\end{figure}

The final fit for this planet (See Fig. \ref{fig:WASP194global}) included photometric data from WASP, HAT, 120-s lightcurves produced by the SPOC for TESS sectors 40, 41, 50, 54, 55, 56, 57, and 60, MuSCAT2 observations in $g$, $r$, $i$, $z_\mathrm{z}$, and one full KeplerCam and OPM observation, as well as the TRES RV data. The planet is found to have a mass of 1.17$\pm$ 0.27 M$_{\text{Jup}}$ and a radius of 1.38 $\pm$ 0.09 R$_{\text{Jup}}$.

\subsubsection{WASP-195}
\begin{figure}
    \includegraphics[width=.9\linewidth]{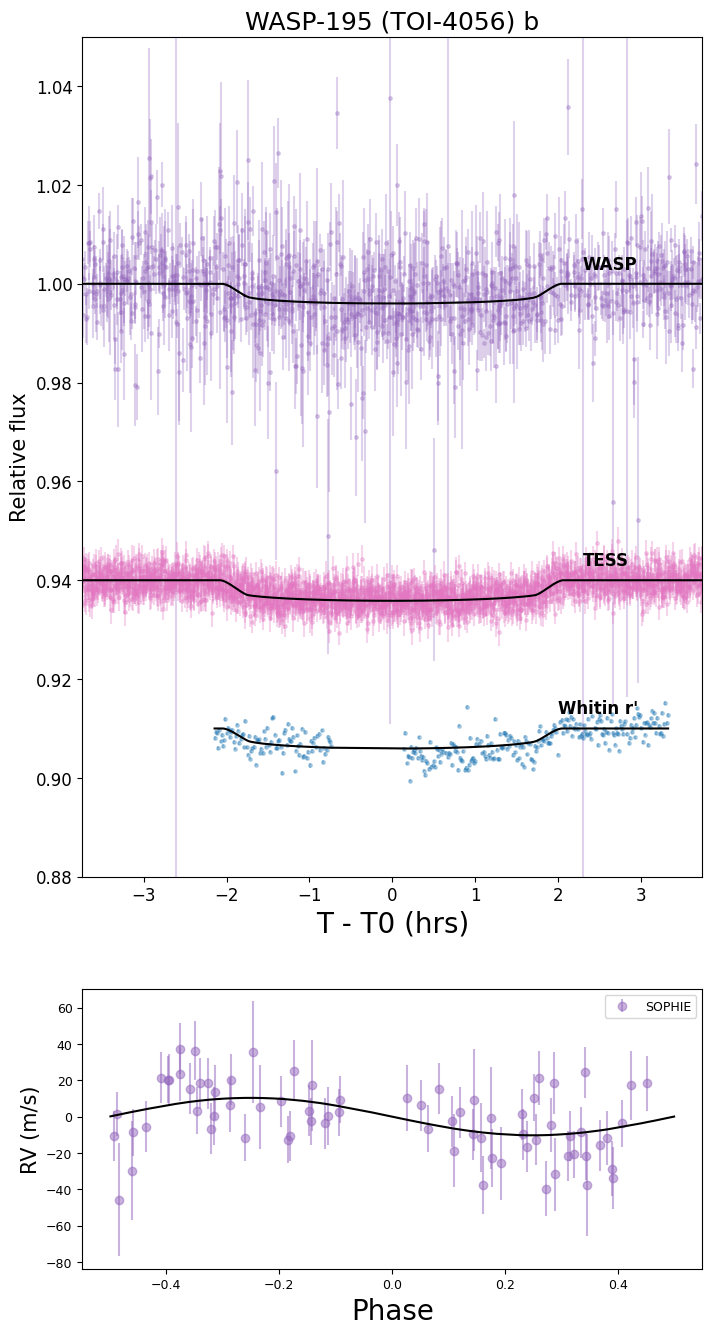}

    \caption{Final joint transit model fit for WASP-195\,b. TESS Sectors 50 and 52 are shown together, and all lightcurves are offset for clarity.}
    \label{fig:WASP195global}
\end{figure}

WASP-195\,b was first flagged as a planet candidate with a 5.05 day orbit by WASP in February 2014. Shortly after, extensive RV follow up observations by SOPHIE began, with 88 observations to date. TESS later observed the star in Sectors 23--25 in the full frame images with a cadence of 30 minutes. The star was again observed by TESS two years later in sectors 50 and 52, this time with 2-minute cutout data, and the transit was identified as a TOI. Centroiding analysis shows the transit source is within 0.076 $\pm$ 2.5" from the star. As part of the TFOP effort, three high-resolution images were taken to identify nearby stars. These observations were able to rule out a companion within 6.5 mag at 0.5". Finally, a photometric lightcurve with a Sloan-\textit{r'} filter was taken by Whitin Observatory. While conditions were cloudy leading to a temporary stop in observations during the transit, the egress was clearly observed.
We fit a transit model using WASP, TESS sectors 50 and 52, and Whitin photometry along with SOPHIE RV measurements (Fig. \ref{fig:WASP195global}). The fit reveals a puffy planet with radius of 0.92 $\pm$ 0.05 R$_{\text{Jup}}$ but a mass of only 0.10$\pm$ 0.03 M$_{\text{Jup}}$. We discuss this curious system in more detail in Section \ref{sec:discussion}. 

\subsubsection{WASP-197}
WASP-197\,b was identified as a planet candidate by \cite{Schanche2019_machine_learning}, and was soon after identified as a TOI by SPOC following TESS observations in Sector 48. There is a nearby star (DR3 734156214654777984) identified by Gaia with a separation of 6.7". This star has an estimated $\Delta$mag 5.6 in the TESS bandpass. To check for possible additional nearby companions, several high-contrast imaging facilities observed the star, establishing a limit of $\Delta$6.8 mag at 0.5" in the K band. Combined centroiding analysis of TESS data disfavors this as a transit source, with an offset of 0.362 $\pm$ 2.6".

\begin{figure}
    \includegraphics[width=0.8\linewidth]{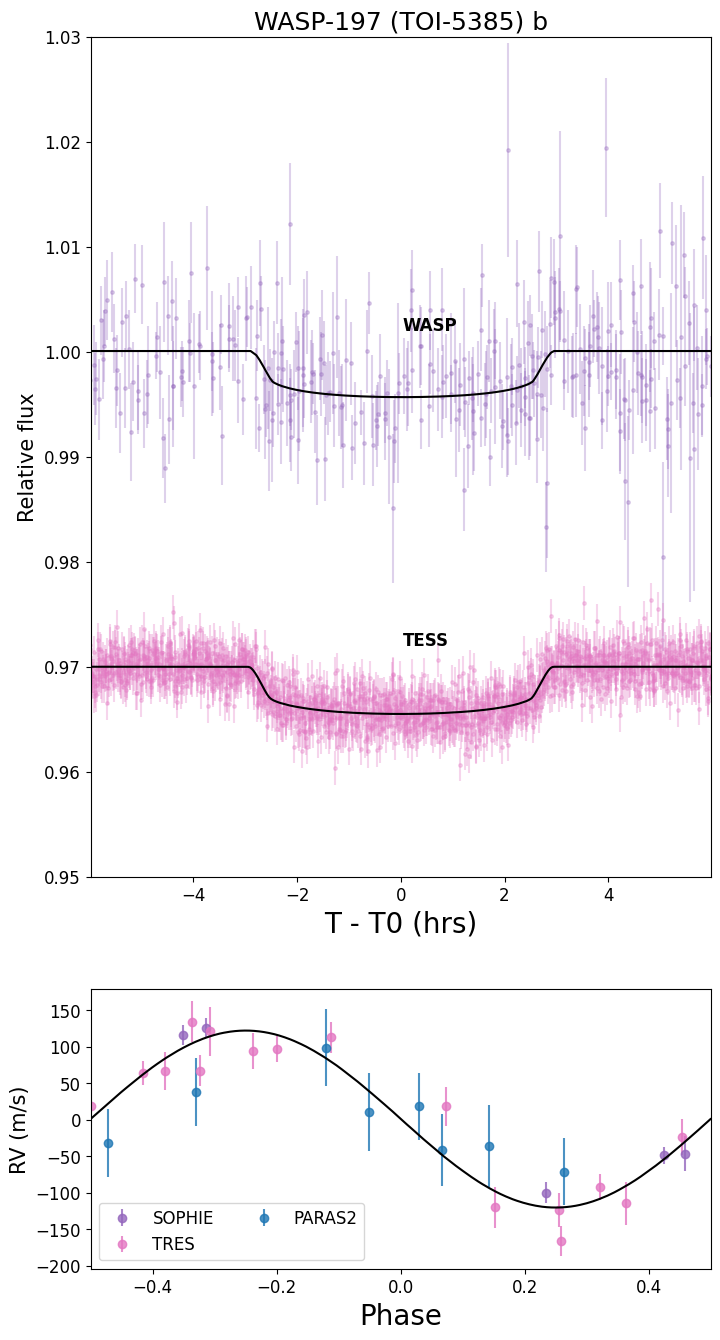}
    \caption{Final joint transit model fit for WASP-197\,b. WASP and TESS lightcurves are offset for clarity.}
    \label{fig:WASP197global}
\end{figure}

Follow up undertaken by three facilities, SOPHIE, TRES, and PARAS-2 showed radial velocity measurements in phase with the transit, allowing us to measure the mass of 1.27 $\pm$ 0.25 M$_{\text{Jup}}$, establishing the transiting object as a planet (Fig. \ref{fig:WASP197global}).



\section{Discussion } \label{sec:discussion}

\begin{figure} 
    \includegraphics[width=0.8\linewidth]{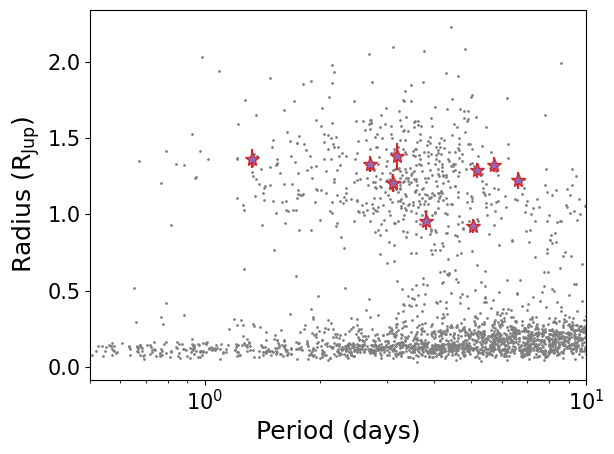}
    \caption{Planet period and radius. Gray points show the known planet population, while red stars show the 9 new planets presented here. }
    \label{fig:final_per_rad}
\end{figure}

\begin{figure}
    \includegraphics[width=0.9\linewidth]{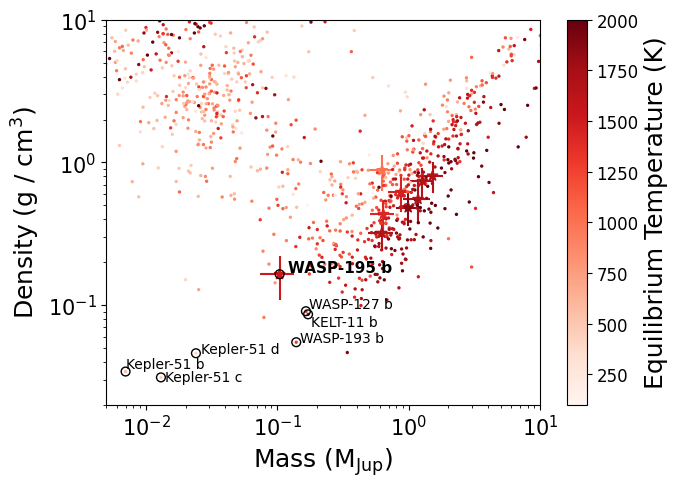}
    \caption{Planet mass and density. The color of the points represent the planets' equilibrium temperatures, with darker colors representing higher temperatures. Small points indicate the known population, with errorbars omitted for visual clarity. The large stars show the 9 planets from this work. The `puffy' exoplanets discussed in Section \ref{sec:discussion} are circled for reference.} 
    \label{fig:final_Mp_dens_Teq}
\end{figure}

The nine giant planets presented here encompass a range of properties exhibited by the hot Jupiter population at large. All planets have periods under 7 days, and radii near that of Jupiter. Figures \ref{fig:final_per_rad} and \ref{fig:final_Mp_dens_Teq} show these planets in the context of the known population contained in the NASA Exoplanet Archive (accessed Oct 7, 2024). 

In particular, Fig. \ref{fig:final_per_rad} shows that the planets are consistent with known hot-Jupiters in terms of their radii and periods. Fig. \ref{fig:final_Mp_dens_Teq} demonstrates the planets reported here are generally consistent with the long-noted trend that hotter, more irradiated planets tend to have larger radii and therefore lower densities \citep{Demory2011}. 

Fig. \ref{fig:final_Mp_dens_Teq} also illustrates that the reported planets show typical mass/radius relationships with the exception of one planet. This outlier is WASP-195\,b which, with a radius of 0.91 R$_{\text{Jup}}$ but a mass of only 0.11 M$_{\text{Jup}}$, has one of the lowest densities among the currently known exoplanets.

Several low-density planets including KELT-11\,b WASP-193\,b, and WASP-127\,b show similar low densities to that of WASP-195\,b. However, these host stars are either approaching the end of their main sequence lifetimes or are already evolving onto the red giant phase. The inflated radii in these systems are at least partially attributed to reinflation which occurs as the levels of irradiation reaching the planets increases during the giant branch evolution stage. 

The stellar models indicate that WASP-195 is a young star (0.75 $\pm$0.55 Gyr). This means that the planet may still be cooling and contracting after the planet's formation. The expected timescale for this contraction is on the order of 1 Gyr \citep{Owen2016}.  Observations of systems of various ages have supported the connection between age and inflation, with a noted trend that puffier low-mass but Jupiter radius planets tend to be found around younger stars, whereas denser planets are found around older stars \citep{Karalis2024}. \cite{LibbyRoberts2020} modeled two similarly puffy sub-Neptune planets orbiting the young ($\approx$0.5 Gyr) star Kepler-51, and found that the planets are likely undergoing contraction and mass loss, with predicted final densities approaching densities more in line with the sub-Neptune population at large. Further modeling of WASP-195\,b's atmospheric evolution may provide evidence for a similar fate.

In addition to contraction, migration may play a role in puffing up the atmosphere. It is proposed that the jovian planets likely formed beyond the ice line, where fast cooling of the envelope could allow even low-mass cores to accrete a large H/He atmosphere \citep{Lee2016}. As the planets migrate inwards, they are exposed to increased stellar radiation, further inflating the atmosphere \citep{Lous2020}.  


Detailed spectral observations of WASP-195\,b, combined with interior modeling, will help to determine constraints on the core and atmospheric mass ratio, providing insight into the possible formation scenarios for the planet. If indeed the planet has a substantial H/He core, the atmosphere may be expected to be undergoing significant mass loss that could be measured. Observations of helium lines surrounding the transit of WASP-107\,b showed a significant absorption following the end of the transit, suggesting the atmosphere is actively being lost with a comet-like tail \citep{Spake2021}.

The eccentricity of the hot Jupiter population has important implications on the dominant formation mechanism (e.g. \cite{Dawson2015}). While hot Jupiters with periods less than 3 days show mostly circular orbits, moderate eccentricities are observed in some giant planets in orbits between 3 and 10 days (See \cite{Dawson2018} for a review). We do not find strong support for eccentricity for any of the planets we fit here and therefore fix all eccentricities to 0 in our model. However, it is possible that additional RV monitoring could provide more refined measurements on the eccentricities.

\section{Conclusion} \label{sec:conclusion}
This paper presents the characterization of nine planets (WASP-102\,b, WASP-116\,b, WASP-149\,b, WASP-154\,b, WASP-155\,b, WASP-188\,b, WASP-194\,b/HAT-P-71\,b, WASP-195\,b, and WASP-197\,b) orbiting FGK stars with periods under 7 days. The planets were identified as candidates by transits observed by WASP and later characterized with ground-based radial velocity measurements. The planet parameters were determined by jointly fitting photometric and radial velocity measurements from a variety of sources, including TESS photometry. In addition, high-resolution imaging data was obtained for all stars to identify and account for any previously unresolved nearby companions. 

All of the host stars are relatively bright, with Gaia magnitudes less than 12.8, allowing for a variety of ground-based follow-up and characterization. The new planets provide additional samples to test our understanding of the demographics of the hot Jupiter population.  While the majority of the reported planets show characteristics typical of the currently known hot Jupiter population, the masses determined from the radial velocity measurements reveal the noticeably low density of WASP-195\,b. This suggests that this planet may be better characterized as a young, puffy member of the sub-Neptune population that is still undergoing contraction and mass loss. Insights into the atmospheric properties gleaned from observations with facilities such as JWST could provide the context to understand the observed low density. Further, this planet, along with WASP-149\,b, have high Transmission Spectroscopy Metric values (153 $\pm$ 48 and 189 $\pm$ 43), making them promising candidates for such atmospheric follow-up.

\section{Acknowledgements}

The material is based upon work supported by NASA under award number 80GSFC24M0006.

HP acknowledges support by the Spanish Ministry of Science and Innovation with the Ramon y Cajal fellowship number RYC2021-031798-I. Funding from the University of La Laguna and the Spanish Ministry of Universities is acknowledged. KKM acknowledges support from the New York Community Trust Fund for Astrophysical Research. The postdoctoral fellowship of KB is funded by F.R.S.-FNRS grant T.0109.20 and by the Francqui Foundation. ML acknowledges support of the Swiss National Science Foundation under grant number PCEFP2\_194576. The contribution of ML has been carried out within the framework of the NCCR PlanetS supported by the Swiss National Science Foundation under grant 51NF40\_205606. KAC acknowledges support from the TESS mission via subaward s3449 from MIT. DRC acknowledges partial support from NASA Grant 18-2XRP18\_2-0007. I.A.S. and P.A.B. acknowledge the support of M.V. Lomonosov Moscow State University Program of Development. S.G.S acknowledges the support from FCT through Investigador FCT contract nr. CEECIND/00826/2018 and  POPH/FSE (EC). GB acknowledges the support from NASA grant 80NSSC22K0315.

Funding for the TESS mission is provided by NASA's Science Mission Directorate. We acknowledge the use of public TESS data from pipelines at the TESS Science Office and at the TESS Science Processing Operations Center. This research has made use of the the NASA Exoplanet Archive and the Exoplanet Follow-up Observation Program (ExoFOP; DOI: 10.26134/ExoFOP5) website, which are operated by the California Institute of Technology, under contract with the National Aeronautics and Space Administration under the Exoplanet Exploration Program. Resources supporting this work were provided by the NASA High-End Computing (HEC) Program through the NASA Advanced Supercomputing (NAS) Division at Ames Research Center for the production of the SPOC data products. This paper includes data collected by the TESS mission that are publicly available from the Mikulski Archive for Space Telescopes (MAST).

Some of the observations in this paper made use of the High-Resolution Imaging instrument Zorro and were obtained under Gemini LLP Proposal Number: GN/S-2021A-LP-105. Zorro was funded by the NASA Exoplanet Exploration Program and built at the NASA Ames Research Center by Steve B. Howell, Nic Scott, Elliott P. Horch, and Emmett Quigley. Zorro was mounted on the Gemini South telescope of the international Gemini Observatory, a program of NSF’s OIR Lab, which is managed by the Association of Universities for Research in Astronomy (AURA) under a cooperative agreement with the National Science Foundation. on behalf of the Gemini partnership: the National Science Foundation (United States), National Research Council (Canada), Agencia Nacional de Investigación y Desarrollo (Chile), Ministerio de Ciencia, Tecnología e Innovación (Argentina), Ministério da Ciência, Tecnologia, Inovações e Comunicações (Brazil), and Korea Astronomy and Space Science Institute (Republic of Korea).

(Some of the) Observations in the paper made use of the NN-EXPLORE Exoplanet and Stellar Speckle Imager (NESSI). NESSI was funded by the NASA Exoplanet Exploration Program and the NASA Ames Research Center. NESSI was built at the Ames Research Center by Steve B. Howell, Nic Scott, Elliott P. Horch, and Emmett Quigley.

We are grateful to PRL-DOS (Department of Space, Government of India), as well as the Director, PRL, for their generous support. Their support has been instrumental in funding the PARAS-2 spectrograph for our exoplanet discovery project. We express our gratitude to all the Mount Abu Observatory staff and PARAS-2 instrument team for their invaluable assistance throughout the observations.

This work makes use of observations from the LCOGT network. Part of the LCOGT telescope time was granted by NOIRLab through the Mid-Scale Innovations Program (MSIP). MSIP is funded by NSF. This paper is based on observations made with the Las Cumbres Observatory’s education network telescopes that were upgraded through generous support from the Gordon and Betty Moore Foundation.

Based on observations obtained at the Hale Telescope, Palomar Observatory, as part of a collaborative agreement between the Caltech Optical Observatories and the Jet Propulsion Laboratory [operated by Caltech for NASA].

We thank the Observatoire de Haute-Provence (CNRS) staff for its support. This work was supported by CNES and the ``Programme National de Plan\'etologie'' (PNP).

TRAPPIST-South is funded by the Belgian Fund for Scientific Research (Fond National de la Recherche Scientifique, FNRS) under the grant FRFC 2.5.594.09.F, with the participation of the Swiss National Science Fundation (SNF). MG and EJ are F.R.S.-FNRS Research Directors.



\bibliographystyle{aasjournal} 
\bibliography{ref}

\begin{appendix}


\section{Final Model Corner Plots} \label{app:corner_plots}

\begin{figure}[h!]\label{fig:Cornerplot_WASP102}
    \includegraphics[width=0.9\linewidth]{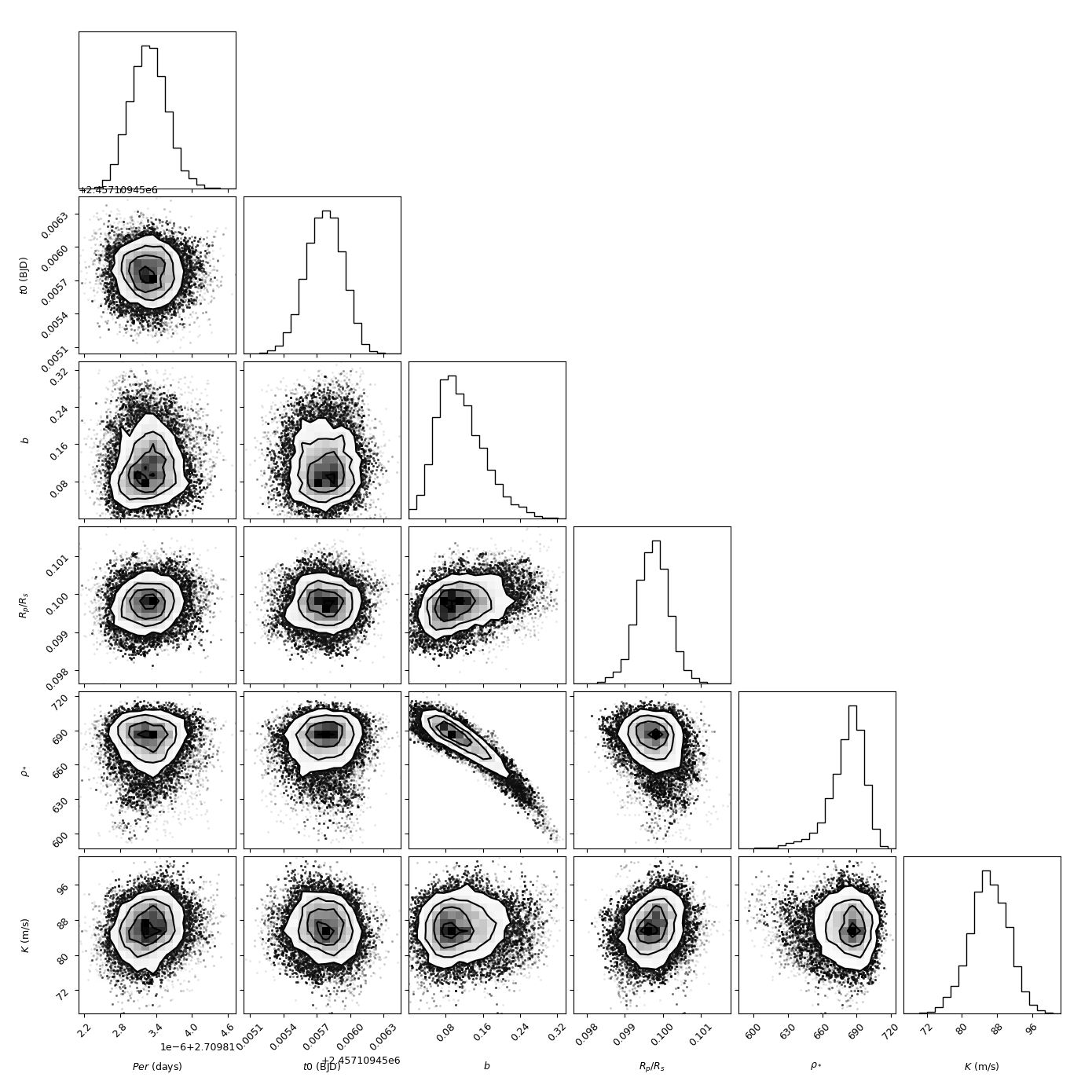}
    \caption{Corner plot showing posteriors for the fitted planet parameters for WASP-102\,b}
\end{figure}

\begin{figure}\label{fig:Cornerplot_WASP116}
    \includegraphics[width=0.9\linewidth]{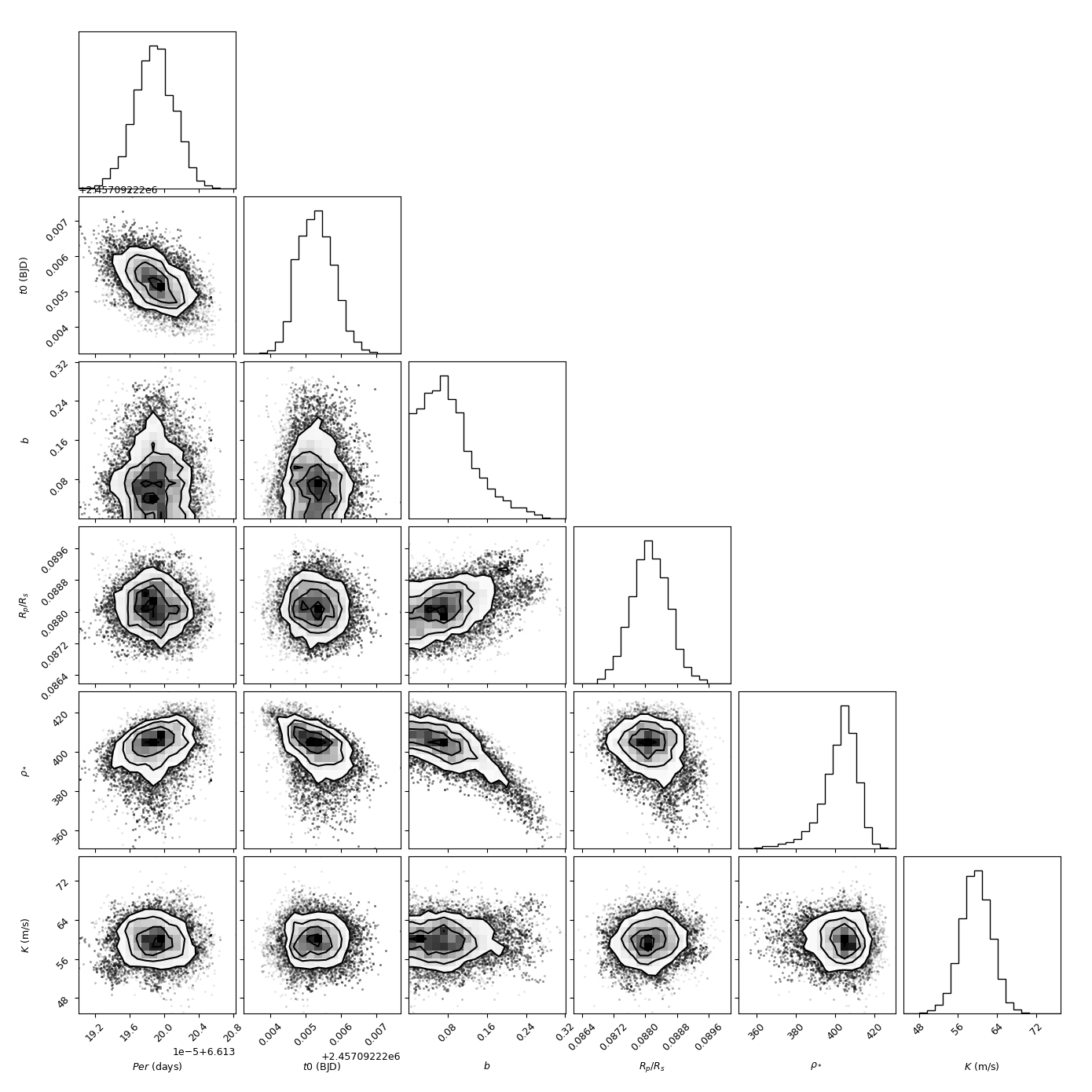}
    \caption{Corner plot showing posteriors for the fitted planet parameters for WASP-116\,b}
\end{figure}
\newpage

\begin{figure}\label{fig:Cornerplot_WASP149}
    \includegraphics[width=0.9\linewidth]{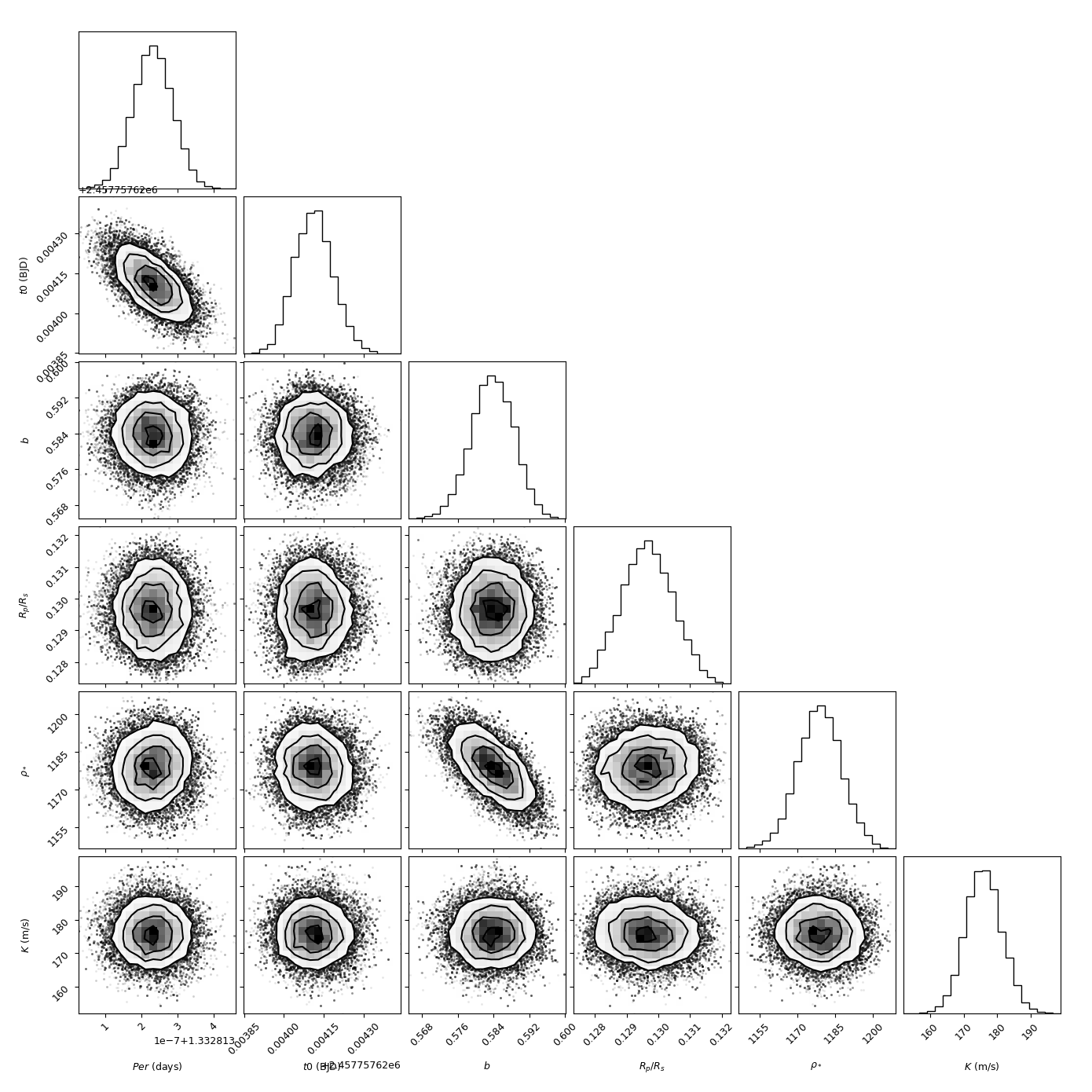}
    \caption{Corner plot showing posteriors for the fitted planet parameters for WASP-149\,b}
\end{figure}
\newpage

\begin{figure}\label{fig:Cornerplot_WASP154}
    \includegraphics[width=0.9\linewidth]{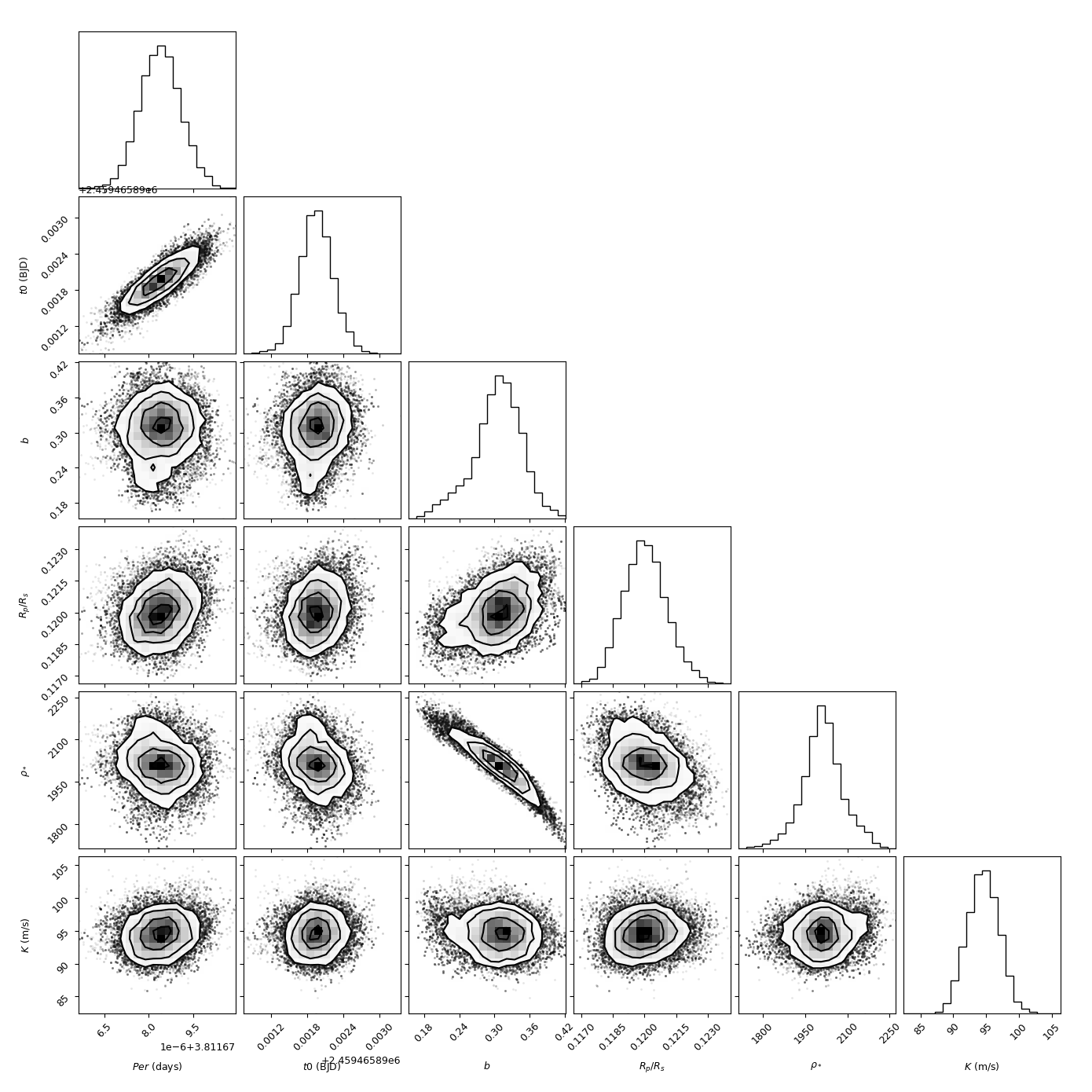}
    \caption{Corner plot showing posteriors for the fitted planet parameters for WASP-154\,b}
\end{figure}
\newpage

\begin{figure}\label{fig:Cornerplot_WASP155}
    \includegraphics[width=0.9\linewidth]{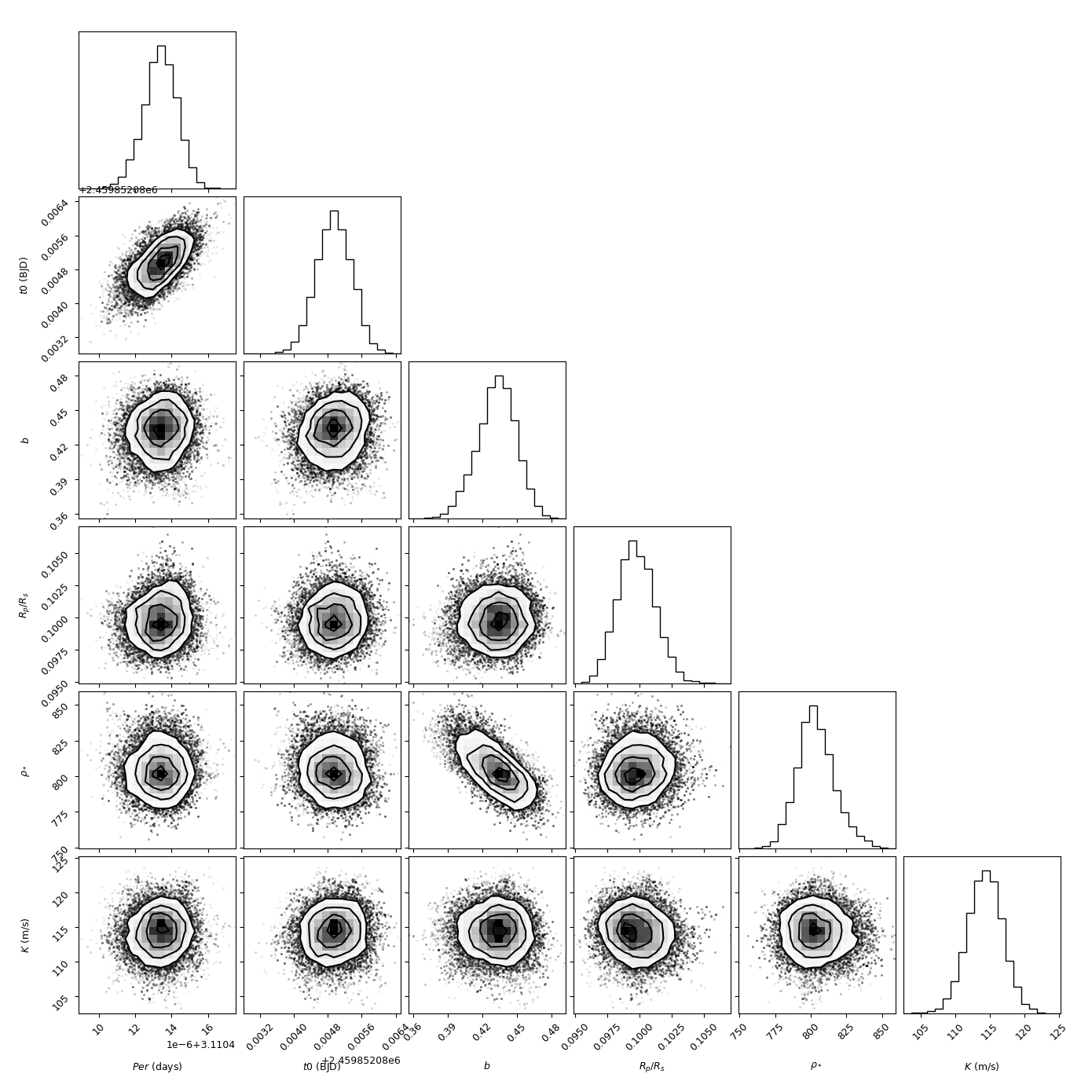}
    \caption{Corner plot showing posteriors for the fitted planet parameters for WASP-155\,b}
\end{figure}
\newpage

\begin{figure}\label{fig:Cornerplot_WASP188}
    \includegraphics[width=0.9\linewidth]{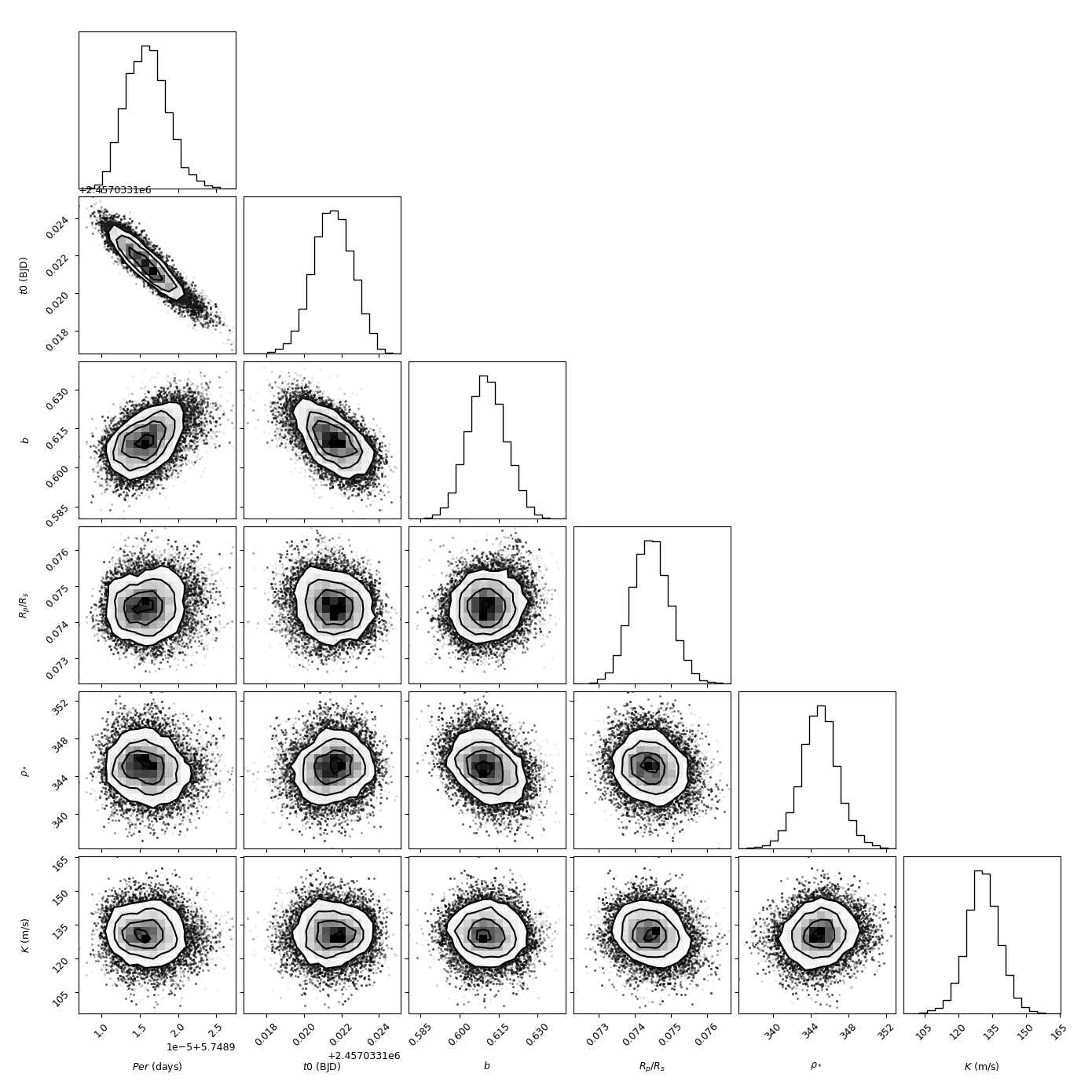}
    \caption{Corner plot showing posteriors for the fitted planet parameters for WASP-188\,b}
\end{figure}
\newpage

\begin{figure}\label{fig:Cornerplot_WASP194}
    \includegraphics[width=0.9\linewidth]{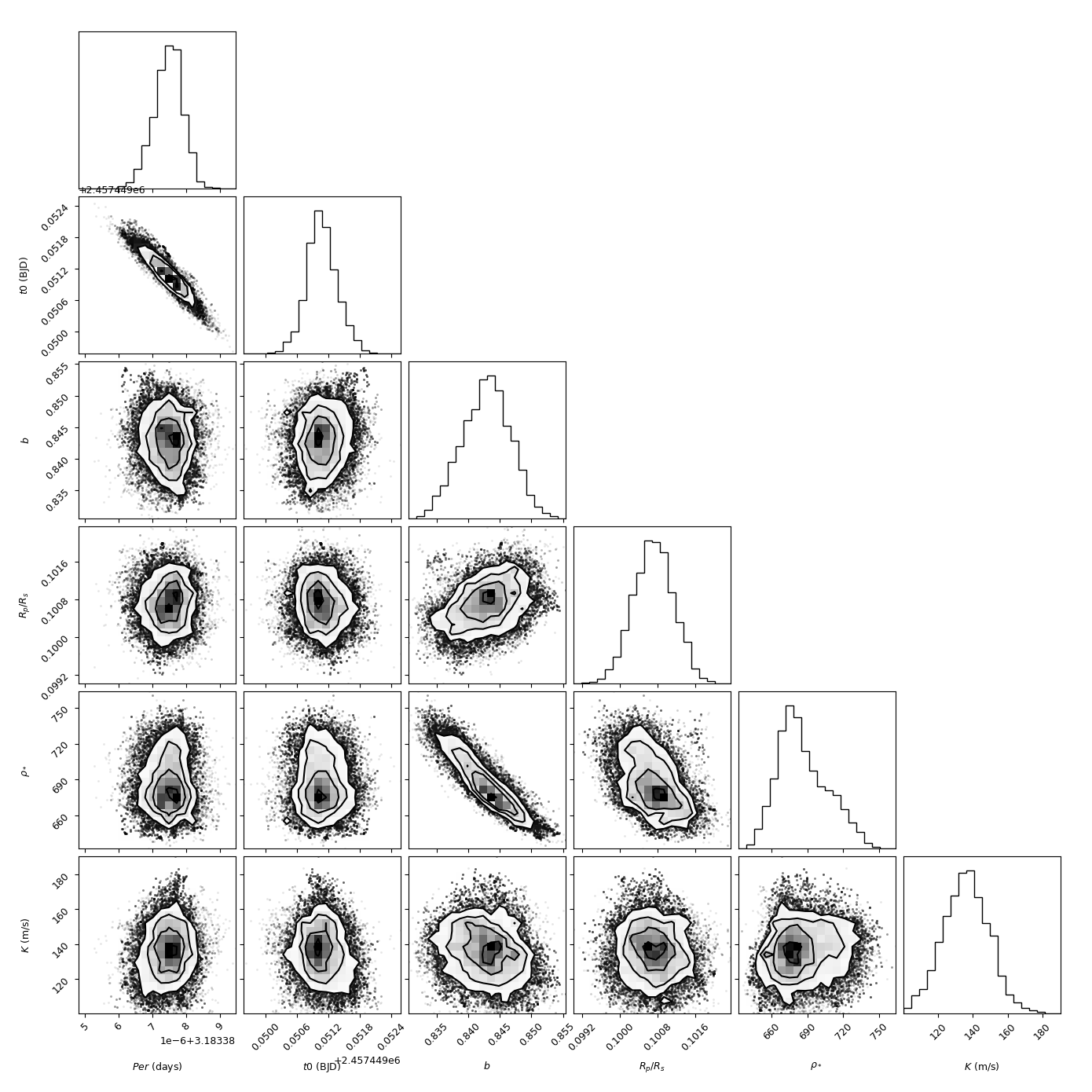}
    \caption{Corner plot showing posteriors for the fitted planet parameters for WASP-194\,b}
\end{figure}
\newpage

\begin{figure}\label{fig:Cornerplot_WASP195}
    \includegraphics[width=0.9\linewidth]{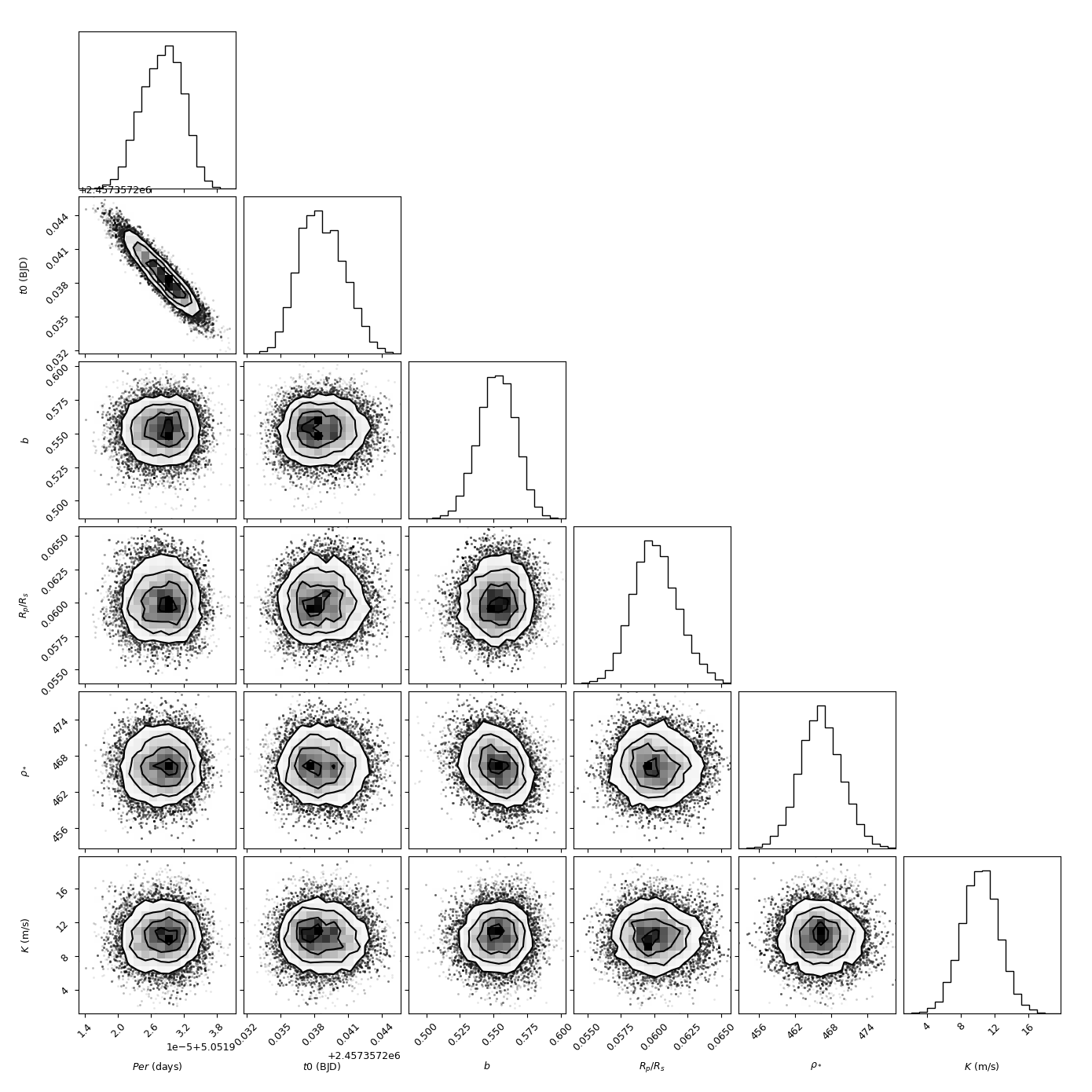}
    \caption{Corner plot showing posteriors for the fitted planet parameters for WASP-195\,b}
\end{figure}
\newpage

\begin{figure}\label{fig:Cornerplot_WASP197}
    \includegraphics[width=0.9\linewidth]{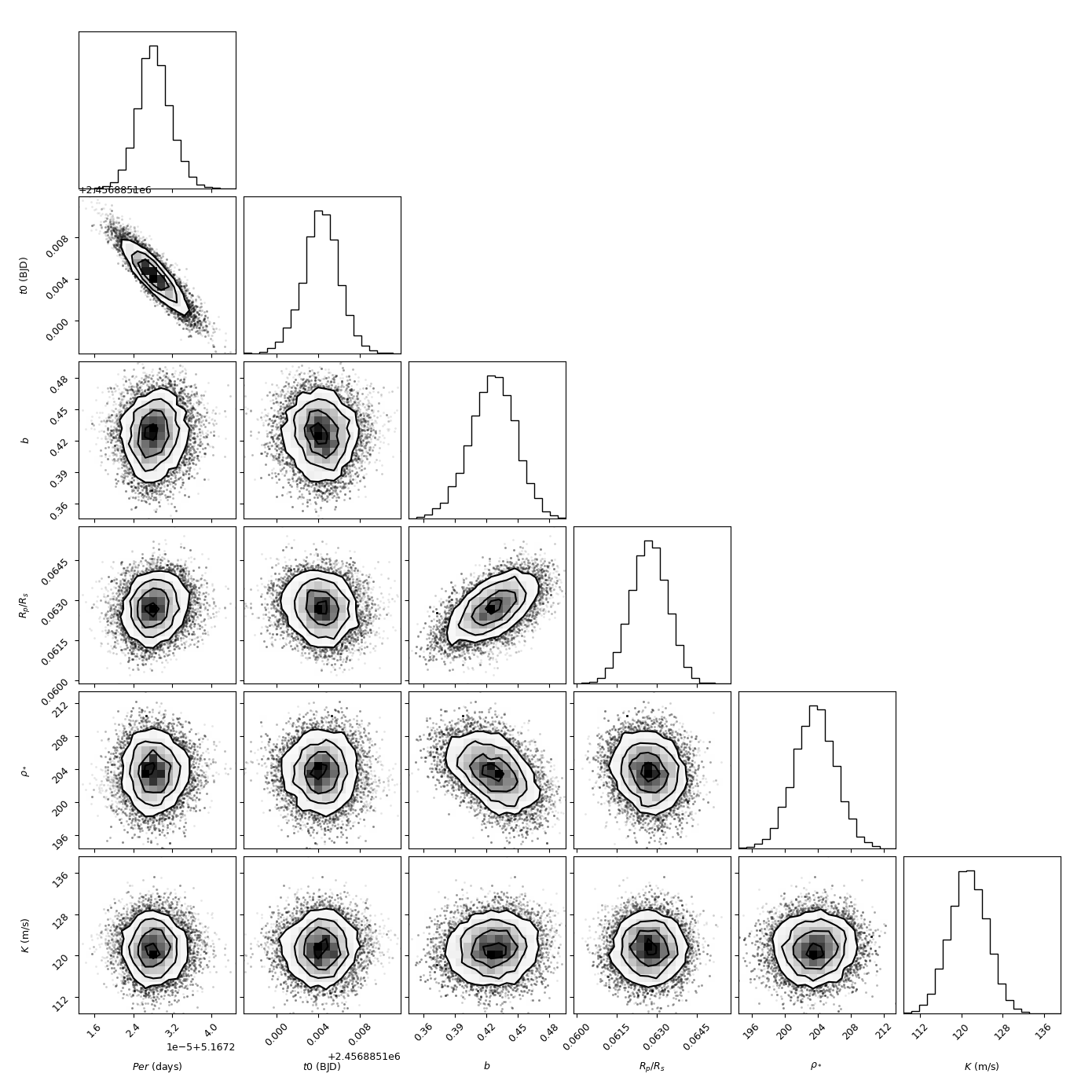}
    \caption{Corner plot showing posteriors for the fitted planet parameters for WASP-197\,b}
\end{figure}

\clearpage

\section{Full list of Model Parameter Priors and Posteriors} \label{app:prior_posterior}
This appendix shows the model parameters along with the prior and posterior values used in the global juliet model. $\mathcal{N}$ priors indicate normal priors, $\mathcal{U}$ indicate uniform priors, and and ln$\mathcal{N}$ indicate log normal distributions. See the juliet documentation\footnote{https://juliet.readthedocs.io/en/latest/user/priorsnparameters.html} for further details on the parameters.

\begin{table*}[]
    \centering
    \caption{Transit fit parameters for WASP-102 (TOI-5385) }
    \begin{tabular}{l c c }
    \hline
    \hline
    Parameter & Prior & Fit Value  \\
    \hline
P\_p1 & $\mathcal{N}$(2.7098094867,0.0002372) & 2.7098132718$^{3.016e-07}_{3.04e-07}$ \\
t0\_p1 & $\mathcal{N}$(2457109.458301,0.01) & 2457109.4557701275$^{0.0001614387}_{0.0001695896}$ \\
b\_p1 & $\mathcal{U}$(0.0,1.0) & 0.1053561292$^{0.0586122843}_{0.0416958994}$ \\
p\_p1 & $\mathcal{N}$(0.09999466,0.1) & 0.0997457844$^{0.0003999763}_{0.0004126752}$ \\
rho & $\mathcal{N}$(585,100) & 684.03713543$^{11.4262294054}_{16.4163205423}$ \\
q1\_TESS\_WASP & $\mathcal{N}$(0.303,0.05) & 0.21933795$^{0.027931697}_{0.0348288851}$ \\
q2\_TESS\_WASP & $\mathcal{N}$(0.383,0.05) & 0.3462643271$^{0.0348421143}_{0.0367554598}$ \\
q1\_EULER1 & $\mathcal{N}$(0.418,0.05) & 0.429942767$^{0.029506276}_{0.027943585}$ \\
q2\_EULER1 & $\mathcal{N}$(0.397,0.05) & 0.3841748256$^{0.0359090282}_{0.0392379723}$ \\
q1\_EULER2 & $\mathcal{N}$(0.310,0.05) & 0.3093626479$^{0.0335669181}_{0.0381012247}$ \\
q2\_EULER2 & $\mathcal{N}$(0.385,0.05) & 0.3803861446$^{0.0384892142}_{0.0355197112}$ \\
q1\_TRAPPIST1\_TRAPPIST2\_TRAPPIST3 & $\mathcal{N}$(0.30349081,0.05) & 0.3537181002$^{0.0307973873}_{0.031618477}$ \\
q2\_TRAPPIST1\_TRAPPIST2\_TRAPPIST3 & $\mathcal{N}$(0.38273734,0.05) & 0.4151080521$^{0.0348155047}_{0.0306933365}$ \\
mdilution\_TESS & $\mathcal{N}$(1.0,0.01) & 0.9869754783$^{0.007289593}_{0.0084533815}$ \\
mflux\_TESS & $\mathcal{N}$(0.0,0.01) & -0.0001342711$^{6.345e-05}_{6.44869e-05}$ \\
sigma\_w\_TESS & ln$\mathcal{N}$(1e-06,1000000.0) & 0.0032181223$^{0.985238121}_{0.0031955477}$ \\
mdilution\_WASP & $\mathcal{N}$(1.0,0.01) & 0.9973126254$^{0.0086343868}_{0.0083784906}$ \\
mflux\_WASP & $\mathcal{N}$(0.0,0.01) & -0.0005717013$^{0.000109142}_{0.000103488}$ \\
sigma\_w\_WASP & ln$\mathcal{N}$(1e-06,1000000.0) & 3297.6306980352$^{257.6755241177}_{232.7289360615}$ \\
mdilution\_EULER1 & $\mathcal{N}$(1.0,0.01) & 1.0074198988$^{0.0073958711}_{0.0061536879}$ \\
mflux\_EULER1 & $\mathcal{N}$(0.0,0.01) & 0.0033593069$^{0.0011068267}_{0.0010334332}$ \\
sigma\_w\_EULER1 & ln$\mathcal{N}$(1e-06,1000000.0) & 439.6844804777$^{108.9069736517}_{109.7533915897}$ \\
theta0\_EULER1 & $\mathcal{U}$(-100,100) & 0.0019320809$^{0.0006875206}_{0.0006326413}$ \\
mdilution\_EULER2 & $\mathcal{N}$(1.0,0.01) & 0.9919475645$^{0.0068671264}_{0.0080144526}$ \\
mflux\_EULER2 & $\mathcal{N}$(0.0,0.01) & -0.0019202767$^{0.0010189849}_{0.0011670411}$ \\
sigma\_w\_EULER2 & ln$\mathcal{N}$(1e-06,1000000.0) & 993.9563169572$^{92.5340345654}_{92.4165864685}$ \\
theta0\_EULER2 & $\mathcal{U}$(-100,100) & -0.0009325131$^{0.0006287706}_{0.0007489514}$ \\
mdilution\_TRAPPIST1 & $\mathcal{N}$(1.0,0.01) & 1.0223432356$^{0.0090554794}_{0.0080193752}$ \\
mflux\_TRAPPIST1 & $\mathcal{N}$(0.0,0.01) & -0.0023889443$^{0.0004810746}_{0.0004375614}$ \\
sigma\_w\_TRAPPIST1 & ln$\mathcal{N}$(1e-06,1000000.0) & 1222.9486938427$^{89.2332773471}_{87.1378057286}$ \\
theta0\_TRAPPIST1 & $\mathcal{U}$(-100,100) & -0.0010845235$^{0.0002741073}_{0.000255327}$ \\
mdilution\_TRAPPIST2 & $\mathcal{N}$(1.0,0.01) & 1.0030891272$^{0.009156801}_{0.0076036294}$ \\
mflux\_TRAPPIST2 & $\mathcal{N}$(0.0,0.01) & 0.0031180731$^{0.0008245315}_{0.0009136841}$ \\
sigma\_w\_TRAPPIST2 & ln$\mathcal{N}$(1e-06,1000000.0) & 3392.037458069$^{115.011594554}_{109.4153474344}$ \\
theta0\_TRAPPIST2 & $\mathcal{U}$(-100,100) & 0.0021688568$^{0.0004704801}_{0.0005199252}$ \\
mdilution\_TRAPPIST3 & $\mathcal{N}$(1.0,0.01) & 0.9883415703$^{0.0080450539}_{0.009753062}$ \\
mflux\_TRAPPIST3 & $\mathcal{N}$(0.0,0.01) & -0.0002547559$^{0.0005762263}_{0.0005222762}$ \\
sigma\_w\_TRAPPIST3 & ln$\mathcal{N}$(1e-06,1000000.0) & 1181.1269937596$^{124.79770027}_{132.2603944195}$ \\
theta0\_TRAPPIST3 & $\mathcal{U}$(-100,100) & 0.0008109566$^{0.0003061372}_{0.0002816134}$ \\
GP\_sigma\_TESS & ln$\mathcal{N}$(1e-06,1000000.0) & 0.0004986827$^{4.34615e-05}_{3.77943e-05}$ \\
GP\_rho\_TESS & ln$\mathcal{N}$(0.001,1.0) & 0.2231794777$^{0.0492960748}_{0.0376425229}$ \\
K\_p1 & $\mathcal{U}$(0,200) & 86.1550827174$^{4.3719118508}_{4.2811423179}$ \\
sigma\_w\_SOPHIE & ln$\mathcal{N}$(0.001,100.0) & 0.1051695987$^{1.2480695501}_{0.0976563169}$ \\
mu\_SOPHIE & $\mathcal{U}$(-200,200) & -1.3939161961$^{3.8244593735}_{4.108173765}$ \\
sigma\_w\_CORALIE & ln$\mathcal{N}$(0.001,100.0) & 0.0941341935$^{1.5615415191}_{0.0883157186}$ \\
mu\_CORALIE & $\mathcal{U}$(-200,200) & -24.0994459181$^{5.3171361087}_{5.4130000819}$ \\
         \hline
    \end{tabular}
    \label{tab:wasp102_fit}
\end{table*}

\begin{table*}[]
    \centering
    \caption{Transit fit parameters for WASP-116 (TOI-4672) }
    \begin{tabular}{l c c }
    \hline
    \hline
    Parameter & Prior & Fit Value  \\
    \hline
P\_p1 & $\mathcal{N}$(6.6132100844,1.65e-05) & 6.613198842$^{2.3278e-06}_{2.2558e-06}$ \\
t0\_p1 & $\mathcal{N}$(2457092.2226665,0.0010837) & 2457092.225277469$^{0.0005252287}_{0.0004949705}$ \\
b\_p1 & $\mathcal{U}$(0.0,1.0) & 0.0733744461$^{0.0606058223}_{0.0469612814}$ \\
p\_p1 & $\mathcal{N}$(0.09992291,0.02) & 0.0881072894$^{0.0004746954}_{0.0004564043}$ \\
rho & $\mathcal{N}$(607.71887813,58.91378521) & 403.6238923096$^{6.7583778605}_{9.8183430686}$ \\
q1\_TESS\_WASP & $\mathcal{N}$(0.283,0.02) & 0.2791791798$^{0.0143123032}_{0.0156518219}$ \\
q2\_TESS\_WASP & $\mathcal{N}$(0.370,0.02) & 0.3724246494$^{0.0168694526}_{0.0171594726}$ \\
q1\_TRAPPIST1\_TRAPPIST2 & $\mathcal{N}$(0.283,0.02) & 0.2809726589$^{0.0173688667}_{0.0170102394}$ \\
q2\_TRAPPIST1\_TRAPPIST2 & $\mathcal{N}$(0.370,0.02) & 0.3670315305$^{0.0187266439}_{0.017734133}$ \\
q1\_EULER & $\mathcal{N}$(0.288,0.02) & 0.2797900789$^{0.0159002299}_{0.0163383575}$ \\
q2\_EULER & $\mathcal{N}$(0.373,0.02) & 0.369418756$^{0.0159332132}_{0.0148226754}$ \\
q1\_HAL\_SAAO & $\mathcal{N}$(0.611,0.02) & 0.5994599508$^{0.0159155633}_{0.0164825789}$ \\
q2\_HAL\_SAAO & $\mathcal{N}$(0.413,0.02) & 0.4079540301$^{0.0150308334}_{0.0176518397}$ \\
mdilution\_WASP & $\mathcal{N}$(1.0,0.001) & 1.0000415273$^{0.00089174}_{0.0007989643}$ \\
mflux\_WASP & $\mathcal{N}$(0.0,0.01) & -0.0006658179$^{9.02379e-05}_{8.01019e-05}$ \\
sigma\_w\_WASP & ln$\mathcal{N}$(1e-06,1000.0) & 685.1802867495$^{266.7787211153}_{685.1678613175}$ \\
mdilution\_TESS & $\mathcal{N}$(1.0,0.001) & 0.9999780885$^{0.0008752141}_{0.0008644195}$ \\
mflux\_TESS & $\mathcal{N}$(0.0,0.01) & -0.0002021315$^{0.0001502431}_{0.0001639391}$ \\
sigma\_w\_TESS & ln$\mathcal{N}$(1e-06,1000.0) & 0.004233758$^{1.5109907248}_{0.0042123966}$ \\
mdilution\_TRAPPIST1 & $\mathcal{N}$(1.0,0.001) & 1.0001525916$^{0.0009452329}_{0.0009208354}$ \\
mflux\_TRAPPIST1 & $\mathcal{N}$(0.0,0.01) & 0.0003496223$^{9.96407e-05}_{9.46914e-05}$ \\
sigma\_w\_TRAPPIST1 & ln$\mathcal{N}$(1e-06,1000.0) & 0.0034504608$^{1.1824371614}_{0.0034355267}$ \\
mdilution\_TRAPPIST2 & $\mathcal{N}$(1.0,0.001) & 1.0000045291$^{0.0008247634}_{0.0008730036}$ \\
mflux\_TRAPPIST2 & $\mathcal{N}$(0.0,0.01) & 0.0005315789$^{0.0001379424}_{0.0001223301}$ \\
sigma\_w\_TRAPPIST2 & ln$\mathcal{N}$(1e-06,1000.0) & 0.0522742857$^{11.8513610506}_{0.0520541481}$ \\
mdilution\_EULER & $\mathcal{N}$(1.0,0.001) & 0.9997683332$^{0.0008312397}_{0.0007886571}$ \\
mflux\_EULER & $\mathcal{N}$(0.0,0.01) & -0.0005353237$^{8.76211e-05}_{8.13065e-05}$ \\
sigma\_w\_EULER & ln$\mathcal{N}$(1e-06,1000.0) & 658.5067501712$^{46.5267532983}_{50.7867917849}$ \\
mdilution\_HAL & $\mathcal{N}$(1.0,0.001) & 0.9998149927$^{0.0009841547}_{0.0008798776}$ \\
mflux\_HAL & $\mathcal{N}$(0.0,0.01) & 0.0005074767$^{0.0002515873}_{0.0002506774}$ \\
sigma\_w\_HAL & ln$\mathcal{N}$(1e-06,1000.0) & 0.0139611156$^{16.521771125}_{0.0139211505}$ \\
mdilution\_SAAO & $\mathcal{N}$(1.0,0.001) & 0.9997253071$^{0.0008622338}_{0.0008604492}$ \\
mflux\_SAAO & $\mathcal{N}$(0.0,0.01) & -0.0003073304$^{0.0002876818}_{0.0002893822}$ \\
sigma\_w\_SAAO & ln$\mathcal{N}$(1e-06,1000.0) & 0.2416750305$^{59.2776922997}_{0.2410538744}$ \\
GP\_sigma\_TESS & ln$\mathcal{N}$(1e-06,1000000.0) & 0.0006115378$^{0.000134656}_{8.5962e-05}$ \\
GP\_rho\_TESS & ln$\mathcal{N}$(0.001,1000.0) & 0.7971313002$^{0.2222778602}_{0.151839757}$ \\
K\_p1 & $\mathcal{N}$(59,10) & 59.6707913485$^{3.0912484731}_{3.0083637762}$ \\
sigma\_w\_SOPHIE & ln$\mathcal{N}$(0.001,100.0) & 0.1806688938$^{2.8986812707}_{0.1741668283}$ \\
mu\_SOPHIE & $\mathcal{U}$(-200,200) & -4.9228153887$^{3.5642382067}_{3.5367170902}$ \\
sigma\_w\_CORALIE & ln$\mathcal{N}$(0.001,100.0) & 0.097427033$^{2.9084275797}_{0.0927210478}$ \\
mu\_CORALIE & $\mathcal{U}$(-200,200) & 12.660738583$^{3.2374412662}_{2.7506861279}$ \\
         \hline
    \end{tabular}
    \label{tab:wasp116_fit}
\end{table*}

\begin{table*}[]
    \centering
    \caption{Transit fit parameters for WASP-149 (TOI-6101) }
    \begin{tabular}{l c c }
    \hline
    \hline
    Parameter & Prior & Fit Value  \\
    \hline
P\_p1 & $\mathcal{N}$(1.332813008459,1.3788792e-05) & 1.3328130092$^{5.52e-08}_{5.39e-08}$ \\
t0\_p1 & $\mathcal{N}$(2457757.6244090004,0.1) & 2457757.624497969$^{7.51209e-05}_{7.60215e-05}$ \\
b\_p1 & $\mathcal{U}$(0.0,1.0) & 0.5835166593$^{0.0046661293}_{0.0048666077}$ \\
p\_p1 & $\mathcal{N}$(0.1300540659,0.0010281845) & 0.1297456937$^{0.0008016409}_{0.0008765418}$ \\
rho & $\mathcal{N}$(1175.08511114,9.13955086) & 1180.0194788408$^{8.9599139213}_{8.7260978722}$ \\
q1\_TESS\_WASP & $\mathcal{N}$(0.318,0.05) & 0.270895631$^{0.0270627515}_{0.0259859114}$ \\
q2\_TESS\_WASP & $\mathcal{N}$(0.379,0.05) & 0.3549526667$^{0.0437989941}_{0.0452509322}$ \\
q1\_TRAPPIST1\_TRAPPIST2\_TRAPPIST3\_ &  & \\
\hspace{5mm}EULER1\_EULER2 & $\mathcal{N}$(0.260,0.05) & 0.189180733$^{0.0250380464}_{0.0251293464}$  \\
q2\_TRAPPIST1\_TRAPPIST2\_TRAPPIST3\_ &  &  \\
\hspace{5mm}EULER1\_EULER2 & $\mathcal{N}$(0.369,0.05) & 0.3561080556$^{0.0463833477}_{0.0457868637}$ \\
mdilution\_TESS & $\mathcal{N}$(1.0,0.1) & 1.0578899028$^{0.0142450709}_{0.0143513999}$ \\
mflux\_TESS & $\mathcal{N}$(0.0,0.1) & -1.65668e-05$^{0.0001490573}_{0.0001430253}$ \\
sigma\_w\_TESS & ln$\mathcal{N}$(1e-06,1000000.0) & 0.0096442937$^{2.9472649219}_{0.0096224959}$ \\
mdilution\_WASP & $\mathcal{N}$(1.0,0.1) & 0.735481089$^{0.024283581}_{0.0252017233}$ \\
mflux\_WASP & $\mathcal{N}$(0.0,0.1) & -0.0008646972$^{0.0001173553}_{0.0001161409}$ \\
sigma\_w\_WASP & ln$\mathcal{N}$(1e-06,1000000.0) & 3734.5131957602$^{94.1675743834}_{95.8023613129}$ \\
mdilution\_TRAPPIST1 & $\mathcal{N}$(1.0,0.1) & 1.0538283109$^{0.0233824644}_{0.022557328}$ \\
mflux\_TRAPPIST1 & $\mathcal{N}$(0.0,0.1) & 4.3203e-05$^{0.0002681413}_{0.000248785}$ \\
sigma\_w\_TRAPPIST1 & ln$\mathcal{N}$(1e-06,1000000.0) & 0.0451939612$^{28.6891182684}_{0.0451604464}$ \\
mdilution\_TRAPPIST2 & $\mathcal{N}$(1.0,0.1) & 1.055652015$^{0.0214697}_{0.0207438846}$ \\
mflux\_TRAPPIST2 & $\mathcal{N}$(0.0,0.1) & -3.14633e-05$^{0.0001857395}_{0.0001779116}$ \\
sigma\_w\_TRAPPIST2 & ln$\mathcal{N}$(1e-06,1000000.0) & 0.1063324204$^{76.6875042517}_{0.106273571}$ \\
mdilution\_TRAPPIST3 & $\mathcal{N}$(1.0,0.1) & 1.0379617888$^{0.0233202609}_{0.02196689}$ \\
mflux\_TRAPPIST3 & $\mathcal{N}$(0.0,0.1) & 0.000205915$^{0.0002022297}_{0.0002171354}$ \\
sigma\_w\_TRAPPIST3 & ln$\mathcal{N}$(1e-06,1000000.0) & 2043.1705688414$^{195.0607290461}_{192.5356096911}$ \\
mdilution\_EULER1 & $\mathcal{N}$(1.0,0.1) & 1.1222511713$^{0.0164283222}_{0.015874568}$ \\
mflux\_EULER1 & $\mathcal{N}$(0.0,0.1) & 1.47144e-05$^{0.0001011284}_{9.7066e-05}$ \\
sigma\_w\_EULER1 & ln$\mathcal{N}$(1e-06,1000000.0) & 699.4015766652$^{85.6120859629}_{82.9893094337}$ \\
mdilution\_EULER2 & $\mathcal{N}$(1.0,0.1) & 1.0117260997$^{0.0206790912}_{0.0195573584}$ \\
mflux\_EULER2 & $\mathcal{N}$(0.0,0.1) & -9.24558e-05$^{0.0001867384}_{0.0001747451}$ \\
sigma\_w\_EULER2 & ln$\mathcal{N}$(1e-06,1000000.0) & 1907.2440205565$^{114.2373831761}_{115.7427292772}$ \\
GP\_sigma\_TESS & ln$\mathcal{N}$(1e-06,1000000.0) & 0.0005592876$^{7.43102e-05}_{6.15795e-05}$ \\
GP\_rho\_TESS & ln$\mathcal{N}$(0.001,1.0) & 0.9146426772$^{0.0598204972}_{0.0951031504}$ \\
K\_p1 & $\mathcal{U}$(150,200) & 175.2500871719$^{5.1915694996}_{5.3126938307}$ \\
sigma\_w\_SOPHIE & ln$\mathcal{N}$(0.001,100.0) & 16.7875628912$^{5.2110615552}_{4.4524398921}$ \\
mu\_SOPHIE & $\mathcal{U}$(-200,200) & -2.7044596201$^{5.3551685373}_{5.4468647299}$ \\
sigma\_w\_CORALIE & ln$\mathcal{N}$(0.001,100.0) & 0.1875257064$^{5.123189558}_{0.1814593819}$ \\
mu\_CORALIE & $\mathcal{U}$(-200,200) & 63.9970215637$^{5.4318380929}_{5.460781203}$ \\
         \hline
    \end{tabular}
    \label{tab:wasp149_fit}
\end{table*}

\begin{table*}[]
    \centering
    \caption{Transit fit parameters for WASP-154 (TOI-5288) }
    \begin{tabular}{l c c }
    \hline
    \hline
    Parameter & Prior & Fit Value  \\
    \hline
P\_p1 & $\mathcal{N}$(3.8116804430,0.000504) & 3.8116783822$^{6.908e-07}_{6.743e-07}$ \\
t0\_p1 & $\mathcal{N}$(2459465.891746,0.0006765) & 2459465.891948188$^{0.0002746694}_{0.0002680938}$ \\
b\_p1 & $\mathcal{U}$(0.0,1.0) & 0.3079167421$^{0.036380917}_{0.0429981457}$ \\
p\_p1 & $\mathcal{N}$(0.12346573,0.1) & 0.1199964174$^{0.0010403615}_{0.000997486}$ \\
rho & $\mathcal{N}$(2023.21377086,100.0) & 2012.3066079352$^{67.1871979084}_{65.6210029869}$ \\
q1\_TESS\_WASP & $\mathcal{U}$(0.289,0.489) & 0.4508430928$^{0.0250717994}_{0.0345814997}$ \\
q2\_TESS\_WASP & $\mathcal{U}$(0.316,0.516) & 0.4643389169$^{0.0319860462}_{0.0353741473}$ \\
q1\_TRAPPISTbb & $\mathcal{U}$(0.289,0.489) & 0.4467170518$^{0.0219964229}_{0.0235593611}$ \\
q2\_TRAPPISTbb & $\mathcal{U}$(0.316,0.516) & 0.4403261339$^{0.0383202716}_{0.0437316507}$ \\
q1\_TRAPPISTz & $\mathcal{U}$(0.230,0.430) & 0.3727096137$^{0.0361168074}_{0.0429967262}$ \\
q2\_TRAPPISTz & $\mathcal{U}$(0.303,0.503) & 0.442698557$^{0.0397696639}_{0.0483749525}$ \\
q1\_EULER & $\mathcal{U}$(0.289,0.489) & 0.4370624249$^{0.0328803101}_{0.0417739976}$ \\
q2\_EULER & $\mathcal{U}$(0.316,0.516) & 0.3676555532$^{0.0445835858}_{0.0328535363}$ \\
q1\_NITES & $\mathcal{U}$(0.307,0.507) & 0.4383466081$^{0.0376749833}_{0.0397752618}$ \\
q2\_NITES & $\mathcal{U}$(0.318,0.518) & 0.4520099281$^{0.0396044246}_{0.0443759221}$ \\
q1\_CTIO & $\mathcal{U}$(0.307,0.507) & 0.3891820637$^{0.0517931612}_{0.0476066309}$ \\
q2\_CTIO & $\mathcal{U}$(0.318,0.518) & 0.3960693662$^{0.0470003171}_{0.0452605473}$ \\
q1\_BRIERFIELD & $\mathcal{U}$(0.432,0.632) & 0.5190209907$^{0.0462877469}_{0.0482946094}$ \\
q2\_BRIERFIELD & $\mathcal{U}$(0.343,0.543) & 0.4534912047$^{0.0510994337}_{0.0547121647}$ \\
mdilution\_TESS & $\mathcal{N}$(1.0,0.01) & 0.9991367237$^{0.007315032}_{0.0085205727}$ \\
mflux\_TESS & $\mathcal{N}$(0.0,0.001) & -0.000199694$^{0.000164997}_{0.0001601702}$ \\
sigma\_w\_TESS & ln$\mathcal{N}$(1e-06,1000000.0) & 0.4089728993$^{18.1793802923}_{0.4053961051}$ \\
mdilution\_WASP & $\mathcal{N}$(1.0,0.1) & 0.8871989634$^{0.0494515904}_{0.0619016074}$ \\
mflux\_WASP & $\mathcal{N}$(0.0,0.001) & -0.0017351131$^{0.0001721779}_{0.0002057879}$ \\
sigma\_w\_WASP & ln$\mathcal{N}$(1e-06,1000000.0) & 7853.7282011593$^{271.8309925099}_{263.9202967764}$ \\
mdilution\_TRAPPISTbb & $\mathcal{N}$(1.0,0.1) & 1.1049684941$^{0.0227859535}_{0.0226281123}$ \\
mflux\_TRAPPISTbb & $\mathcal{N}$(0.0,0.001) & 0.0001990382$^{0.0001268673}_{0.0001228831}$ \\
sigma\_w\_TRAPPISTbb & ln$\mathcal{N}$(1e-06,1000000.0) & 0.1525203245$^{24.2542857104}_{0.1520499573}$ \\
mdilution\_TRAPPISTz & $\mathcal{N}$(1.0,0.1) & 1.0442023402$^{0.0222077029}_{0.0208510699}$ \\
mflux\_TRAPPISTz & $\mathcal{N}$(0.0,0.001) & 0.0002780083$^{0.0001782695}_{0.000194189}$ \\
sigma\_w\_TRAPPISTz & ln$\mathcal{N}$(1e-06,1000000.0) & 2263.3515205752$^{161.0368474278}_{150.9062948736}$ \\
mdilution\_EULER & $\mathcal{N}$(1.0,0.1) & 1.0499844634$^{0.0298515172}_{0.0277088423}$ \\
mflux\_EULER & $\mathcal{N}$(0.0,0.001) & 1.21629e-05$^{0.0003294141}_{0.0003234106}$ \\
sigma\_w\_EULER & ln$\mathcal{N}$(1e-06,1000000.0) & 0.2799482303$^{32.2414631249}_{0.2790876711}$ \\
mdilution\_NITES & $\mathcal{N}$(1.0,0.1) & 1.0407940189$^{0.0266478238}_{0.0279623029}$ \\
mflux\_NITES & $\mathcal{N}$(0.0,0.001) & 0.000135819$^{0.0002311654}_{0.0002104355}$ \\
sigma\_w\_NITES & ln$\mathcal{N}$(1e-06,1000000.0) & 3899.3428450207$^{148.5958358891}_{148.0711927536}$ \\
mdilution\_CTIO & $\mathcal{N}$(1.0,0.1) & 1.1085143496$^{0.0737869117}_{0.0520910063}$ \\
mflux\_CTIO & $\mathcal{N}$(0.0,0.001) & 0.0006985495$^{0.0006932311}_{0.0005729919}$ \\
sigma\_w\_CTIO & ln$\mathcal{N}$(1e-06,1000000.0) & 4521.7123314514$^{831.9575791167}_{938.9757950716}$ \\
mdilution\_BRIERFIELD & $\mathcal{N}$(1.0,0.1) & 1.0305988532$^{0.0330892101}_{0.033656751}$ \\
mflux\_BRIERFIELD & $\mathcal{N}$(0.0,0.001) & -0.0001391221$^{0.0002558455}_{0.0002755107}$ \\
sigma\_w\_BRIERFIELD & ln$\mathcal{N}$(1e-06,1000000.0) & 0.0091565533$^{5.708662326}_{0.0091204784}$ \\
GP\_sigma\_TESS & ln$\mathcal{N}$(1e-06,1000000.0) & 0.0009293299$^{0.0001505228}_{0.0001139022}$ \\
GP\_rho\_TESS & ln$\mathcal{N}$(0.001,1000.0) & 0.3508402243$^{0.0954216921}_{0.0663824698}$ \\
GP\_sigma\_CTIO & ln$\mathcal{N}$(1e-06,1000000.0) & 0.0071097148$^{0.0011650882}_{0.0009924534}$ \\
GP\_rho\_CTIO & ln$\mathcal{N}$(0.001,1000.0) & 0.008861773$^{0.0034875278}_{0.0029849736}$ \\
K\_p1 & $\mathcal{U}$(50,150) & 94.5473462087$^{2.4564572389}_{2.4392762168}$ \\
sigma\_w\_SOPHIE & ln$\mathcal{N}$(0.001,100.0) & 0.0856111545$^{0.4971383}_{0.073097678}$ \\
mu\_SOPHIE & $\mathcal{U}$(-200,200) & -2.1985169476$^{2.0785764233}_{1.9866222946}$ \\
         \hline
    \end{tabular}
    \label{tab:wasp154_fit}
\end{table*}

\begin{table*}[]
    \centering
    \caption{Transit fit parameters for WASP-155 (TOI-6135) }
    \begin{tabular}{l c c }
    \hline
    \hline
    Parameter & Prior & Fit Value  \\
    \hline
P\_p1 & $\mathcal{N}$(3.11042,0.00022) & 3.1104134$^{8.879e-07}_{9.155e-07}$ \\
t0\_p1 & $\mathcal{N}$(2459852.086444,0.01) & 2459852.0849425416$^{0.0004197014}_{0.0004093861}$ \\
b\_p1 & $\mathcal{U}$(0.0,1.0) & 0.4333351767$^{0.0156862161}_{0.0176369792}$ \\
p\_p1 & $\mathcal{N}$(0.0995,0.0034) & 0.099700355$^{0.0014972551}_{0.0013828619}$ \\
rho & $\mathcal{N}$(805.9586,15.0268) & 802.7357362942$^{13.8519729832}_{11.3992426628}$ \\
q1\_TESS\_WASP & $\mathcal{N}$(0.319,0.05) & 0.3141291411$^{0.0399185398}_{0.0391568684}$ \\
q2\_TESS\_WASP & $\mathcal{N}$(0.384,0.05) & 0.3757595253$^{0.0407183443}_{0.0434760151}$ \\
q1\_NITES & $\mathcal{N}$(0.437,0.05) & 0.4147779192$^{0.041584317}_{0.0437695251}$ \\
q2\_NITES & $\mathcal{N}$(0.400,0.05) & 0.3916757073$^{0.0445596393}_{0.0430080976}$ \\
q1\_MUSCATg & $\mathcal{N}$(0.644,0.05) & 0.6331485031$^{0.0458291376}_{0.0438849086}$ \\
q2\_MUSCATg & $\mathcal{N}$(0.454,0.05) & 0.4647771294$^{0.0406456309}_{0.042585879}$ \\
q1\_MUSCATr & $\mathcal{N}$(0.437,0.05) & 0.4419663675$^{0.0433662961}_{0.0427008823}$ \\
q2\_MUSCATr & $\mathcal{N}$(0.400,0.05) & 0.3978450437$^{0.0464718358}_{0.04163652}$ \\
q1\_MUSCATz & $\mathcal{N}$(0.326,0.05) & 0.3201599757$^{0.0422236996}_{0.0454615577}$ \\
q2\_MUSCATz & $\mathcal{N}$(0.387,0.05) & 0.3971976227$^{0.0424763127}_{0.0427466223}$ \\
mdilution\_TESS & (1.0,0.1,0.0,1.0) & 0.927697933$^{0.0309997303}_{0.0315731731}$ \\
mflux\_TESS & $\mathcal{N}$(0.0,0.1) & -0.0001718234$^{0.0003487486}_{0.0001933453}$ \\
sigma\_w\_TESS & ln$\mathcal{N}$(1e-06,1000000.0) & 0.0122692979$^{2.3173766316}_{0.0122099816}$ \\
mdilution\_WASP & (0.916,0.1,0.0,1.0) & 0.5993747625$^{0.0546426813}_{0.0549601353}$ \\
mflux\_WASP & $\mathcal{N}$(0.0,0.1) & -0.00071012$^{0.0001755394}_{0.0001566108}$ \\
sigma\_w\_WASP & ln$\mathcal{N}$(1e-06,1000000.0) & 3795.7667541617$^{157.9155887041}_{159.7255438737}$ \\
mdilution\_NITES & (0.916,0.1,0.0,1.0) & 0.9522658038$^{0.0284601533}_{0.0339996981}$ \\
mflux\_NITES & $\mathcal{N}$(0.0,0.1) & 0.0002372748$^{0.0002333442}_{0.0002209218}$ \\
sigma\_w\_NITES & ln$\mathcal{N}$(1e-06,1000000.0) & 4003.4902174354$^{167.0721043406}_{157.2582626918}$ \\
mdilution\_MUSCATg & (0.916,0.1,0.0,1.0) & 0.8917166293$^{0.0472077669}_{0.0459084796}$ \\
mflux\_MUSCATg & $\mathcal{N}$(0.0,0.1) & 0.0004511538$^{0.0004150193}_{0.0004068175}$ \\
sigma\_w\_MUSCATg & ln$\mathcal{N}$(1e-06,1000000.0) & 0.1097556831$^{20.5590174074}_{0.1096374254}$ \\
mdilution\_MUSCATr & (0.916,0.1,0.0,1.0) & 0.9407066261$^{0.0345003271}_{0.0417266182}$ \\
mflux\_MUSCATr & $\mathcal{N}$(0.0,0.1) & 8.43069e-05$^{0.0003020602}_{0.00030498}$ \\
sigma\_w\_MUSCATr & ln$\mathcal{N}$(1e-06,1000000.0) & 0.0058569285$^{2.7449341919}_{0.0058331897}$ \\
mdilution\_MUSCATz & (0.916,0.1,0.0,1.0) & 0.9186730411$^{0.0488050476}_{0.0571463159}$ \\
mflux\_MUSCATz & $\mathcal{N}$(0.0,0.1) & 0.0001265011$^{0.0005364691}_{0.0005114088}$ \\
sigma\_w\_MUSCATz & ln$\mathcal{N}$(1e-06,1000000.0) & 0.0938672872$^{24.7356671397}_{0.0937253338}$ \\
GP\_sigma\_TESS & ln$\mathcal{N}$(1e-06,1000000.0) & 0.0001894911$^{0.0004814822}_{0.0001462942}$ \\
GP\_rho\_TESS & ln$\mathcal{N}$(0.001,1000.0) & 96.5025320943$^{265.3190482537}_{81.2918704758}$ \\
K\_p1 & $\mathcal{U}$(100,200) & 114.2896329155$^{2.5023397902}_{2.5271992191}$ \\
sigma\_w\_SOPHIE & ln$\mathcal{N}$(0.001,100.0) & 0.1170253708$^{1.3259236328}_{0.1090850863}$ \\
mu\_SOPHIE & $\mathcal{U}$(-200,200) & -52.1663213506$^{1.83844793}_{1.8508810164}$ \\
         \hline
    \end{tabular}
    \label{tab:wasp155_fit}
\end{table*}

\begin{table*}[]
    \centering
    \caption{Transit fit parameters for WASP-188 (TOI-5190) }
    \begin{tabular}{l c c }
    \hline
    \hline
    Parameter & Prior & Fit Value  \\
    \hline
P\_p1 & $\mathcal{N}$(5.748917201934258,4.2e-05) & 5.7489161127$^{2.7438e-06}_{2.565e-06}$ \\
t0\_p1 & $\mathcal{N}$(2457033.120404399,0.0017694) & 2457033.121413108$^{0.0009945962}_{0.0009807926}$ \\
b\_p1 & $\mathcal{U}$(0.0,1.0) & 0.6074296137$^{0.0067745482}_{0.0063080088}$ \\
p\_p1 & $\mathcal{N}$(0.07768,0.0184) & 0.0742208863$^{0.0005318704}_{0.0005231851}$ \\
rho & $\mathcal{N}$(345.0557595147527,2.3418538432640594) & 345.9705219663$^{1.7141308865}_{1.6859888434}$ \\
q1\_TESS40\_TESS53\_TESS54\_WASP & $\mathcal{N}$(0.259,0.05) & 0.2289771238$^{0.0361105131}_{0.0428138282}$ \\
q2\_TESS40\_TESS53\_TESS54\_WASP & $\mathcal{N}$(0.368,0.05) & 0.3798064871$^{0.0387253331}_{0.036769472}$ \\
q1\_MUSCATg & $\mathcal{N}$(0.604,0.05) & 0.6285965117$^{0.038965377}_{0.0326317441}$ \\
q2\_MUSCATg & $\mathcal{N}$(0.394,0.05) & 0.3975796374$^{0.0414779106}_{0.0366971882}$ \\
q1\_MUSCATr & $\mathcal{N}$(0.369,0.05) & 0.3657404018$^{0.0370329641}_{0.0373043871}$ \\
q2\_MUSCATr & $\mathcal{N}$(0.378,0.05) & 0.3616491205$^{0.0390312054}_{0.0387238245}$ \\
q1\_KEPCAM\_MUSCATi & $\mathcal{N}$(0.260,0.05) & 0.2234622075$^{0.029883121}_{0.0334628777}$ \\
q2\_KEPCAM\_MUSCATi & $\mathcal{N}$(0.373,0.05) & 0.356876811$^{0.0321280885}_{0.0374079926}$ \\
q1\_MUSCATz & $\mathcal{N}$(0.191,0.05) & 0.2184447984$^{0.0345477192}_{0.0283029986}$ \\
q2\_MUSCATz & $\mathcal{N}$(0.352,0.05) & 0.3241213395$^{0.0313575845}_{0.0348325453}$ \\
mdilution\_TESS40 & $\mathcal{N}$(1.0,0.01) & 0.9974709266$^{0.0083103907}_{0.0086349227}$ \\
mflux\_TESS40 & $\mathcal{N}$(0.0,0.01) & 0.0009584689$^{0.00102669}_{0.0009626982}$ \\
sigma\_w\_TESS40 & ln$\mathcal{N}$(1e-06,1000.0) & 0.0397525854$^{5.1610893437}_{0.0396230588}$ \\
mdilution\_TESS53 & $\mathcal{N}$(1.0,0.01) & 1.0066620366$^{0.0071544809}_{0.0062952578}$ \\
mflux\_TESS53 & $\mathcal{N}$(0.0,0.01) & -3.20733e-05$^{7.01139e-05}_{7.18449e-05}$ \\
sigma\_w\_TESS53 & ln$\mathcal{N}$(1e-06,1000.0) & 0.1215819339$^{10.1837581907}_{0.1210353649}$ \\
mdilution\_TESS54 & $\mathcal{N}$(1.0,0.01) & 1.0055613144$^{0.0070207369}_{0.0066427755}$ \\
mflux\_TESS54 & $\mathcal{N}$(0.0,0.01) & -5.3104e-06$^{9.73785e-05}_{9.46533e-05}$ \\
sigma\_w\_TESS54 & ln$\mathcal{N}$(1e-06,1000.0) & 0.0548347402$^{7.348142699}_{0.0547076887}$ \\
mdilution\_WASP & $\mathcal{N}$(0.99212,0.01) & 0.9902592043$^{0.0084341726}_{0.0090946603}$ \\
mflux\_WASP & $\mathcal{N}$(0.0,0.01) & -0.0003496665$^{9.87948e-05}_{9.32022e-05}$ \\
sigma\_w\_WASP & ln$\mathcal{N}$(1e-06,1000.0) & 0.2865960746$^{12.4931084273}_{0.2840711348}$ \\
mdilution\_MUSCATg & $\mathcal{N}$(0.99212,0.01) & 0.9943829945$^{0.0081698098}_{0.007235579}$ \\
mflux\_MUSCATg & $\mathcal{N}$(0.0,0.01) & 0.0042490347$^{0.002035752}_{0.002314719}$ \\
sigma\_w\_MUSCATg & ln$\mathcal{N}$(1e-06,1000.0) & 5.0849005878$^{178.4726573916}_{5.0636190514}$ \\
mdilution\_MUSCATr & $\mathcal{N}$(0.99212,0.01) & 0.9928558307$^{0.0073993393}_{0.0074815637}$ \\
mflux\_MUSCATr & $\mathcal{N}$(0.0,0.01) & 0.0080319739$^{0.0016369839}_{0.0022585189}$ \\
sigma\_w\_MUSCATr & ln$\mathcal{N}$(1e-06,1000.0) & 333.7509116232$^{74.7193176193}_{72.0327384419}$ \\
mdilution\_MUSCATi & $\mathcal{N}$(0.99212,0.01) & 0.9920765135$^{0.0075220513}_{0.0083060513}$ \\
mflux\_MUSCATi & $\mathcal{N}$(0.0,0.01) & -0.0052653228$^{0.0023059586}_{0.0019112161}$ \\
sigma\_w\_MUSCATi & ln$\mathcal{N}$(1e-06,1000.0) & 6.0373621361$^{309.0030207724}_{6.0302843349}$ \\
mdilution\_KEPCAM & $\mathcal{N}$(0.99212,0.01) & 0.9982448831$^{0.0075705768}_{0.0063729514}$ \\
mflux\_KEPCAM & $\mathcal{N}$(0.0,0.01) & -0.0068925876$^{0.0006201709}_{0.0006109368}$ \\
sigma\_w\_KEPCAM & ln$\mathcal{N}$(1e-06,1000.0) & 826.2335892879$^{80.2729333717}_{76.3803828886}$ \\
mdilution\_MUSCATz & $\mathcal{N}$(0.99212,0.01) & 0.9894752157$^{0.0077055845}_{0.0080742892}$ \\
mflux\_MUSCATz & $\mathcal{N}$(0.0,0.01) & -0.0013354336$^{0.0027376358}_{0.002645851}$ \\
sigma\_w\_MUSCATz & ln$\mathcal{N}$(1e-06,1000.0) & 0.0438161923$^{45.9402539998}_{0.04377306}$ \\
GP\_rho\_TESS40 & ln$\mathcal{N}$(0.001,1000.0) & 1.9403926331$^{0.7570831994}_{0.4529567866}$ \\
GP\_rho\_TESS53 & ln$\mathcal{N}$(0.001,1000.0) & 0.3184509644$^{0.0777400865}_{0.0603454775}$ \\
GP\_rho\_TESS54 & ln$\mathcal{N}$(0.001,1000.0) & 0.4367461343$^{0.1245250315}_{0.0922206454}$ \\
GP\_sigma\_TESS40 & ln$\mathcal{N}$(1e-06,1000000.0) & 0.0028277147$^{0.00107437}_{0.0006314182}$ \\
GP\_sigma\_TESS53 & ln$\mathcal{N}$(1e-06,1000000.0) & 0.0003887932$^{4.95426e-05}_{4.12134e-05}$ \\
GP\_sigma\_TESS54 & ln$\mathcal{N}$(1e-06,1000000.0) & 0.0004737353$^{6.20709e-05}_{4.87706e-05}$ \\
         \hline
    \end{tabular}
    \label{tab:wasp188_fit}
\end{table*}

\begin{table*}[]
\centering
    \caption{Transit fit parameters for WASP-188 (TOI-5190) continued }
    \begin{tabular}{l c c }
    \hline
    \hline
    Parameter & Prior & Fit Value  \\
    \hline
theta0\_KEPCAM & $\mathcal{U}$(-100,100) & -0.0022403181$^{0.0005307109}_{0.0005204109}$ \\
theta0\_MUSCATg & $\mathcal{U}$(-100,100) & 0.0030174096$^{0.0018977572}_{0.0021617063}$ \\
theta0\_MUSCATr & $\mathcal{U}$(-100,100) & 0.0071347146$^{0.0015533885}_{0.0021241689}$ \\
theta0\_MUSCATi & $\mathcal{U}$(-100,100) & -0.0052740663$^{0.0021727221}_{0.0018236678}$ \\
theta0\_MUSCATz & $\mathcal{U}$(-100,100) & -0.0016313579$^{0.0025905155}_{0.0025133789}$ \\
K\_p1 & $\mathcal{N}$(130,10) & 124.634032159$^{5.8228927452}_{6.4888593133}$ \\
sigma\_w\_SOPHIE & ln$\mathcal{N}$(0.001,1000.0) & 0.516522383$^{5.436624353}_{0.4953261305}$ \\
mu\_SOPHIE & $\mathcal{U}$(-200,200) & -23.5118582181$^{9.9913427459}_{9.9009076157}$ \\
         \hline
    \end{tabular}
    \label{tab:wasp188_fit2}
\end{table*}

\begin{table*}[]
    \centering
    \caption{Transit fit parameters for WASP-194 (TOI-3791) }
    \begin{tabular}{l c c }
    \hline
    \hline
    Parameter & Prior & Fit Value  \\
    \hline
P\_p1 & $\mathcal{N}$(3.183388539182917,2.3e-06) & 3.1833874926$^{3.762e-07}_{4.588e-07}$ \\
t0\_p1 & $\mathcal{N}$(2457449.0508702304,0.1) & 2457449.051061772$^{0.0003043022}_{0.0002522403}$ \\
b\_p1 & $\mathcal{U}$(0.0,1.0) & 0.8429506623$^{0.0037330556}_{0.0040255196}$ \\
p\_p1 & $\mathcal{N}$(0.09002749,0.1) & 0.1007140291$^{0.0004226804}_{0.0004284749}$ \\
rho & $\mathcal{N}$(650.14169915,38.63056624) & 682.6560915578$^{27.1377403937}_{16.1001539218}$ \\
q1\_TESS40\_TESS41\_TESS50\_TESS54\_TESS55\_ & & \\
\hspace{5mm}TESS56\_TESS57\_TESS60\_WASP\_HAT & $\mathcal{N}$(0.246,0.05) & 0.25906315$^{0.0245295124}_{0.0240763093}$ \\
q2\_TESS40\_TESS41\_TESS50\_TESS54\_TESS55\_ & & \\
\hspace{5mm}TESS56\_TESS57\_TESS60\_WASP\_HAT & $\mathcal{N}$(0.370,0.05) & 0.38859783$^{0.0353646148}_{0.0315729309}$ \\
q1\_MUSCATg & $\mathcal{N}$(0.563,0.05) & 0.5535069329$^{0.0299199293}_{0.0352088849}$ \\
q2\_MUSCATg & $\mathcal{N}$(0.405,0.05) & 0.3809804629$^{0.0270038259}_{0.0335331991}$ \\
q1\_MUSCATr & $\mathcal{N}$(0.354,0.05) & 0.3291457088$^{0.023070262}_{0.0278952791}$ \\
q2\_MUSCATr & $\mathcal{N}$(0.383,0.05) & 0.3672481764$^{0.0343517867}_{0.0350506874}$ \\
q1\_MUSCATi\_KEPCAM & $\mathcal{N}$(0.252,0.05) & 0.288458254$^{0.0375505403}_{0.032009377}$ \\
q2\_MUSCATi\_KEPCAM & $\mathcal{N}$(0.374,0.05) & 0.3687287104$^{0.0339083947}_{0.0324903596}$ \\
q1\_MUSCATz & $\mathcal{N}$(0.186,0.05) & 0.1340368587$^{0.0351562437}_{0.0320914713}$ \\
q2\_MUSCATz & $\mathcal{N}$(0.362,0.05) & 0.3673985872$^{0.0374629459}_{0.0358287801}$ \\
q1\_RC & $\mathcal{N}$(0.246,0.05) & 0.2598394104$^{0.0344362427}_{0.0336325625}$ \\
q2\_RC & $\mathcal{N}$(0.370,0.05) & 0.4009241498$^{0.0336729195}_{0.0239114623}$ \\
mdilution\_TESS40 & $\mathcal{N}$(1.0,0.01) & 1.0115725656$^{0.0078273844}_{0.0060174802}$ \\
mflux\_TESS40 & $\mathcal{N}$(0.0,0.01) & -0.0001133939$^{4.11725e-05}_{5.81341e-05}$ \\
sigma\_w\_TESS40 & ln$\mathcal{N}$(1e-06,1000000.0) & 0.0013385746$^{0.1922723505}_{0.0013197798}$ \\
mdilution\_TESS41 & $\mathcal{N}$(1.0,0.01) & 1.0026542679$^{0.0048082138}_{0.0045947426}$ \\
mflux\_TESS41 & $\mathcal{N}$(0.0,0.01) & -7.29876e-05$^{6.30299e-05}_{5.76457e-05}$ \\
sigma\_w\_TESS41 & ln$\mathcal{N}$(1e-06,1000000.0) & 0.7911904358$^{12.7080418345}_{0.7761192793}$ \\
mdilution\_TESS50 & $\mathcal{N}$(1.0,0.01) & 1.0011513998$^{0.0060181408}_{0.0059329434}$ \\
mflux\_TESS50 & $\mathcal{N}$(0.0,0.01) & -0.0001035053$^{0.0001020113}_{0.0001026979}$ \\
sigma\_w\_TESS50 & ln$\mathcal{N}$(1e-06,1000000.0) & 0.0044277466$^{0.1580059802}_{0.0043465869}$ \\
mdilution\_TESS54 & $\mathcal{N}$(1.0,0.01) & 1.0052723389$^{0.0058172621}_{0.0062122071}$ \\
mflux\_TESS54 & $\mathcal{N}$(0.0,0.01) & -9.68801e-05$^{5.07361e-05}_{5.31442e-05}$ \\
sigma\_w\_TESS54 & ln$\mathcal{N}$(1e-06,1000000.0) & 0.0307983841$^{3.4609689272}_{0.0306273475}$ \\
mdilution\_TESS55 & $\mathcal{N}$(1.0,0.01) & 1.0044928793$^{0.0075414072}_{0.0054064807}$ \\
mflux\_TESS55 & $\mathcal{N}$(0.0,0.01) & -0.0001593391$^{8.55669e-05}_{9.84732e-05}$ \\
sigma\_w\_TESS55 & ln$\mathcal{N}$(1e-06,1000000.0) & 0.0066784527$^{1.1659062996}_{0.0066210453}$ \\
mdilution\_TESS56 & $\mathcal{N}$(1.0,0.01) & 1.0077518221$^{0.0067047839}_{0.0052136509}$ \\
mflux\_TESS56 & $\mathcal{N}$(0.0,0.01) & -0.0001396791$^{7.36097e-05}_{7.48166e-05}$ \\
sigma\_w\_TESS56 & ln$\mathcal{N}$(1e-06,1000000.0) & 0.0493973228$^{1.9651868246}_{0.047965496}$ \\
mdilution\_TESS57 & $\mathcal{N}$(1.0,0.01) & 0.9985743825$^{0.0070256899}_{0.0075867635}$ \\
mflux\_TESS57 & $\mathcal{N}$(0.0,0.01) & -4.13419e-05$^{0.0001176962}_{0.0001103043}$ \\
sigma\_w\_TESS57 & ln$\mathcal{N}$(1e-06,1000000.0) & 0.0448987182$^{2.8246069534}_{0.0444556648}$ \\
mdilution\_TESS60 & $\mathcal{N}$(1.0,0.01) & 1.0031811757$^{0.0066976857}_{0.006333771}$ \\
mflux\_TESS60 & $\mathcal{N}$(0.0,0.01) & -1.26398e-05$^{0.0001463868}_{0.0001725705}$ \\
sigma\_w\_TESS60 & ln$\mathcal{N}$(1e-06,1000000.0) & 0.018630976$^{1.4519330083}_{0.0182682889}$ \\
mdilution\_WASP & $\mathcal{N}$(1.0,0.01) & 0.9776611587$^{0.0071322531}_{0.0091926927}$ \\
mflux\_WASP & $\mathcal{N}$(0.0,0.01) & -7.64003e-05$^{3.79773e-05}_{3.94747e-05}$ \\
sigma\_w\_WASP & ln$\mathcal{N}$(1e-06,1000000.0) & 2981.7815927603$^{48.0120080129}_{44.2857633519}$ \\
mdilution\_HAT & $\mathcal{N}$(1.0,0.01) & 0.9752653977$^{0.010700023}_{0.0106573354}$ \\
mflux\_HAT & $\mathcal{N}$(0.0,0.01) & -0.0002944227$^{3.72699e-05}_{3.90598e-05}$ \\
         \hline
    \end{tabular}
    \label{tab:wasp194_fit}
\end{table*}

\begin{table*}[]
    \centering
    \caption{Transit fit parameters for WASP-194 (TOI-3791) continued }
    \begin{tabular}{l c c }
    \hline
    \hline
    Parameter & Prior & Fit Value   \\
    \hline
sigma\_w\_HAT & ln$\mathcal{N}$(1e-06,1000000.0) & 3785.983829801$^{40.3682173179}_{36.3910057287}$ \\
mdilution\_MUSCATg & $\mathcal{N}$(1.0,0.01) & 1.0039618325$^{0.0058207159}_{0.0059842346}$ \\
mflux\_MUSCATg & $\mathcal{N}$(0.0,0.01) & -0.0001228768$^{7.16935e-05}_{6.94372e-05}$ \\
sigma\_w\_MUSCATg & ln$\mathcal{N}$(1e-06,1000000.0) & 0.0059458624$^{2.2894163782}_{0.0059187072}$ \\
mdilution\_MUSCATr & $\mathcal{N}$(1.0,0.01) & 1.0035131349$^{0.0060243359}_{0.0065880126}$ \\
mflux\_MUSCATr & $\mathcal{N}$(0.0,0.01) & -0.0001079051$^{8.42942e-05}_{7.59203e-05}$ \\
sigma\_w\_MUSCATr & ln$\mathcal{N}$(1e-06,1000000.0) & 0.1661120465$^{13.1321583399}_{0.1641644405}$ \\
mdilution\_MUSCATi & $\mathcal{N}$(1.0,0.01) & 1.0030352669$^{0.0071457174}_{0.006420786}$ \\
mflux\_MUSCATi & $\mathcal{N}$(0.0,0.01) & 9.4407e-06$^{9.64724e-05}_{8.19043e-05}$ \\
sigma\_w\_MUSCATi & ln$\mathcal{N}$(1e-06,1000000.0) & 0.0020262409$^{0.2038179045}_{0.001989026}$ \\
mdilution\_MUSCATz & $\mathcal{N}$(1.0,0.01) & 1.0060283973$^{0.0064737596}_{0.0053299057}$ \\
mflux\_MUSCATz & $\mathcal{N}$(0.0,0.01) & 4.81993e-05$^{9.26552e-05}_{9.73896e-05}$ \\
sigma\_w\_MUSCATz & ln$\mathcal{N}$(1e-06,1000000.0) & 0.4564373337$^{20.3369950767}_{0.450609579}$ \\
mdilution\_KEPCAM & $\mathcal{N}$(1.0,0.01) & 0.9907603159$^{0.0052110136}_{0.0071972537}$ \\
mflux\_KEPCAM & $\mathcal{N}$(0.0,0.01) & -0.0038562468$^{7.9648e-05}_{8.89333e-05}$ \\
sigma\_w\_KEPCAM & ln$\mathcal{N}$(1e-06,1000000.0) & 1733.0238654106$^{63.0598012658}_{59.1012464557}$ \\
mdilution\_RC & $\mathcal{N}$(1.0,0.01) & 1.0005873457$^{0.0060525252}_{0.0063811364}$ \\
mflux\_RC & $\mathcal{N}$(0.0,0.01) & -0.004166384$^{0.0001903637}_{0.0002012398}$ \\
sigma\_w\_RC & ln$\mathcal{N}$(1e-06,1000000.0) & 1.2616273065$^{77.9104499899}_{1.2532169879}$ \\
GP\_sigma\_TESS40 & ln$\mathcal{N}$(1e-06,1000000.0) & 0.0003713836$^{3.80178e-05}_{3.32244e-05}$ \\
GP\_sigma\_TESS41 & ln$\mathcal{N}$(1e-06,1000000.0) & 0.0004036356$^{3.99019e-05}_{3.68657e-05}$ \\
GP\_sigma\_TESS50 & ln$\mathcal{N}$(1e-06,1000000.0) & 0.000694798$^{5.26686e-05}_{4.66942e-05}$ \\
GP\_sigma\_TESS54 & ln$\mathcal{N}$(1e-06,1000000.0) & 0.0003447457$^{3.45178e-05}_{3.07932e-05}$ \\
GP\_sigma\_TESS55 & ln$\mathcal{N}$(1e-06,1000000.0) & 0.0006414169$^{5.44758e-05}_{4.42095e-05}$ \\
GP\_sigma\_TESS56 & ln$\mathcal{N}$(1e-06,1000000.0) & 0.0005651272$^{3.34656e-05}_{2.82403e-05}$ \\
GP\_sigma\_TESS57 & ln$\mathcal{N}$(1e-06,1000000.0) & 0.0009772096$^{8.53691e-05}_{7.6696e-05}$ \\
GP\_sigma\_TESS60 & ln$\mathcal{N}$(1e-06,1000000.0) & 0.0009653446$^{9.08928e-05}_{7.38158e-05}$ \\
GP\_rho\_TESS40\_TESS41\_TESS50\_TESS54\_TESS55\_& & \\
\hspace{5mm}TESS56\_TESS57\_TESS60 & ln$\mathcal{N}$(0.001,1.0) & 0.3742177427$^{0.0273326698}_{0.020758208}$ \\
K\_p1 & $\mathcal{U}$(100,200) & 135.8978150862$^{13.4237902265}_{13.2638712553}$ \\
sigma\_w\_TRES & ln$\mathcal{N}$(0.001,100.0) & 42.5138069837$^{8.3080669683}_{7.5470359516}$ \\
mu\_TRES & $\mathcal{U}$(-200,200) & 11.585333702$^{9.9478278437}_{10.4815289283}$ \\
         \hline
    \end{tabular}
    \label{tab:wasp194_fit2}
\end{table*}

\begin{table*}[]
    \centering
    \caption{Transit fit parameters for WASP-195 (TOI-4056) }
    \begin{tabular}{l c c }
    \hline
    \hline
    Parameter & Prior & Fit Value   \\
    \hline
P\_p1 & $\mathcal{N}$(5.0519235985,0.0008289764) & 5.0519281663$^{3.5593e-06}_{3.9969e-06}$ \\
t0\_p1 & $\mathcal{N}$(2457357.2360845003,0.1) & 2457357.2385544833$^{0.0022035227}_{0.0018485114}$ \\
b\_p1 & $\mathcal{U}$(0.0,1.0) & 0.5526703443$^{0.0133935926}_{0.0140694981}$ \\
p\_p1 & $\mathcal{N}$(0.0593474131,0.1) & 0.0600184558$^{0.0017849997}_{0.0015720535}$ \\
rho & $\mathcal{N}$(466.4162218447,3.6014589812) & 466.2154235185$^{3.5497071467}_{3.3338643553}$ \\
q1\_TESS50\_TESS52\_WASP & $\mathcal{N}$(0.292,0.05) & 0.3070968875$^{0.0457653937}_{0.0457838902}$ \\
q2\_TESS50\_TESS52\_WASP & $\mathcal{N}$(0.375,0.05) & 0.3812066628$^{0.0472670431}_{0.0476334301}$ \\
q1\_WHITIN & $\mathcal{N}$(0.407,0.05) & 0.4052567739$^{0.0477646198}_{0.0448733205}$ \\
q2\_WHITIN & $\mathcal{N}$(0.386,0.05) & 0.400170146$^{0.0503461931}_{0.0484196575}$ \\
mdilution\_TESS50 & $\mathcal{N}$(1.0,0.1) & 0.9690138553$^{0.0578012441}_{0.0559783603}$ \\
mflux\_TESS50 & $\mathcal{N}$(0.0,0.1) & -0.0001092109$^{8.62658e-05}_{8.41074e-05}$ \\
sigma\_w\_TESS50 & ln$\mathcal{N}$(1e-06,1000000.0) & 0.0193919509$^{5.3345182257}_{0.019366495}$ \\
mdilution\_TESS52 & $\mathcal{N}$(1.0,0.1) & 0.9768484276$^{0.0555557618}_{0.0584797865}$ \\
mflux\_TESS52 & $\mathcal{N}$(0.0,0.1) & -0.0001431263$^{7.20808e-05}_{7.33578e-05}$ \\
sigma\_w\_TESS52 & ln$\mathcal{N}$(1e-06,1000000.0) & 647.2288091785$^{55.7974156471}_{58.7718198562}$ \\
mdilution\_WHITIN & $\mathcal{N}$(1.0,0.1) & 1.1077453377$^{0.0700572976}_{0.0669737588}$ \\
mflux\_WHITIN & $\mathcal{N}$(0.0,0.1) & 0.0011333281$^{0.0002151377}_{0.0001994051}$ \\
sigma\_w\_WHITIN & ln$\mathcal{N}$(1e-06,1000000.0) & 2141.4918250882$^{87.152651389}_{79.1006428919}$ \\
mdilution\_WASP & $\mathcal{N}$(1.0,0.1) & 0.862631045$^{0.0690907318}_{0.0671240383}$ \\
mflux\_WASP & $\mathcal{N}$(0.0,0.1) & -0.0003130898$^{5.37702e-05}_{5.91666e-05}$ \\
sigma\_w\_WASP & ln$\mathcal{N}$(1e-06,1000000.0) & 3210.1236515619$^{50.0704681741}_{52.3957566765}$ \\
GP\_sigma\_TESS50\_TESS52 & ln$\mathcal{N}$(1e-06,1000000.0) & 0.0003632071$^{3.68545e-05}_{3.18405e-05}$ \\
GP\_rho\_TESS50\_TESS52 & ln$\mathcal{N}$(0.001,1.0) & 0.4008553132$^{0.0981633925}_{0.0735554783}$ \\
K\_p1 & $\mathcal{U}$(0,100) & 10.3151027974$^{2.1587409776}_{2.190100253}$ \\
sigma\_w\_SOPHIE & ln$\mathcal{N}$(0.001,100.0) & 0.1188327729$^{2.0169439363}_{0.113923275}$ \\
mu\_SOPHIE & $\mathcal{U}$(-200,200) & 0.6263250229$^{1.7465653617}_{1.7538345359}$ \\
         \hline
    \end{tabular}
    \label{tab:wasp195_fit}
\end{table*}

\begin{table*}[]
    \centering
    \caption{Transit fit parameters for WASP-197 (TOI-5385) }
    \begin{tabular}{l c c }
    \hline
    \hline
    Parameter & Prior & Fit Value   \\
    \hline
P\_p1 & $\mathcal{N}$(5.1673975424,0.0001424990) & 5.1672282173$^{3.374e-06}_{3.0872e-06}$ \\
t0\_p1 & $\mathcal{N}$(2456885.10442669,0.00984619) & 2456885.1042765197$^{0.0016615191}_{0.0017282381}$ \\
b\_p1 & $\mathcal{U}$(0.0,1.0) & 0.4264762673$^{0.0205639081}_{0.0223141525}$ \\
p\_p1 & $\mathcal{N}$(0.06215864,0.00519078) & 0.062700973$^{0.000682823}_{0.0006896528}$ \\
rho & $\mathcal{N}$(203.52063911,2.61883175) & 203.6550035014$^{2.4926411968}_{2.4644417042}$ \\
q1\_TESS\_WASP & $\mathcal{N}$(0.299,0.1) & 0.3257858184$^{0.0777625008}_{0.0714184397}$ \\
q2\_TESS\_WASP & $\mathcal{N}$(0.386,0.1) & 0.3499384359$^{0.0855411108}_{0.0902647773}$ \\
mdilution\_TESS & $\mathcal{N}$(0.998,0.002) & 0.9984501233$^{0.0018980785}_{0.0019561675}$ \\
mflux\_TESS & $\mathcal{N}$(0.0,0.01) & -7.88686e-05$^{8.8607e-05}_{7.40073e-05}$ \\
sigma\_w\_TESS & ln$\mathcal{N}$(1e-06,1000.0) & 0.0139151907$^{7.2455937123}_{0.0138901541}$ \\
mdilution\_WASP & $\mathcal{N}$(0.998,0.002) & 0.9976577903$^{0.0019101858}_{0.0019059635}$ \\
mflux\_WASP & $\mathcal{N}$(0.0,0.01) & -0.0004532348$^{7.78511e-05}_{7.94655e-05}$ \\
sigma\_w\_WASP & ln$\mathcal{N}$(1e-06,1000.0) & 978.620196656$^{15.9397311334}_{35.1501727972}$ \\
GP\_sigma\_TESS & ln$\mathcal{N}$(1e-06,1000000.0) & 0.000230365$^{0.0001013371}_{4.53201e-05}$ \\
GP\_rho\_TESS & ln$\mathcal{N}$(0.001,1000.0) & 1.09614481$^{0.8389861418}_{0.4109940314}$ \\
K\_p1 & $\mathcal{N}$(122.953702,4.593494) & 121.3203297164$^{3.7030518462}_{3.5892211382}$ \\
sigma\_w\_SOPHIE & ln$\mathcal{N}$(0.001,100.0) & 0.1513259654$^{5.4327495333}_{0.1459950066}$ \\
mu\_SOPHIE & $\mathcal{U}$(-50,50) & 46.9693679127$^{2.2481617569}_{4.6622015493}$ \\
sigma\_w\_TRES & ln$\mathcal{N}$(0.001,100.0) & 8.1608091289$^{14.8214451441}_{8.1249628613}$ \\
mu\_TRES & $\mathcal{U}$(-50,50) & -41.696254999$^{5.8718991926}_{4.7864628847}$ \\
sigma\_w\_PARAS2 & ln$\mathcal{N}$(0.001,100.0) & 42.4235117966$^{17.1233814473}_{12.2313255668}$ \\
mu\_PARAS2 & $\mathcal{U}$(-50,50) & 10.5980154432$^{15.9283900112}_{14.9167009835}$ \\
         \hline
    \end{tabular}
    \label{tab:wasp197_fit}
\end{table*}

\clearpage

\section{High Resolution Imaging of Host Stars} \label{app:highres}
This appendix shows high resolution images used to identify nearby companion stars.

\begin{figure}[!hbt]
    \centering
    \includegraphics[width=.4\linewidth]{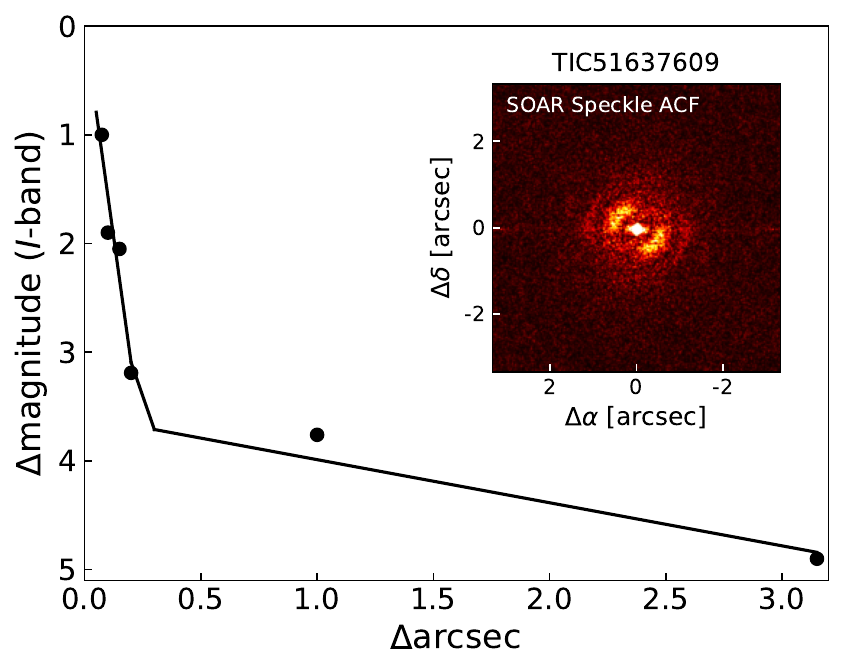}
    \caption{High Resolution images of WASP-102 taken by HRCam}
    \label{fig:CC_WASP102_HRCam}
\end{figure}

\begin{figure}[!htbp]
    \centering
    \begin{minipage}{0.4\linewidth}
        \centering
        \includegraphics[width=\linewidth]{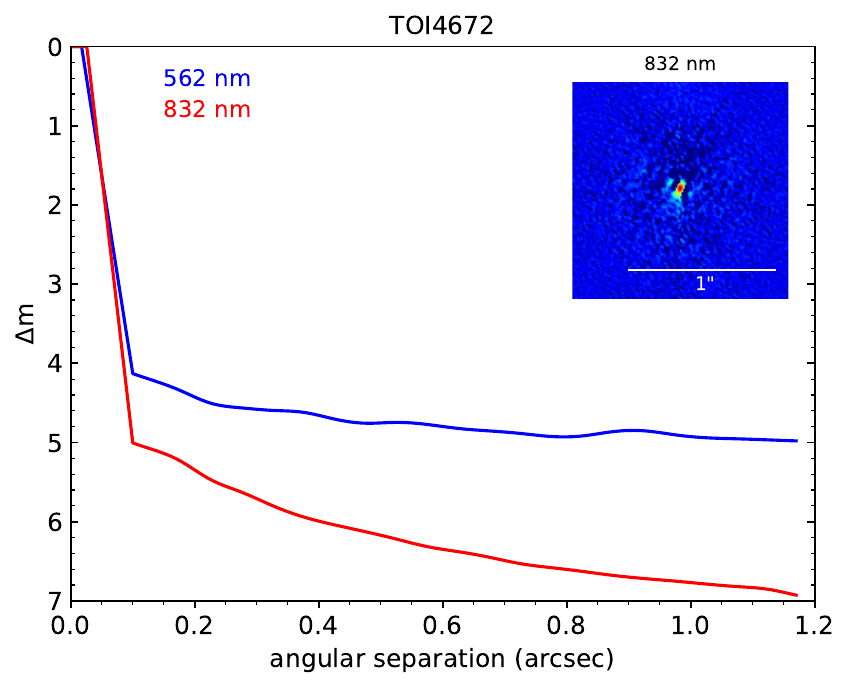}
    \end{minipage}\hfill
    \begin{minipage}{0.4\linewidth}
        \centering
        \includegraphics[width=\linewidth]{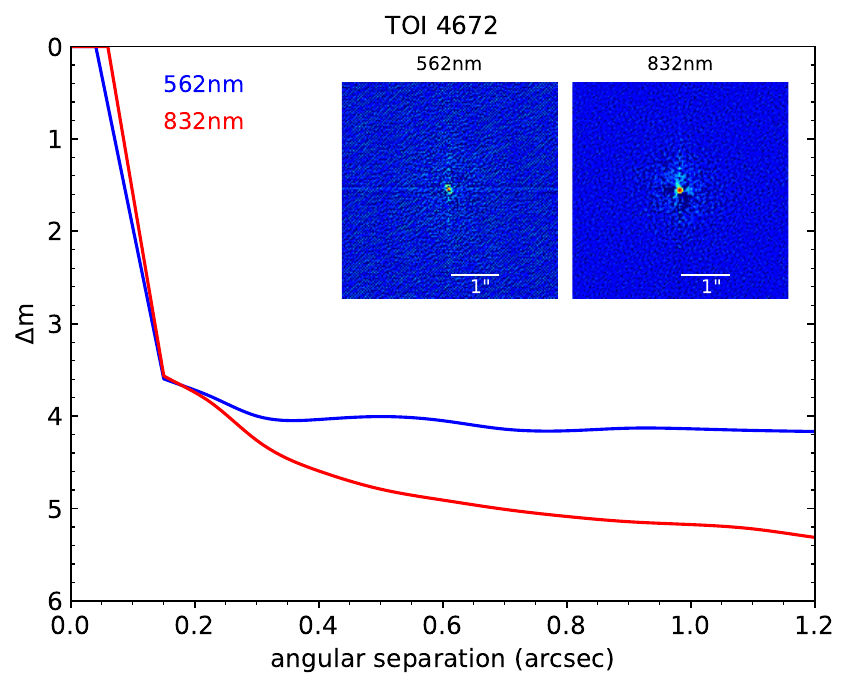}
    \end{minipage}
    \caption{High resolution images of WASP-116 taken by Zorro (left) and NESSI (right)}
    \label{fig:CC_WASP116}
\end{figure}

\begin{figure}[!htbp]
    \centering
    \includegraphics[width=.45\linewidth]{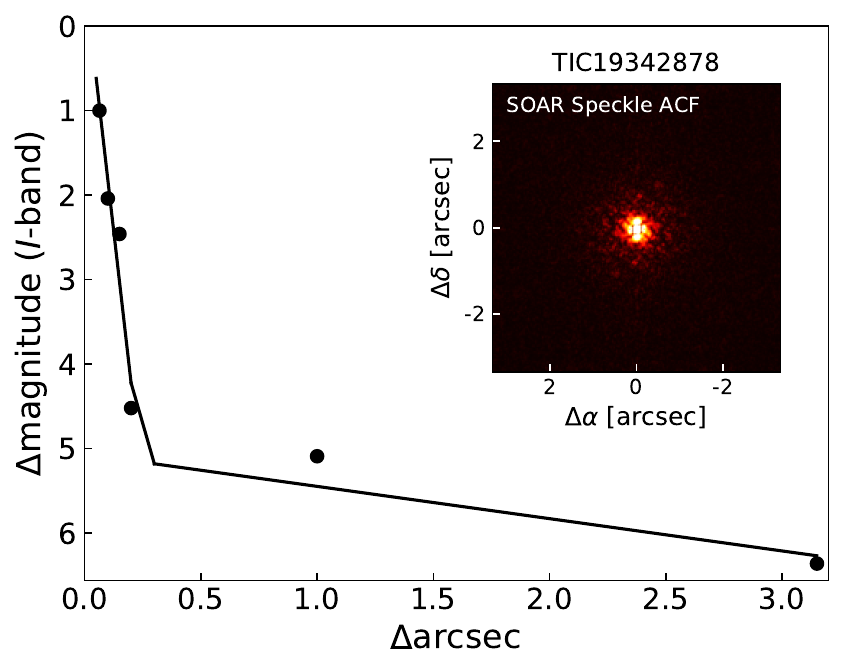}
    \caption{High Resolution images of WASP-149 taken by HRCam. }
    \label{fig:CC_WASP149_SOAR}
\end{figure}

\begin{figure}[!htbp]
    \centering
    \includegraphics[width=.45\linewidth]{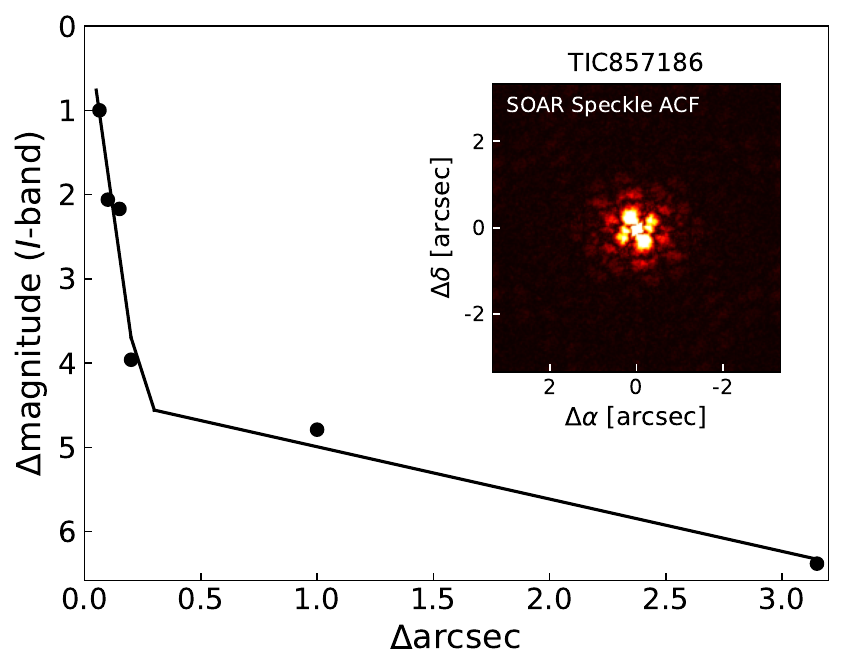}
    \caption{High Resolution images of WASP-154 taken by HRCam}
    \label{fig:CC_WASP154_HRCam}
\end{figure}

\begin{figure}[!htbp]
    \centering
    \begin{minipage}{0.45\linewidth}
        \centering
        \includegraphics[width=\linewidth]{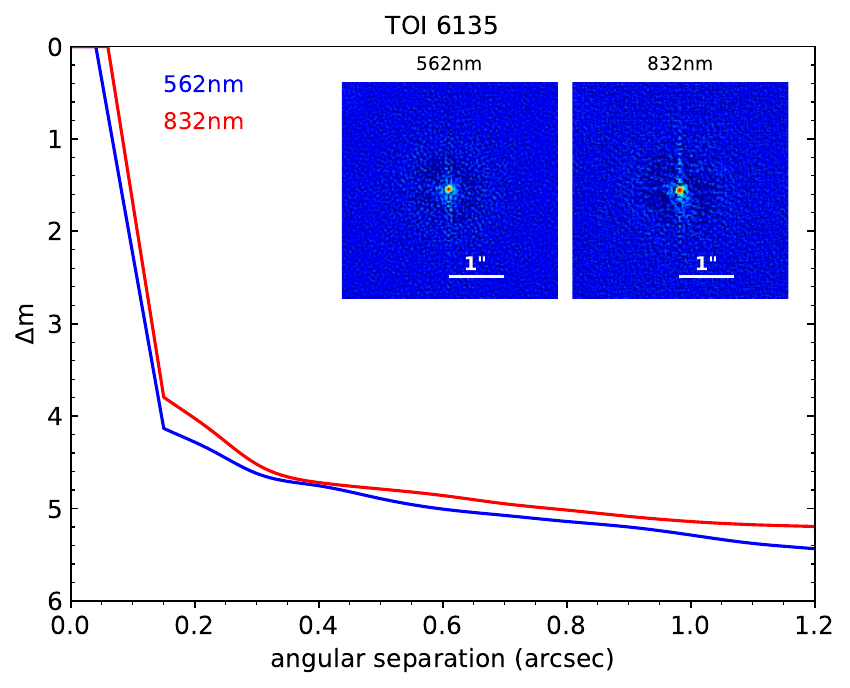}
    \end{minipage}\hfill
    \begin{minipage}{0.45\linewidth}
        \centering
        \includegraphics[width=\linewidth]{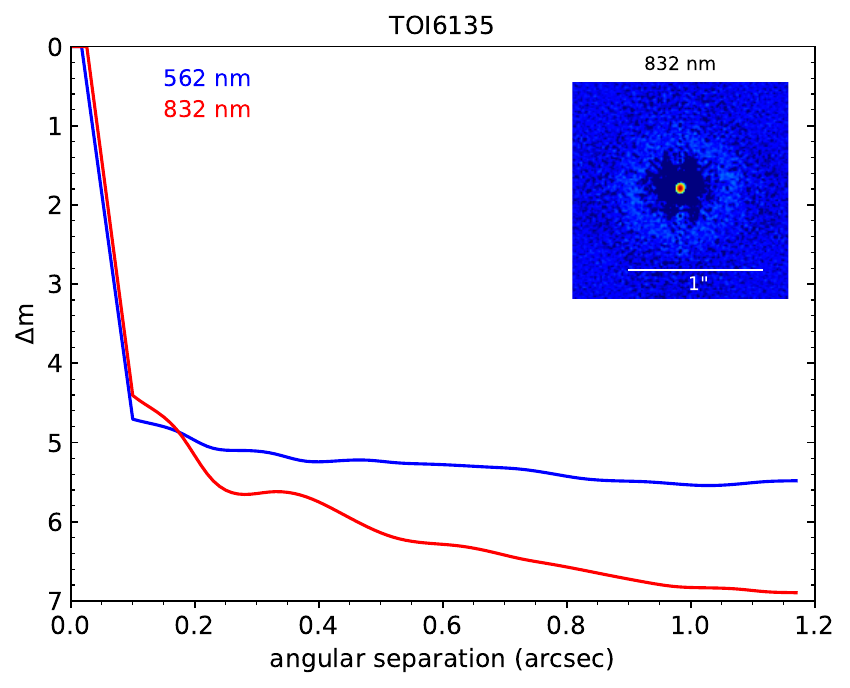}
    \end{minipage}
    \caption{High resolution images of WASP-155 taken by NESSI (left) and Alopeke (right)}
    \label{fig:CC_WASP155}
\end{figure}

\begin{figure}[!htbp]
    \centering
    \includegraphics[width=.45\linewidth]{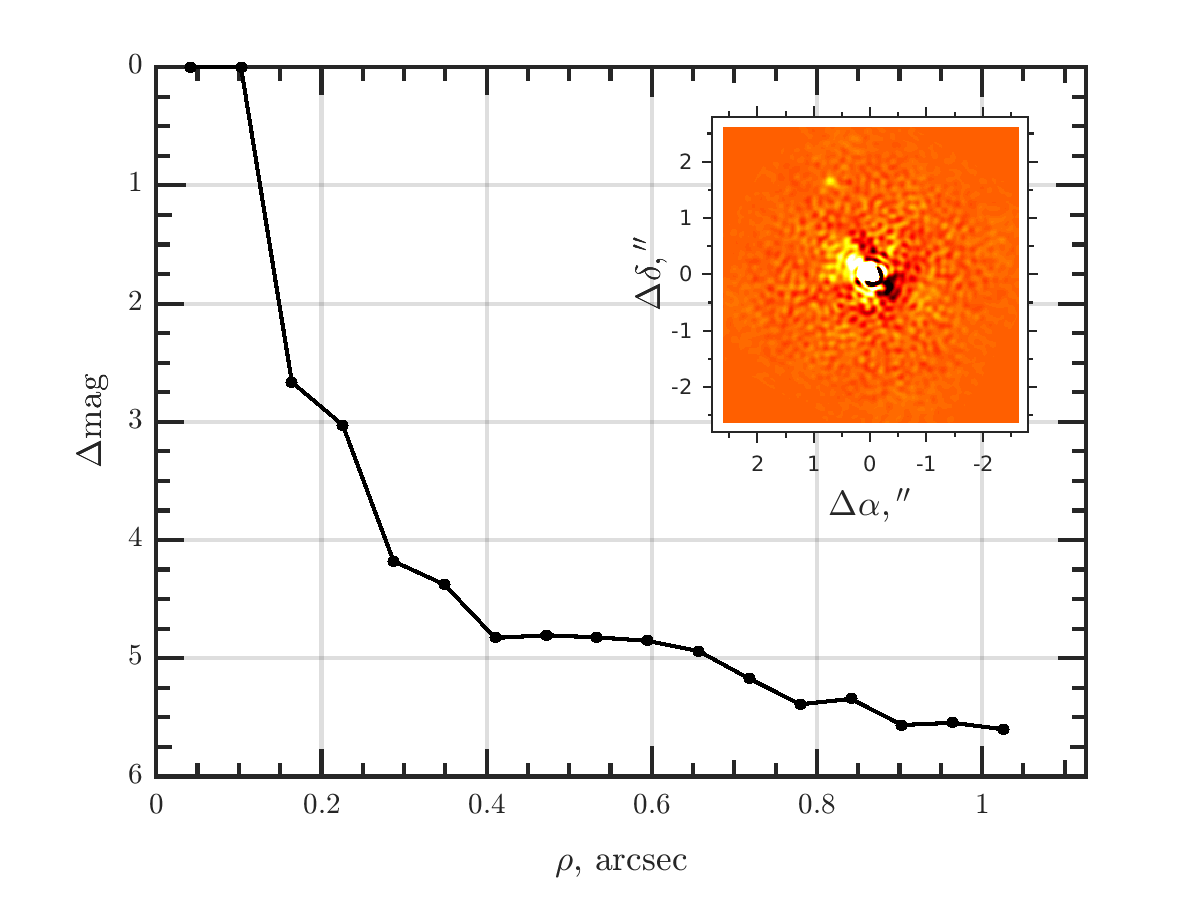}
    \caption{High Resolution images of WASP-188 taken by SAI}
    \label{fig:CC_WASP188_SAI}
\end{figure}

\begin{figure}[!htbp]
    \centering
    \includegraphics[width=.45\linewidth]{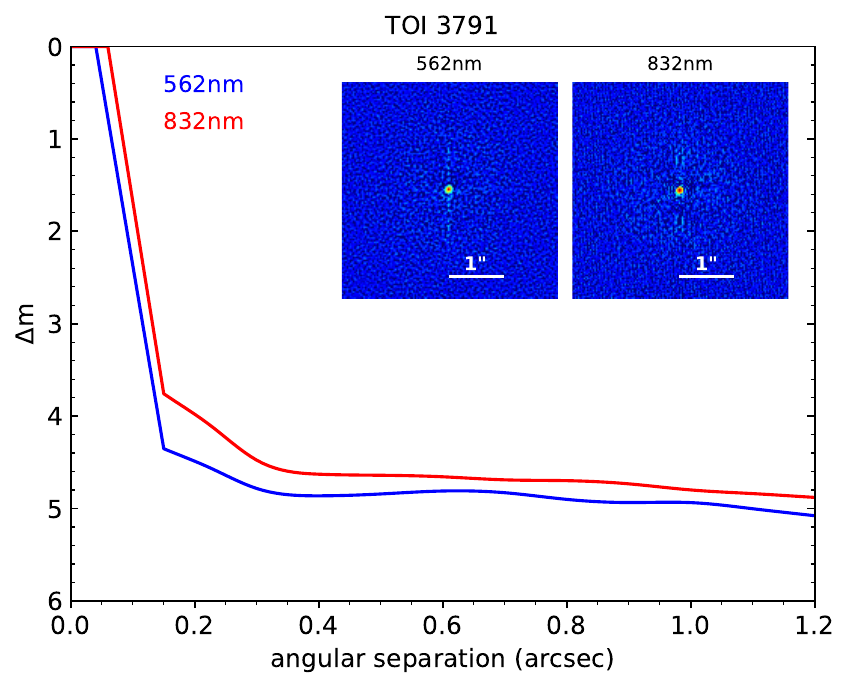}
    \caption{High Resolution images of WASP-194 taken by NESSI at 562 and 832~nm. }
    \label{fig:TOI3791_NESSI}
\end{figure}

\begin{figure}[!htbp]
    \centering
    \begin{minipage}{0.45\linewidth}
        \centering
        \includegraphics[width=\linewidth]{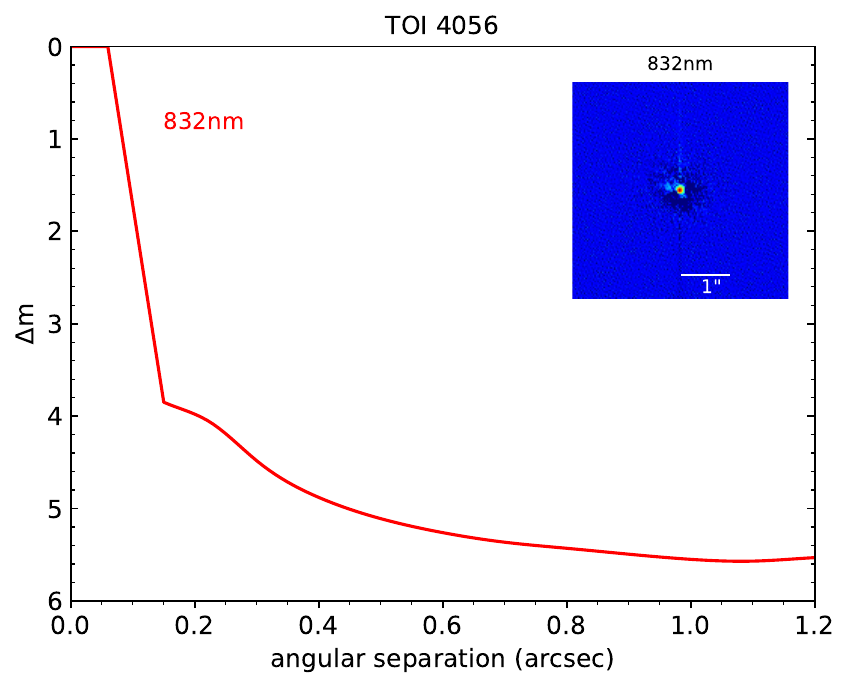}
    \end{minipage}\hfill
    \begin{minipage}{0.45\linewidth}
        \centering
        \includegraphics[width=\linewidth]{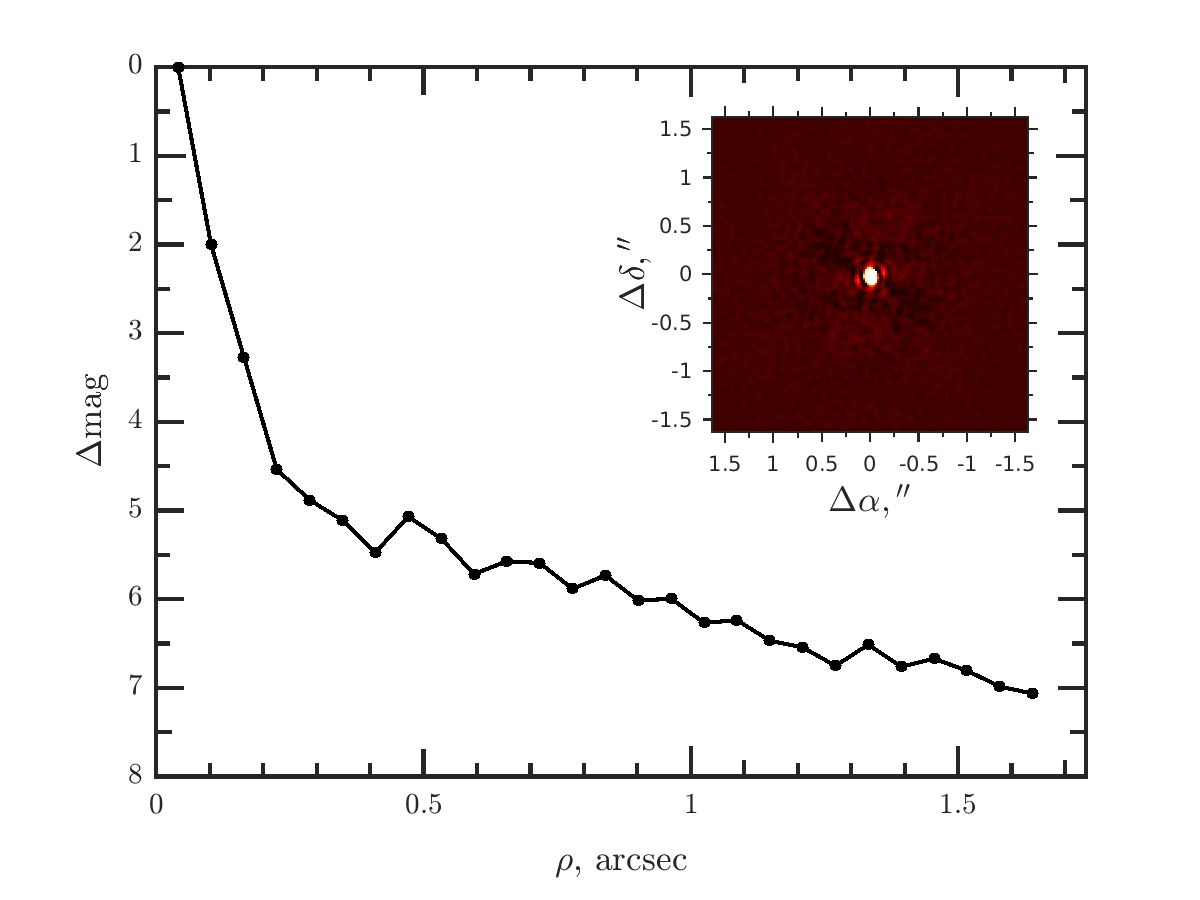}
    \end{minipage}
    \caption{High Resolution images of WASP-195 taken by NESSI (left) and SAI (right)}
    \label{fig:CC_WASP195}
\end{figure}

\begin{figure}[!htbp]
    \centering
    \begin{minipage}{0.45\linewidth}
        \centering
        \includegraphics[width=\linewidth]{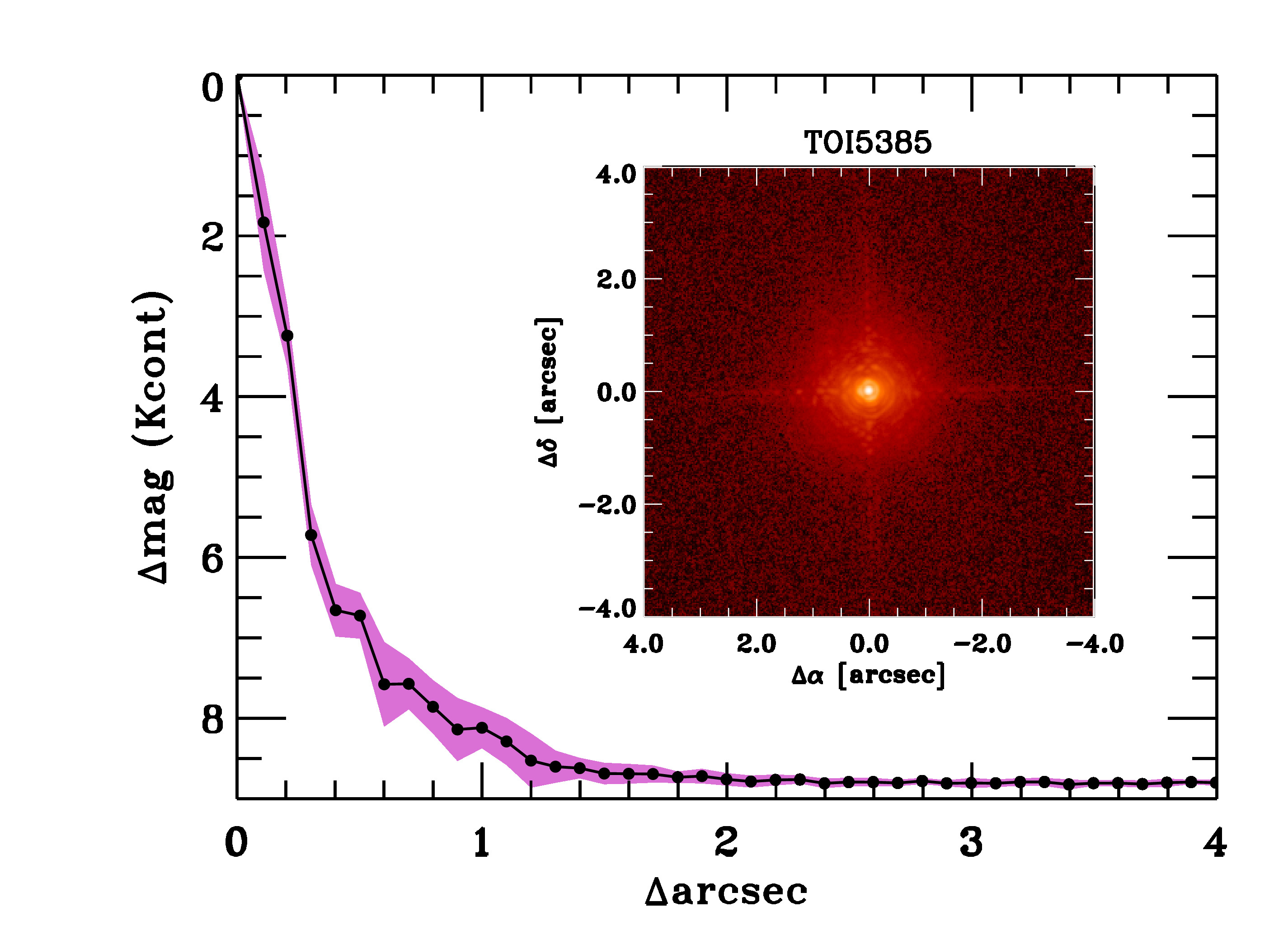}
    \end{minipage}\hfill
    \begin{minipage}{0.45\linewidth}
        \centering
        \includegraphics[width=\linewidth]{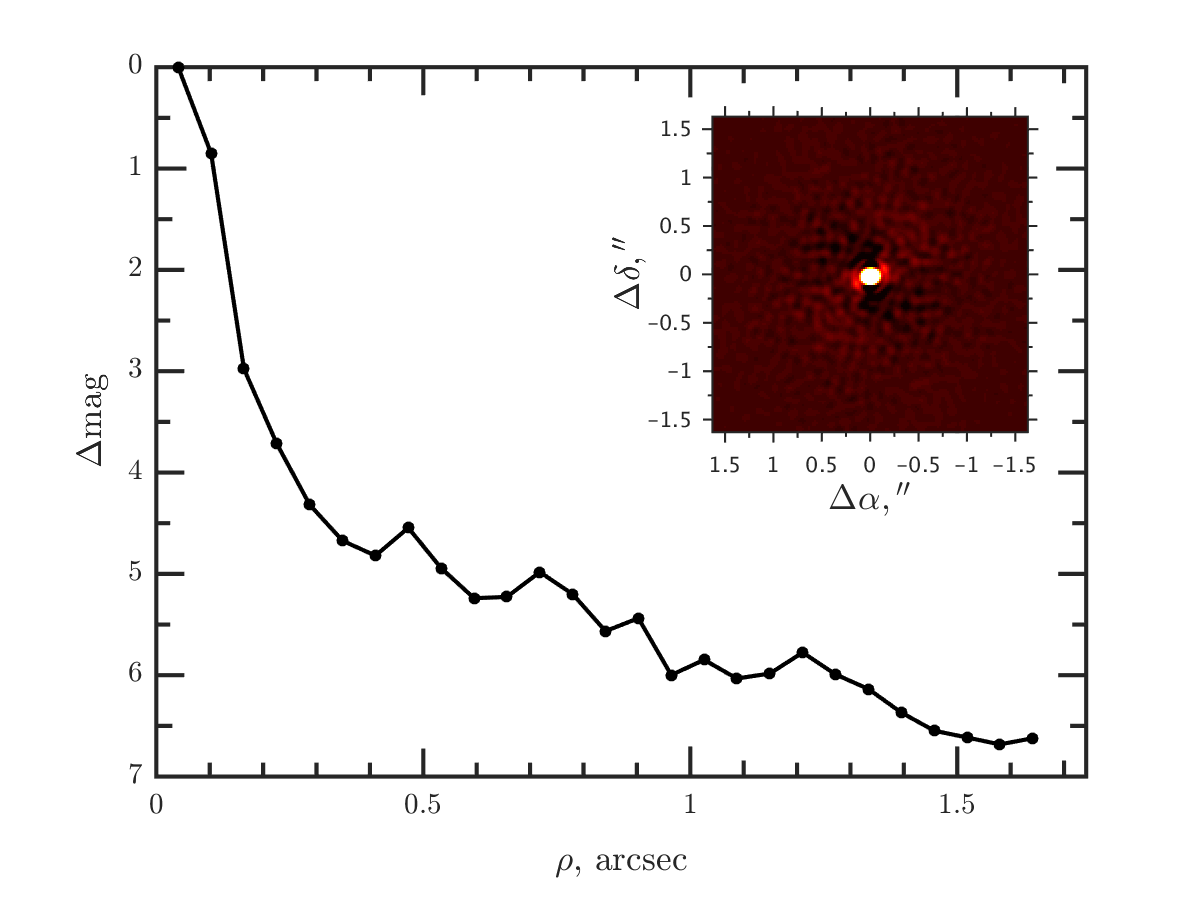}
    \end{minipage}
    \caption{High Resolution images of WASP-197 taken by PHARO (left) and SAI (right)}
    \label{fig:CC_WASP197}
\end{figure}

\end{appendix}

\end{document}